\documentclass[traditabstract,longauth]{aa}
\usepackage{txfonts}
\usepackage{graphicx}
\usepackage[nonamebreak]{natbib}
\usepackage{color}
\usepackage{lscape}
\usepackage{fixltx2e}
\usepackage{ifthen}
\usepackage[breaklinks,colorlinks,citecolor=blue]{hyperref}
\usepackage{afterpage}
\usepackage{verbatim}
\usepackage[usenames,dvipsnames,svgnames,table]{xcolor}
\usepackage{siunitx} % For handling units
\usepackage{soul}
\usepackage[normalem]{ulem}
\usepackage{csquotes}
\usepackage[yyyymmdd,hhmmss]{datetime}
\usepackage{hyperref}
\bibpunct{(}{)}{;}{a}{}{,} %
\def\setsymbol#1#2{\expandafter\def\csname #1\endcsname{#2}}
\def\getsymbol#1{\csname #1\endcsname}

\def\Planck{\textit{Planck}}

\def\all2013resultspapers{\nocite{planck2013-p01, planck2013-p02, planck2013-p02a, planck2013-p02d, planck2013-p02b, planck2013-p03, planck2013-p03c, planck2013-p03f, planck2013-p03d, planck2013-p03e, planck2013-p01a, planck2013-p06, planck2013-p03a, planck2013-pip88, planck2013-p08, planck2013-p11, planck2013-p12, planck2013-p13, planck2013-p14, planck2013-p15, planck2013-p05b, planck2013-p17, planck2013-p09, planck2013-p09a, planck2013-p20, planck2013-p19, planck2013-pipaberration, planck2013-p05, planck2013-p05a, planck2013-pip56, planck2013-p06b}}

\newbox\tablebox    \newdimen\tablewidth
\def\leaderfil{\leaders\hbox to 5pt{\hss.\hss}\hfil}
\def\endPlancktable{\tablewidth=\columnwidth 
    $$\hss\copy\tablebox\hss$$
    \vskip-\lastskip\vskip -2pt}
\def\endPlancktablewide{\tablewidth=\textwidth 
    $$\hss\copy\tablebox\hss$$
    \vskip-\lastskip\vskip -2pt}
\def\tablenote#1 #2\par{\begingroup \parindent=0.8em
    \abovedisplayshortskip=0pt\belowdisplayshortskip=0pt
    \noindent
    $$\hss\vbox{\hsize\tablewidth \hangindent=\parindent \hangafter=1 \noindent
    \hbox to \parindent{$^#1$\hss}\strut#2\strut\par}\hss$$
    \endgroup}
\def\doubleline{\vskip 3pt\hrule \vskip 1.5pt \hrule \vskip 5pt}

\def\L2{\ifmmode L_2\else $L_2$\fi}

\def\DeltaT{\ifmmode \Delta T\else $\Delta T$\fi}
\def\deltat{\ifmmode \Delta t\else $\Delta t$\fi}
\def\fknee{\ifmmode f_{\rm knee}\else $f_{\rm knee}$\fi}
\def\Fmax{\ifmmode F_{\rm max}\else $F_{\rm max}$\fi}
\def\solar{\ifmmode{\rm M}_{\mathord\odot}\else${\rm M}_{\mathord\odot}$\fi}
\def\Msolar{\ifmmode{\rm M}_{\mathord\odot}\else${\rm M}_{\mathord\odot}$\fi}
\def\Lsolar{\ifmmode{\rm L}_{\mathord\odot}\else${\rm L}_{\mathord\odot}$\fi}

\def\inv{\ifmmode^{-1}\else$^{-1}$\fi}
\def\mo{\ifmmode^{-1}\else$^{-1}$\fi}
\def\sup#1{\ifmmode ^{\rm #1}\else $^{\rm #1}$\fi}
\def\expo#1{\ifmmode \times 10^{#1}\else $\times 10^{#1}$\fi}
\def\,{\thinspace}
\def\lsim{\mathrel{\raise .4ex\hbox{\rlap{$<$}\lower 1.2ex\hbox{$\sim$}}}}
\def\gsim{\mathrel{\raise .4ex\hbox{\rlap{$>$}\lower 1.2ex\hbox{$\sim$}}}}

\def\simprop{\mathrel{\raise .4ex\hbox{\rlap{$\propto$}\lower 1.2ex\hbox{$\sim$}}}}
\def\deg{\ifmmode^\circ\else$^\circ$\fi}
\def\pdeg{\ifmmode $\setbox0=\hbox{$^{\circ}$}\rlap{\hskip.11\wd0 .}$^{\circ}
          \else \setbox0=\hbox{$^{\circ}$}\rlap{\hskip.11\wd0 .}$^{\circ}$\fi}
\def\arcs{\ifmmode {^{\scriptstyle\prime\prime}}
          \else $^{\scriptstyle\prime\prime}$\fi}
\def\arcm{\ifmmode {^{\scriptstyle\prime}}
          \else $^{\scriptstyle\prime}$\fi}
\newdimen\sa  \newdimen\sb
\def\parcs{\sa=.07em \sb=.03em
     \ifmmode \hbox{\rlap{.}}^{\scriptstyle\prime\kern -\sb\prime}\hbox{\kern -\sa}
     \else \rlap{.}$^{\scriptstyle\prime\kern -\sb\prime}$\kern -\sa\fi}
\def\parcm{\sa=.08em \sb=.03em
     \ifmmode \hbox{\rlap{.}\kern\sa}^{\scriptstyle\prime}\hbox{\kern-\sb}
     \else \rlap{.}\kern\sa$^{\scriptstyle\prime}$\kern-\sb\fi}
\def\ra[#1 #2 #3.#4]{#1\sup{h}#2\sup{m}#3\sup{s}\llap.#4}
\def\dec[#1 #2 #3.#4]{#1\deg#2\arcm#3\arcs\llap.#4}
\def\deco[#1 #2 #3]{#1\deg#2\arcm#3\arcs}
\def\rra[#1 #2]{#1\sup{h}#2\sup{m}}

\def\dots{\relax\ifmmode \ldots\else $\ldots$\fi}
\def\WHzsr{\ifmmode $W\,Hz\mo\,sr\mo$\else W\,Hz\mo\,sr\mo\fi}
\def\mHz{\ifmmode $\,mHz$\else \,mHz\fi}
\def\GHz{\ifmmode $\,GHz$\else \,GHz\fi}
\def\mKs{\ifmmode $\,mK\,s$^{1/2}\else \,mK\,s$^{1/2}$\fi}
\def\muKs{\ifmmode \,\mu$K\,s$^{1/2}\else \,$\mu$K\,s$^{1/2}$\fi}
\def\muKRJs{\ifmmode \,\mu$K$_{\rm RJ}$\,s$^{1/2}\else \,$\mu$K$_{\rm RJ}$\,s$^{1/2}$\fi}
\def\muKHz{\ifmmode \,\mu$K\,Hz$^{-1/2}\else \,$\mu$K\,Hz$^{-1/2}$\fi}
\def\MJysr{\ifmmode \,$MJy\,sr\mo$\else \,MJy\,sr\mo\fi}
\def\MJysrmK{\ifmmode \,$MJy\,sr\mo$\,mK$_{\rm CMB}\mo\else \,MJy\,sr\mo\,mK$_{\rm CMB}\mo$\fi}
\def\microns{\ifmmode \,\mu$m$\else \,$\mu$m\fi}

\def\muK{\ifmmode \,\mu$K$\else \,$\mu$\hbox{K}\fi}
\def\microK{\ifmmode \,\mu$K$\else \,$\mu$\hbox{K}\fi}
\def\muW{\ifmmode \,\mu$W$\else \,$\mu$\hbox{W}\fi}
\def\kms{\ifmmode $\,km\,s$^{-1}\else \,km\,s$^{-1}$\fi}
\def\kmsMpc{\ifmmode $\,\kms\,Mpc\mo$\else \,\kms\,Mpc\mo\fi}
\setsymbol{LFI:center:frequency:70GHz:units}{70.3\,GHz}
\setsymbol{LFI:center:frequency:44GHz:units}{44.1\,GHz}
\setsymbol{LFI:center:frequency:30GHz:units}{28.5\,GHz}

\setsymbol{LFI:center:frequency:70GHz}{70.3}
\setsymbol{LFI:center:frequency:44GHz}{44.1}
\setsymbol{LFI:center:frequency:30GHz}{28.5}

\setsymbol{LFI:center:frequency:LFI18:Rad:M:units}{71.7\GHz}
\setsymbol{LFI:center:frequency:LFI19:Rad:M:units}{67.5\GHz}
\setsymbol{LFI:center:frequency:LFI20:Rad:M:units}{69.2\GHz}
\setsymbol{LFI:center:frequency:LFI21:Rad:M:units}{70.4\GHz}
\setsymbol{LFI:center:frequency:LFI22:Rad:M:units}{71.5\GHz}
\setsymbol{LFI:center:frequency:LFI23:Rad:M:units}{70.8\GHz}
\setsymbol{LFI:center:frequency:LFI24:Rad:M:units}{44.4\GHz}
\setsymbol{LFI:center:frequency:LFI25:Rad:M:units}{44.0\GHz}
\setsymbol{LFI:center:frequency:LFI26:Rad:M:units}{43.9\GHz}
\setsymbol{LFI:center:frequency:LFI27:Rad:M:units}{28.3\GHz}
\setsymbol{LFI:center:frequency:LFI28:Rad:M:units}{28.8\GHz}
\setsymbol{LFI:center:frequency:LFI18:Rad:S:units}{70.1\GHz}
\setsymbol{LFI:center:frequency:LFI19:Rad:S:units}{69.6\GHz}
\setsymbol{LFI:center:frequency:LFI20:Rad:S:units}{69.5\GHz}
\setsymbol{LFI:center:frequency:LFI21:Rad:S:units}{69.5\GHz}
\setsymbol{LFI:center:frequency:LFI22:Rad:S:units}{72.8\GHz}
\setsymbol{LFI:center:frequency:LFI23:Rad:S:units}{71.3\GHz}
\setsymbol{LFI:center:frequency:LFI24:Rad:S:units}{44.1\GHz}
\setsymbol{LFI:center:frequency:LFI25:Rad:S:units}{44.1\GHz}
\setsymbol{LFI:center:frequency:LFI26:Rad:S:units}{44.1\GHz}
\setsymbol{LFI:center:frequency:LFI27:Rad:S:units}{28.5\GHz}
\setsymbol{LFI:center:frequency:LFI28:Rad:S:units}{28.2\GHz}

\setsymbol{LFI:center:frequency:LFI18:Rad:M}{71.7}
\setsymbol{LFI:center:frequency:LFI19:Rad:M}{67.5}
\setsymbol{LFI:center:frequency:LFI20:Rad:M}{69.2}
\setsymbol{LFI:center:frequency:LFI21:Rad:M}{70.4}
\setsymbol{LFI:center:frequency:LFI22:Rad:M}{71.5}
\setsymbol{LFI:center:frequency:LFI23:Rad:M}{70.8}
\setsymbol{LFI:center:frequency:LFI24:Rad:M}{44.4}
\setsymbol{LFI:center:frequency:LFI25:Rad:M}{44.0}
\setsymbol{LFI:center:frequency:LFI26:Rad:M}{43.9}
\setsymbol{LFI:center:frequency:LFI27:Rad:M}{28.3}
\setsymbol{LFI:center:frequency:LFI28:Rad:M}{28.8}
\setsymbol{LFI:center:frequency:LFI18:Rad:S}{70.1}
\setsymbol{LFI:center:frequency:LFI19:Rad:S}{69.6}
\setsymbol{LFI:center:frequency:LFI20:Rad:S}{69.5}
\setsymbol{LFI:center:frequency:LFI21:Rad:S}{69.5}
\setsymbol{LFI:center:frequency:LFI22:Rad:S}{72.8}
\setsymbol{LFI:center:frequency:LFI23:Rad:S}{71.3}
\setsymbol{LFI:center:frequency:LFI24:Rad:S}{44.1}
\setsymbol{LFI:center:frequency:LFI25:Rad:S}{44.1}
\setsymbol{LFI:center:frequency:LFI26:Rad:S}{44.1}
\setsymbol{LFI:center:frequency:LFI27:Rad:S}{28.5}
\setsymbol{LFI:center:frequency:LFI28:Rad:S}{28.2}

\setsymbol{LFI:white:noise:sensitivity:70GHz:units}{134.7\muKs}
\setsymbol{LFI:white:noise:sensitivity:44GHz:units}{164.7\muKs}
\setsymbol{LFI:white:noise:sensitivity:30GHz:units}{143.4\muKs}

\setsymbol{LFI:white:noise:sensitivity:70GHz}{134.7}
\setsymbol{LFI:white:noise:sensitivity:44GHz}{164.7}
\setsymbol{LFI:white:noise:sensitivity:30GHz}{143.4}

\setsymbol{LFI:white:noise:sensitivity:LFI18:Rad:M:units}{512.0\muKs}
\setsymbol{LFI:white:noise:sensitivity:LFI19:Rad:M:units}{581.4\muKs}
\setsymbol{LFI:white:noise:sensitivity:LFI20:Rad:M:units}{590.8\muKs}
\setsymbol{LFI:white:noise:sensitivity:LFI21:Rad:M:units}{455.2\muKs}
\setsymbol{LFI:white:noise:sensitivity:LFI22:Rad:M:units}{492.0\muKs}
\setsymbol{LFI:white:noise:sensitivity:LFI23:Rad:M:units}{507.7\muKs}
\setsymbol{LFI:white:noise:sensitivity:LFI24:Rad:M:units}{462.2\muKs}
\setsymbol{LFI:white:noise:sensitivity:LFI25:Rad:M:units}{413.6\muKs}
\setsymbol{LFI:white:noise:sensitivity:LFI26:Rad:M:units}{478.6\muKs}
\setsymbol{LFI:white:noise:sensitivity:LFI27:Rad:M:units}{277.7\muKs}
\setsymbol{LFI:white:noise:sensitivity:LFI28:Rad:M:units}{312.3\muKs}
\setsymbol{LFI:white:noise:sensitivity:LFI18:Rad:S:units}{465.7\muKs}
\setsymbol{LFI:white:noise:sensitivity:LFI19:Rad:S:units}{555.6\muKs}
\setsymbol{LFI:white:noise:sensitivity:LFI20:Rad:S:units}{623.2\muKs}
\setsymbol{LFI:white:noise:sensitivity:LFI21:Rad:S:units}{564.1\muKs}
\setsymbol{LFI:white:noise:sensitivity:LFI22:Rad:S:units}{534.4\muKs}
\setsymbol{LFI:white:noise:sensitivity:LFI23:Rad:S:units}{542.4\muKs}
\setsymbol{LFI:white:noise:sensitivity:LFI24:Rad:S:units}{399.2\muKs}
\setsymbol{LFI:white:noise:sensitivity:LFI25:Rad:S:units}{392.6\muKs}
\setsymbol{LFI:white:noise:sensitivity:LFI26:Rad:S:units}{418.6\muKs}
\setsymbol{LFI:white:noise:sensitivity:LFI27:Rad:S:units}{302.9\muKs}
\setsymbol{LFI:white:noise:sensitivity:LFI28:Rad:S:units}{285.3\muKs}

\setsymbol{LFI:white:noise:sensitivity:LFI18:Rad:M}{512.0}
\setsymbol{LFI:white:noise:sensitivity:LFI19:Rad:M}{581.4}
\setsymbol{LFI:white:noise:sensitivity:LFI20:Rad:M}{590.8}
\setsymbol{LFI:white:noise:sensitivity:LFI21:Rad:M}{455.2}
\setsymbol{LFI:white:noise:sensitivity:LFI22:Rad:M}{492.0}
\setsymbol{LFI:white:noise:sensitivity:LFI23:Rad:M}{507.7}
\setsymbol{LFI:white:noise:sensitivity:LFI24:Rad:M}{462.2}
\setsymbol{LFI:white:noise:sensitivity:LFI25:Rad:M}{413.6}
\setsymbol{LFI:white:noise:sensitivity:LFI26:Rad:M}{478.6}
\setsymbol{LFI:white:noise:sensitivity:LFI27:Rad:M}{277.7}
\setsymbol{LFI:white:noise:sensitivity:LFI28:Rad:M}{312.3}
\setsymbol{LFI:white:noise:sensitivity:LFI18:Rad:S}{465.7}
\setsymbol{LFI:white:noise:sensitivity:LFI19:Rad:S}{555.6}
\setsymbol{LFI:white:noise:sensitivity:LFI20:Rad:S}{623.2}
\setsymbol{LFI:white:noise:sensitivity:LFI21:Rad:S}{564.1}
\setsymbol{LFI:white:noise:sensitivity:LFI22:Rad:S}{534.4}
\setsymbol{LFI:white:noise:sensitivity:LFI23:Rad:S}{542.4}
\setsymbol{LFI:white:noise:sensitivity:LFI24:Rad:S}{399.2}
\setsymbol{LFI:white:noise:sensitivity:LFI25:Rad:S}{392.6}
\setsymbol{LFI:white:noise:sensitivity:LFI26:Rad:S}{418.6}
\setsymbol{LFI:white:noise:sensitivity:LFI27:Rad:S}{302.9}
\setsymbol{LFI:white:noise:sensitivity:LFI28:Rad:S}{285.3}

\setsymbol{LFI:knee:frequency:70GHz:units}{29.5\mHz}
\setsymbol{LFI:knee:frequency:44GHz:units}{56.2\mHz}
\setsymbol{LFI:knee:frequency:30GHz:units}{113.7\mHz}

\setsymbol{LFI:knee:frequency:70GHz}{29.5}
\setsymbol{LFI:knee:frequency:44GHz}{56.2}
\setsymbol{LFI:knee:frequency:30GHz}{113.7}

\setsymbol{LFI:knee:frequency:LFI18:Rad:M:units}{16.3\mHz}
\setsymbol{LFI:knee:frequency:LFI19:Rad:M:units}{15.1\mHz}
\setsymbol{LFI:knee:frequency:LFI20:Rad:M:units}{18.7\mHz}
\setsymbol{LFI:knee:frequency:LFI21:Rad:M:units}{37.2\mHz}
\setsymbol{LFI:knee:frequency:LFI22:Rad:M:units}{12.7\mHz}
\setsymbol{LFI:knee:frequency:LFI23:Rad:M:units}{34.6\mHz}
\setsymbol{LFI:knee:frequency:LFI24:Rad:M:units}{46.2\mHz}
\setsymbol{LFI:knee:frequency:LFI25:Rad:M:units}{24.9\mHz}
\setsymbol{LFI:knee:frequency:LFI26:Rad:M:units}{67.6\mHz}
\setsymbol{LFI:knee:frequency:LFI27:Rad:M:units}{187.4\mHz}
\setsymbol{LFI:knee:frequency:LFI28:Rad:M:units}{122.2\mHz}
\setsymbol{LFI:knee:frequency:LFI18:Rad:S:units}{17.7\mHz}
\setsymbol{LFI:knee:frequency:LFI19:Rad:S:units}{22.0\mHz}
\setsymbol{LFI:knee:frequency:LFI20:Rad:S:units}{8.7\mHz}
\setsymbol{LFI:knee:frequency:LFI21:Rad:S:units}{25.9\mHz}
\setsymbol{LFI:knee:frequency:LFI22:Rad:S:units}{15.8\mHz}
\setsymbol{LFI:knee:frequency:LFI23:Rad:S:units}{129.8\mHz}
\setsymbol{LFI:knee:frequency:LFI24:Rad:S:units}{100.9\mHz}
\setsymbol{LFI:knee:frequency:LFI25:Rad:S:units}{38.9\mHz}
\setsymbol{LFI:knee:frequency:LFI26:Rad:S:units}{58.9\mHz}
\setsymbol{LFI:knee:frequency:LFI27:Rad:S:units}{104.4\mHz}
\setsymbol{LFI:knee:frequency:LFI28:Rad:S:units}{40.7\mHz}

\setsymbol{LFI:knee:frequency:LFI18:Rad:M}{16.3}
\setsymbol{LFI:knee:frequency:LFI19:Rad:M}{15.1}
\setsymbol{LFI:knee:frequency:LFI20:Rad:M}{18.7}
\setsymbol{LFI:knee:frequency:LFI21:Rad:M}{37.2}
\setsymbol{LFI:knee:frequency:LFI22:Rad:M}{12.7}
\setsymbol{LFI:knee:frequency:LFI23:Rad:M}{34.6}
\setsymbol{LFI:knee:frequency:LFI24:Rad:M}{46.2}
\setsymbol{LFI:knee:frequency:LFI25:Rad:M}{24.9}
\setsymbol{LFI:knee:frequency:LFI26:Rad:M}{67.6}
\setsymbol{LFI:knee:frequency:LFI27:Rad:M}{187.4}
\setsymbol{LFI:knee:frequency:LFI28:Rad:M}{122.2}
\setsymbol{LFI:knee:frequency:LFI18:Rad:S}{17.7}
\setsymbol{LFI:knee:frequency:LFI19:Rad:S}{22.0}
\setsymbol{LFI:knee:frequency:LFI20:Rad:S}{8.7}
\setsymbol{LFI:knee:frequency:LFI21:Rad:S}{25.9}
\setsymbol{LFI:knee:frequency:LFI22:Rad:S}{15.8}
\setsymbol{LFI:knee:frequency:LFI23:Rad:S}{129.8}
\setsymbol{LFI:knee:frequency:LFI24:Rad:S}{100.9}
\setsymbol{LFI:knee:frequency:LFI25:Rad:S}{38.9}
\setsymbol{LFI:knee:frequency:LFI26:Rad:S}{58.9}
\setsymbol{LFI:knee:frequency:LFI27:Rad:S}{104.4}
\setsymbol{LFI:knee:frequency:LFI28:Rad:S}{40.7}

\setsymbol{LFI:slope:70GHz:units}{$-1.03$\mHz}
\setsymbol{LFI:slope:44GHz:units}{$-0.89$\mHz}
\setsymbol{LFI:slope:30GHz:units}{$-0.87$\mHz}

\setsymbol{LFI:slope:70GHz}{$-1.03$}
\setsymbol{LFI:slope:44GHz}{$-0.89$}
\setsymbol{LFI:slope:30GHz}{$-0.87$}

\setsymbol{LFI:slope:LFI18:Rad:M:units}{$-1.04$\mHz}
\setsymbol{LFI:slope:LFI19:Rad:M:units}{$-1.09$\mHz}
\setsymbol{LFI:slope:LFI20:Rad:M:units}{$-0.69$\mHz}
\setsymbol{LFI:slope:LFI21:Rad:M:units}{$-1.56$\mHz}
\setsymbol{LFI:slope:LFI22:Rad:M:units}{$-1.01$\mHz}
\setsymbol{LFI:slope:LFI23:Rad:M:units}{$-0.96$\mHz}
\setsymbol{LFI:slope:LFI24:Rad:M:units}{$-0.83$\mHz}
\setsymbol{LFI:slope:LFI25:Rad:M:units}{$-0.91$\mHz}
\setsymbol{LFI:slope:LFI26:Rad:M:units}{$-0.95$\mHz}
\setsymbol{LFI:slope:LFI27:Rad:M:units}{$-0.87$\mHz}
\setsymbol{LFI:slope:LFI28:Rad:M:units}{$-0.88$\mHz}
\setsymbol{LFI:slope:LFI18:Rad:S:units}{$-1.15$\mHz}
\setsymbol{LFI:slope:LFI19:Rad:S:units}{$-1.00$\mHz}
\setsymbol{LFI:slope:LFI20:Rad:S:units}{$-0.95$\mHz}
\setsymbol{LFI:slope:LFI21:Rad:S:units}{$-0.92$\mHz}
\setsymbol{LFI:slope:LFI22:Rad:S:units}{$-1.01$\mHz}
\setsymbol{LFI:slope:LFI23:Rad:S:units}{$-0.95$\mHz}
\setsymbol{LFI:slope:LFI24:Rad:S:units}{$-0.73$\mHz}
\setsymbol{LFI:slope:LFI25:Rad:S:units}{$-1.16$\mHz}
\setsymbol{LFI:slope:LFI26:Rad:S:units}{$-0.79$\mHz}
\setsymbol{LFI:slope:LFI27:Rad:S:units}{$-0.82$\mHz}
\setsymbol{LFI:slope:LFI28:Rad:S:units}{$-0.91$\mHz}

\setsymbol{LFI:slope:LFI18:Rad:M}{$-1.04$}
\setsymbol{LFI:slope:LFI19:Rad:M}{$-1.09$}
\setsymbol{LFI:slope:LFI20:Rad:M}{$-0.69$}
\setsymbol{LFI:slope:LFI21:Rad:M}{$-1.56$}
\setsymbol{LFI:slope:LFI22:Rad:M}{$-1.01$}
\setsymbol{LFI:slope:LFI23:Rad:M}{$-0.96$}
\setsymbol{LFI:slope:LFI24:Rad:M}{$-0.83$}
\setsymbol{LFI:slope:LFI25:Rad:M}{$-0.91$}
\setsymbol{LFI:slope:LFI26:Rad:M}{$-0.95$}
\setsymbol{LFI:slope:LFI27:Rad:M}{$-0.87$}
\setsymbol{LFI:slope:LFI28:Rad:M}{$-0.88$}
\setsymbol{LFI:slope:LFI18:Rad:S}{$-1.15$}
\setsymbol{LFI:slope:LFI19:Rad:S}{$-1.00$}
\setsymbol{LFI:slope:LFI20:Rad:S}{$-0.95$}
\setsymbol{LFI:slope:LFI21:Rad:S}{$-0.92$}
\setsymbol{LFI:slope:LFI22:Rad:S}{$-1.01$}
\setsymbol{LFI:slope:LFI23:Rad:S}{$-0.95$}
\setsymbol{LFI:slope:LFI24:Rad:S}{$-0.73$}
\setsymbol{LFI:slope:LFI25:Rad:S}{$-1.16$}
\setsymbol{LFI:slope:LFI26:Rad:S}{$-0.79$}
\setsymbol{LFI:slope:LFI27:Rad:S}{$-0.82$}
\setsymbol{LFI:slope:LFI28:Rad:S}{$-0.91$}

\setsymbol{LFI:FWHM:70GHz:units}{13\parcm01}
\setsymbol{LFI:FWHM:44GHz:units}{27\parcm92}
\setsymbol{LFI:FWHM:30GHz:units}{32\parcm65}

\setsymbol{LFI:FWHM:70GHz}{13.01}
\setsymbol{LFI:FWHM:44GHz}{27.92}
\setsymbol{LFI:FWHM:30GHz}{32.65}

\setsymbol{LFI:FWHM:LFI18:units}{13\parcm39}
\setsymbol{LFI:FWHM:LFI19:units}{13\parcm01}
\setsymbol{LFI:FWHM:LFI20:units}{12\parcm75}
\setsymbol{LFI:FWHM:LFI21:units}{12\parcm74}
\setsymbol{LFI:FWHM:LFI22:units}{12\parcm87}
\setsymbol{LFI:FWHM:LFI23:units}{13\parcm27}
\setsymbol{LFI:FWHM:LFI24:units}{22\parcm98}
\setsymbol{LFI:FWHM:LFI25:units}{30\parcm46}
\setsymbol{LFI:FWHM:LFI26:units}{30\parcm31}
\setsymbol{LFI:FWHM:LFI27:units}{32\parcm65}
\setsymbol{LFI:FWHM:LFI28:units}{32\parcm66}

\setsymbol{LFI:FWHM:LFI18}{13.39}
\setsymbol{LFI:FWHM:LFI19}{13.01}
\setsymbol{LFI:FWHM:LFI20}{12.75}
\setsymbol{LFI:FWHM:LFI21}{12.74}
\setsymbol{LFI:FWHM:LFI22}{12.87}
\setsymbol{LFI:FWHM:LFI23}{13.27}
\setsymbol{LFI:FWHM:LFI24}{22.98}
\setsymbol{LFI:FWHM:LFI25}{30.46}
\setsymbol{LFI:FWHM:LFI26}{30.31}
\setsymbol{LFI:FWHM:LFI27}{32.65}
\setsymbol{LFI:FWHM:LFI28}{32.66}

\setsymbol{LFI:FWHM:uncertainty:LFI18:units}{0.170\arcm}
\setsymbol{LFI:FWHM:uncertainty:LFI19:units}{0.174\arcm}
\setsymbol{LFI:FWHM:uncertainty:LFI20:units}{0.170\arcm}
\setsymbol{LFI:FWHM:uncertainty:LFI21:units}{0.156\arcm}
\setsymbol{LFI:FWHM:uncertainty:LFI22:units}{0.164\arcm}
\setsymbol{LFI:FWHM:uncertainty:LFI23:units}{0.171\arcm}
\setsymbol{LFI:FWHM:uncertainty:LFI24:units}{0.652\arcm}
\setsymbol{LFI:FWHM:uncertainty:LFI25:units}{1.075\arcm}
\setsymbol{LFI:FWHM:uncertainty:LFI26:units}{1.131\arcm}
\setsymbol{LFI:FWHM:uncertainty:LFI27:units}{1.266\arcm}
\setsymbol{LFI:FWHM:uncertainty:LFI28:units}{1.287\arcm}

\setsymbol{LFI:FWHM:uncertainty:LFI18}{0.170}
\setsymbol{LFI:FWHM:uncertainty:LFI19}{0.174}
\setsymbol{LFI:FWHM:uncertainty:LFI20}{0.170}
\setsymbol{LFI:FWHM:uncertainty:LFI21}{0.156}
\setsymbol{LFI:FWHM:uncertainty:LFI22}{0.164}
\setsymbol{LFI:FWHM:uncertainty:LFI23}{0.171}
\setsymbol{LFI:FWHM:uncertainty:LFI24}{0.652}
\setsymbol{LFI:FWHM:uncertainty:LFI25}{1.075}
\setsymbol{LFI:FWHM:uncertainty:LFI26}{1.131}
\setsymbol{LFI:FWHM:uncertainty:LFI27}{1.266}
\setsymbol{LFI:FWHM:uncertainty:LFI28}{1.287}

\setsymbol{HFI:center:frequency:100GHz:units}{100\,GHz}
\setsymbol{HFI:center:frequency:143GHz:units}{143\,GHz}
\setsymbol{HFI:center:frequency:217GHz:units}{217\,GHz}
\setsymbol{HFI:center:frequency:353GHz:units}{353\,GHz}
\setsymbol{HFI:center:frequency:545GHz:units}{545\,GHz}
\setsymbol{HFI:center:frequency:857GHz:units}{857\,GHz}

\setsymbol{HFI:center:frequency:100GHz}{100}
\setsymbol{HFI:center:frequency:143GHz}{143}
\setsymbol{HFI:center:frequency:217GHz}{217}
\setsymbol{HFI:center:frequency:353GHz}{353}
\setsymbol{HFI:center:frequency:545GHz}{545}
\setsymbol{HFI:center:frequency:857GHz}{857}

\setsymbol{HFI:Ndetectors:100GHz}{8}
\setsymbol{HFI:Ndetectors:143GHz}{11}
\setsymbol{HFI:Ndetectors:217GHz}{12}
\setsymbol{HFI:Ndetectors:353GHz}{12}
\setsymbol{HFI:Ndetectors:545GHz}{3}
\setsymbol{HFI:Ndetectors:857GHz}{4}

\setsymbol{HFI:FWHM:Maps:100GHz:units}{9\parcm88}
\setsymbol{HFI:FWHM:Maps:143GHz:units}{7\parcm18}
\setsymbol{HFI:FWHM:Maps:217GHz:units}{4\parcm87}
\setsymbol{HFI:FWHM:Maps:353GHz:units}{4\parcm65}
\setsymbol{HFI:FWHM:Maps:545GHz:units}{4\parcm72}
\setsymbol{HFI:FWHM:Maps:857GHz:units}{4\parcm39}
\setsymbol{HFI:FWHM:Maps:100GHz}{9.88}
\setsymbol{HFI:FWHM:Maps:143GHz}{7.18}
\setsymbol{HFI:FWHM:Maps:217GHz}{4.87}
\setsymbol{HFI:FWHM:Maps:353GHz}{4.65}
\setsymbol{HFI:FWHM:Maps:545GHz}{4.72}
\setsymbol{HFI:FWHM:Maps:857GHz}{4.39}

\setsymbol{HFI:beam:ellipticity:Maps:100GHz}{1.15}
\setsymbol{HFI:beam:ellipticity:Maps:143GHz}{1.01}
\setsymbol{HFI:beam:ellipticity:Maps:217GHz}{1.06}
\setsymbol{HFI:beam:ellipticity:Maps:353GHz}{1.05}
\setsymbol{HFI:beam:ellipticity:Maps:545GHz}{1.14}
\setsymbol{HFI:beam:ellipticity:Maps:857GHz}{1.19}

\setsymbol{HFI:FWHM:Mars:100GHz:units}{9\parcm37}
\setsymbol{HFI:FWHM:Mars:143GHz:units}{7\parcm04}
\setsymbol{HFI:FWHM:Mars:217GHz:units}{4\parcm68}
\setsymbol{HFI:FWHM:Mars:353GHz:units}{4\parcm43}
\setsymbol{HFI:FWHM:Mars:545GHz:units}{3\parcm80}
\setsymbol{HFI:FWHM:Mars:857GHz:units}{3\parcm67}

\setsymbol{HFI:FWHM:Mars:100GHz}{9.37}
\setsymbol{HFI:FWHM:Mars:143GHz}{7.04}
\setsymbol{HFI:FWHM:Mars:217GHz}{4.68}
\setsymbol{HFI:FWHM:Mars:353GHz}{4.43}
\setsymbol{HFI:FWHM:Mars:545GHz}{3.80}
\setsymbol{HFI:FWHM:Mars:857GHz}{3.67}

\setsymbol{HFI:beam:ellipticity:Mars:100GHz}{1.18}
\setsymbol{HFI:beam:ellipticity:Mars:143GHz}{1.03}
\setsymbol{HFI:beam:ellipticity:Mars:217GHz}{1.14}
\setsymbol{HFI:beam:ellipticity:Mars:353GHz}{1.09}
\setsymbol{HFI:beam:ellipticity:Mars:545GHz}{1.25}
\setsymbol{HFI:beam:ellipticity:Mars:857GHz}{1.03}

\setsymbol{HFI:CMB:relative:calibration:100GHz}{$\lsim 1\%$}
\setsymbol{HFI:CMB:relative:calibration:143GHz}{$\lsim 1\%$}
\setsymbol{HFI:CMB:relative:calibration:217GHz}{$\lsim 1\%$}
\setsymbol{HFI:CMB:relative:calibration:353GHz}{$\lsim 1\%$}
\setsymbol{HFI:CMB:relative:calibration:545GHz}{}
\setsymbol{HFI:CMB:relative:calibration:857GHz}{}

\setsymbol{HFI:CMB:absolute:calibration:100GHz}{$\lsim 2\%$}
\setsymbol{HFI:CMB:absolute:calibration:143GHz}{$\lsim 2\%$}
\setsymbol{HFI:CMB:absolute:calibration:217GHz}{$\lsim 2\%$}
\setsymbol{HFI:CMB:absolute:calibration:353GHz}{$\lsim 2\%$}
\setsymbol{HFI:CMB:absolute:calibration:545GHz}{}
\setsymbol{HFI:CMB:absolute:calibration:857GHz}{}

\setsymbol{HFI:FIRAS:gain:calibration:accuracy:statistical:100GHz}{}
\setsymbol{HFI:FIRAS:gain:calibration:accuracy:statistical:143GHz}{}
\setsymbol{HFI:FIRAS:gain:calibration:accuracy:statistical:217GHz}{}
\setsymbol{HFI:FIRAS:gain:calibration:accuracy:statistical:353GHz}{2.5\%}
\setsymbol{HFI:FIRAS:gain:calibration:accuracy:statistical:545GHz}{1\%}
\setsymbol{HFI:FIRAS:gain:calibration:accuracy:statistical:857GHz}{0.5\%}

\setsymbol{HFI:FIRAS:gain:calibration:accuracy:systematic:100GHz}{}
\setsymbol{HFI:FIRAS:gain:calibration:accuracy:systematic:143GHz}{}
\setsymbol{HFI:FIRAS:gain:calibration:accuracy:systematic:217GHz}{}
\setsymbol{HFI:FIRAS:gain:calibration:accuracy:systematic:353GHz}{}
\setsymbol{HFI:FIRAS:gain:calibration:accuracy:systematic:545GHz}{7\%}
\setsymbol{HFI:FIRAS:gain:calibration:accuracy:systematic:857GHz}{7\%}

\setsymbol{HFI:FIRAS:zero:point:accuracy:100GHz:units}{0.8\MJysr}
\setsymbol{HFI:FIRAS:zero:point:accuracy:143GHz:units}{}
\setsymbol{HFI:FIRAS:zero:point:accuracy:217GHz:units}{}
\setsymbol{HFI:FIRAS:zero:point:accuracy:353GHz:units}{1.4\MJysr}
\setsymbol{HFI:FIRAS:zero:point:accuracy:545GHz:units}{2.2\MJysr}
\setsymbol{HFI:FIRAS:zero:point:accuracy:857GHz:units}{1.7\MJysr}

\setsymbol{HFI:FIRAS:zero:point:accuracy:100GHz}{0.8}
\setsymbol{HFI:FIRAS:zero:point:accuracy:143GHz}{}
\setsymbol{HFI:FIRAS:zero:point:accuracy:217GHz}{}
\setsymbol{HFI:FIRAS:zero:point:accuracy:353GHz}{1.4}
\setsymbol{HFI:FIRAS:zero:point:accuracy:545GHz}{2.2}
\setsymbol{HFI:FIRAS:zero:point:accuracy:857GHz}{1.7}

\setsymbol{HFI:unit:conversion:100GHz:units}{0.2415\MJysrmK}
\setsymbol{HFI:unit:conversion:143GHz:units}{0.3694\MJysrmK}
\setsymbol{HFI:unit:conversion:217GHz:units}{0.4811\MJysrmK}
\setsymbol{HFI:unit:conversion:353GHz:units}{0.2883\MJysrmK}
\setsymbol{HFI:unit:conversion:545GHz:units}{0.05826\MJysrmK}
\setsymbol{HFI:unit:conversion:857GHz:units}{0.002238\MJysrmK}

\setsymbol{HFI:unit:conversion:100GHz}{0.2415}
\setsymbol{HFI:unit:conversion:143GHz}{0.3694}
\setsymbol{HFI:unit:conversion:217GHz}{0.4811}
\setsymbol{HFI:unit:conversion:353GHz}{0.2883}
\setsymbol{HFI:unit:conversion:545GHz}{0.05826}
\setsymbol{HFI:unit:conversion:857GHz}{0.002238}

\setsymbol{HFI:colour:correction:alpha=-2:V1.01:100GHz}{0.9893}
\setsymbol{HFI:colour:correction:alpha=-2:V1.01:143GHz}{0.9759}
\setsymbol{HFI:colour:correction:alpha=-2:V1.01:217GHz}{1.0007}
\setsymbol{HFI:colour:correction:alpha=-2:V1.01:353GHz}{1.0028}
\setsymbol{HFI:colour:correction:alpha=-2:V1.01:545GHz}{1.0019}
\setsymbol{HFI:colour:correction:alpha=-2:V1.01:857GHz}{0.9889}

\setsymbol{HFI:colour:correction:alpha=0:V1.01:100GHz}{1.0008}
\setsymbol{HFI:colour:correction:alpha=0:V1.01:143GHz}{1.0148}
\setsymbol{HFI:colour:correction:alpha=0:V1.01:217GHz}{0.9909}
\setsymbol{HFI:colour:correction:alpha=0:V1.01:353GHz}{0.9888}
\setsymbol{HFI:colour:correction:alpha=0:V1.01:545GHz}{0.9878}
\setsymbol{HFI:colour:correction:alpha=0:V1.01:857GHz}{1.0014}

\providecommand{\sorthelp}[1]{}

\graphicspath{{./Figures/}}

\newcommand{\lto}{$\mrm{L}_2$ }  
\newcommand{\lton}{$\mrm{L}_2$} 
\newcommand{\mrm}[1]{\mathrm{#1}}

\renewcommand{\d}{d}
\newcommand{\pder}[2]{\frac{\partial #1}{\partial #2}}  % for partial derivatives
\usepackage{bm}
\usepackage{etoolbox}
\makeatletter
\def\dd{\@ifnextchar^{\dd@grabsp}{\dd@{}}}
\def\dd@grabsp^#1#2{\dd@{#1}{#2}}
\def\dd@#1#2{\mathop{d\ifblank{#1}{}{^{#1}}{#2}}}
\let\int@original\int
\def\int{\int@checkfirstsb}
\def\int@checkfirstsb{\@ifnextchar_{\int@checksecondsp}{\int@checkfirstsp}}
\def\int@checkfirstsp{\@ifnextchar^{\int@checksecondsb}{\int@{}{}}}
\def\int@checksecondsp_#1{\@ifnextchar^{\int@grabsp{#1}}{\int@{#1}{}}}
\def\int@checksecondsb^#1{\@ifnextchar_{\int@grabsb{#1}}{\int@{}{#1}}}
\def\int@grabsb#1_#2{\int@{#2}{#1}}
\def\int@grabsp#1^#2{\int@{#1}{#2}}
\def\int@#1#2{\int@original\ifblank{#1}{}{_{#1}}\ifblank{#2}{}{^{#2}}\mathopen{}}
\makeatother

\newcommand{\WMAP}{WMAP\/}
\newcommand{\wmap}{WMAP }
\newcommand{\wmapn}{WMAP}
\newcommand{\act}{ACT }
\newcommand{\actn}{ACT}
\newcommand{\herschel}{\textit{Herschel} }
\newcommand{\herscheln}{\textit{Herschel}}

\begin{document}

\title{\Planck\ intermediate results. LII. \\Planet flux densities}

%This author list corresponds to \title{Author list for A33\_Zodi\_light}
%Prepared by M. Lopez-Caniego (Marcos.Lopez.Caniego@sciops.esa.int), ESAC/ESA
%This version is from Mon Nov 21 08:40:06 2016 CET
%\subtitle{There are 149 co-authors in this list}
\author{\small
Planck Collaboration: Y.~Akrami\inst{47, 78}
\and
M.~Ashdown\inst{54, 5}
\and
J.~Aumont\inst{45}
\and
C.~Baccigalupi\inst{67}
\and
M.~Ballardini\inst{23, 37, 40}
\and
A.~J.~Banday\inst{76, 8}
\and
R.~B.~Barreiro\inst{49}
\and
N.~Bartolo\inst{22, 50}
\and
S.~Basak\inst{67}
\and
K.~Benabed\inst{46, 75}
% ...
}
\institute{\small
APC, AstroParticule et Cosmologie, Universit\'{e} Paris Diderot, CNRS/IN2P3, CEA/lrfu, Observatoire de Paris, Sorbonne Paris Cit\'{e}, 10, rue Alice Domon et L\'{e}onie Duquet, 75205 Paris Cedex 13, France\goodbreak
\and
African Institute for Mathematical Sciences, 6-8 Melrose Road, Muizenberg, Cape Town, South Africa\goodbreak
\and
Agenzia Spaziale Italiana Science Data Center, Via del Politecnico snc, 00133, Roma, Italy\goodbreak
\and
Aix Marseille Univ, CNRS, LAM, Laboratoire d'Astrophysique de Marseille, Marseille, France\goodbreak
\and
Astrophysics Group, Cavendish Laboratory, University of Cambridge, J J Thomson Avenue, Cambridge CB3 0HE, U.K.\goodbreak
\and
Astrophysics \& Cosmology Research Unit, School of Mathematics, Statistics \& Computer Science, University of KwaZulu-Natal, Westville Campus, Private Bag X54001, Durban 4000, South Africa\goodbreak
\and
CITA, University of Toronto, 60 St. George St., Toronto, ON M5S 3H8, Canada\goodbreak
\and
CNRS, IRAP, 9 Av. colonel Roche, BP 44346, F-31028 Toulouse cedex 4, France\goodbreak
\and
California Institute of Technology, Pasadena, California, U.S.A.\goodbreak
\and
Computational Cosmology Center, Lawrence Berkeley National Laboratory, Berkeley, California, U.S.A.\goodbreak
\and
DTU Space, National Space Institute, Technical University of Denmark, Elektrovej 327, DK-2800 Kgs. Lyngby, Denmark\goodbreak
\and
D\'{e}partement de Physique Th\'{e}orique, Universit\'{e} de Gen\`{e}ve, 24, Quai E. Ansermet,1211 Gen\`{e}ve 4, Switzerland\goodbreak
\and
Departamento de Astrof\'{i}sica, Universidad de La Laguna (ULL), E-38206 La Laguna, Tenerife, Spain\goodbreak
\and
Departamento de F\'{\i}sica, Universidad de Oviedo, Avda. Calvo Sotelo s/n, Oviedo, Spain\goodbreak
\and
Department of Astrophysics/IMAPP, Radboud University, P.O. Box 9010, 6500 GL Nijmegen, The Netherlands\goodbreak
\and
Department of Physics \& Astronomy, University of British Columbia, 6224 Agricultural Road, Vancouver, British Columbia, Canada\goodbreak
\and
Department of Physics and Astronomy, Dana and David Dornsife College of Letter, Arts and Sciences, University of Southern California, Los Angeles, CA 90089, U.S.A.\goodbreak
\and
Department of Physics and Astronomy, University College London, London WC1E 6BT, U.K.\goodbreak
\and
Department of Physics and Astronomy, University of Sussex, Brighton BN1 9QH, U.K.\goodbreak
\and
Department of Physics, Gustaf H\"{a}llstr\"{o}min katu 2a, University of Helsinki, Helsinki, Finland\goodbreak
\and
Department of Physics, Princeton University, Princeton, New Jersey, U.S.A.\goodbreak
\and
Dipartimento di Fisica e Astronomia G. Galilei, Universit\`{a} degli Studi di Padova, via Marzolo 8, 35131 Padova, Italy\goodbreak
\and
Dipartimento di Fisica e Astronomia, Alma Mater Studiorum, Universit\`{a} degli Studi di Bologna, Viale Berti Pichat 6/2, I-40127, Bologna, Italy\goodbreak
\and
Dipartimento di Fisica e Scienze della Terra, Universit\`{a} di Ferrara, Via Saragat 1, 44122 Ferrara, Italy\goodbreak
\and
Dipartimento di Fisica, Universit\`{a} La Sapienza, P. le A. Moro 2, Roma, Italy\goodbreak
\and
Dipartimento di Fisica, Universit\`{a} degli Studi di Milano, Via Celoria, 16, Milano, Italy\goodbreak
\and
Dipartimento di Fisica, Universit\`{a} degli Studi di Trieste, via A. Valerio 2, Trieste, Italy\goodbreak
\and
Dipartimento di Matematica, Universit\`{a} di Roma Tor Vergata, Via della Ricerca Scientifica, 1, Roma, Italy\goodbreak
\and
European Space Agency, ESAC, Planck Science Office, Camino bajo del Castillo, s/n, Urbanizaci\'{o}n Villafranca del Castillo, Villanueva de la Ca\~{n}ada, Madrid, Spain\goodbreak
\and
European Space Agency, ESTEC, Keplerlaan 1, 2201 AZ Noordwijk, The Netherlands\goodbreak
\and
Gran Sasso Science Institute, INFN, viale F. Crispi 7, 67100 L'Aquila, Italy\goodbreak
\and
Haverford College Astronomy Department, 370 Lancaster Avenue, Haverford, Pennsylvania, U.S.A.\goodbreak
\and
Helsinki Institute of Physics, Gustaf H\"{a}llstr\"{o}min katu 2, University of Helsinki, Helsinki, Finland\goodbreak
\and
INAF - Osservatorio Astronomico di Padova, Vicolo dell'Osservatorio 5, Padova, Italy\goodbreak
\and
INAF - Osservatorio Astronomico di Roma, via di Frascati 33, Monte Porzio Catone, Italy\goodbreak
\and
INAF - Osservatorio Astronomico di Trieste, Via G.B. Tiepolo 11, Trieste, Italy\goodbreak
\and
INAF/IASF Bologna, Via Gobetti 101, Bologna, Italy\goodbreak
\and
INAF/IASF Milano, Via E. Bassini 15, Milano, Italy\goodbreak
\and
INFN - CNAF, viale Berti Pichat 6/2, 40127 Bologna, Italy\goodbreak
\and
INFN, Sezione di Bologna, viale Berti Pichat 6/2, 40127 Bologna, Italy\goodbreak
\and
INFN, Sezione di Ferrara, Via Saragat 1, 44122 Ferrara, Italy\goodbreak
\and
INFN, Sezione di Roma 1, Universit\`{a} di Roma Sapienza, Piazzale Aldo Moro 2, 00185, Roma, Italy\goodbreak
\and
INFN, Sezione di Roma 2, Universit\`{a} di Roma Tor Vergata, Via della Ricerca Scientifica, 1, Roma, Italy\goodbreak
\and
Imperial College London, Astrophysics group, Blackett Laboratory, Prince Consort Road, London, SW7 2AZ, U.K.\goodbreak
\and
Institut d'Astrophysique Spatiale, CNRS, Univ. Paris-Sud, Universit\'{e} Paris-Saclay, B\^{a}t. 121, 91405 Orsay cedex, France\goodbreak
\and
Institut d'Astrophysique de Paris, CNRS (UMR7095), 98 bis Boulevard Arago, F-75014, Paris, France\goodbreak
\and
Institute of Theoretical Astrophysics, University of Oslo, Blindern, Oslo, Norway\goodbreak
\and
Instituto de Astrof\'{\i}sica de Canarias, C/V\'{\i}a L\'{a}ctea s/n, La Laguna, Tenerife, Spain\goodbreak
\and
Instituto de F\'{\i}sica de Cantabria (CSIC-Universidad de Cantabria), Avda. de los Castros s/n, Santander, Spain\goodbreak
\and
Istituto Nazionale di Fisica Nucleare, Sezione di Padova, via Marzolo 8, I-35131 Padova, Italy\goodbreak
\and
Jet Propulsion Laboratory, California Institute of Technology, 4800 Oak Grove Drive, Pasadena, California, U.S.A.\goodbreak
\and
Jodrell Bank Centre for Astrophysics, Alan Turing Building, School of Physics and Astronomy, The University of Manchester, Oxford Road, Manchester, M13 9PL, U.K.\goodbreak
\and
Kavli Institute for Cosmological Physics, University of Chicago, Chicago, IL 60637, USA\goodbreak
\and
Kavli Institute for Cosmology Cambridge, Madingley Road, Cambridge, CB3 0HA, U.K.\goodbreak
\and
LAL, Universit\'{e} Paris-Sud, CNRS/IN2P3, Orsay, France\goodbreak
\and
LERMA, CNRS, Observatoire de Paris, 61 Avenue de l'Observatoire, Paris, France\goodbreak
\and
LESIA, Observatoire de Paris, CNRS, UPMC, Universit\'{e} Paris-Diderot, 5 Place J. Janssen, 92195 Meudon, France\goodbreak
\and
Laboratoire Traitement et Communication de l'Information, CNRS (UMR 5141) and T\'{e}l\'{e}com ParisTech, 46 rue Barrault F-75634 Paris Cedex 13, France\goodbreak
\and
Laboratoire de Physique Subatomique et Cosmologie, Universit\'{e} Grenoble-Alpes, CNRS/IN2P3, 53, rue des Martyrs, 38026 Grenoble Cedex, France\goodbreak
\and
Laboratoire de Physique Th\'{e}orique, Universit\'{e} Paris-Sud 11 \& CNRS, B\^{a}timent 210, 91405 Orsay, France\goodbreak
\and
Lawrence Berkeley National Laboratory, Berkeley, California, U.S.A.\goodbreak
\and
Low Temperature Laboratory, Department ofÊApplied Physics, Aalto University, Espoo, FI-00076 AALTO, Finland\goodbreak
\and
Max-Planck-Institut f\"{u}r Astrophysik, Karl-Schwarzschild-Str. 1, 85741 Garching, Germany\goodbreak
\and
Mullard Space Science Laboratory, University College London, Surrey RH5 6NT, U.K.\goodbreak
\and
Nicolaus Copernicus Astronomical Center, Polish Academy of Sciences, Bartycka 18, 00-716 Warsaw, Poland\goodbreak
\and
Nordita (Nordic Institute for Theoretical Physics), Roslagstullsbacken 23, SE-106 91 Stockholm, Sweden\goodbreak
\and
SISSA, Astrophysics Sector, via Bonomea 265, 34136, Trieste, Italy\goodbreak
\and
School of Chemistry and Physics, University of KwaZulu-Natal, Westville Campus, Private Bag X54001, Durban, 4000, South Africa\goodbreak
\and
School of Physics and Astronomy, Cardiff University, Queens Buildings, The Parade, Cardiff, CF24 3AA, U.K.\goodbreak
\and
Simon Fraser University, Department of Physics, 8888 University Drive, Burnaby BC, Canada\goodbreak
\and
Sorbonne Universit\'{e}-UPMC, UMR7095, Institut d'Astrophysique de Paris, 98 bis Boulevard Arago, F-75014, Paris, France\goodbreak
\and
Space Sciences Laboratory, University of California, Berkeley, California, U.S.A.\goodbreak
\and
Sub-Department of Astrophysics, University of Oxford, Keble Road, Oxford OX1 3RH, U.K.\goodbreak
\and
The Oskar Klein Centre for Cosmoparticle Physics, Department of Physics,Stockholm University, AlbaNova, SE-106 91 Stockholm, Sweden\goodbreak
\and
UPMC Univ Paris 06, UMR7095, 98 bis Boulevard Arago, F-75014, Paris, France\goodbreak
\and
Universit\'{e} de Toulouse, UPS-OMP, IRAP, F-31028 Toulouse cedex 4, France\goodbreak
\and
University of Granada, Departamento de F\'{\i}sica Te\'{o}rica y del Cosmos, Facultad de Ciencias, Granada, Spain\goodbreak
\and
University of Heidelberg, Institute for Theoretical Physics, Philosophenweg 16, 69120, Heidelberg, Germany\goodbreak
% ...
}
 % TO BE UPDATED

\authorrunning{Planck Collaboration}
\titlerunning{Planet flux densities}

\abstract{
Measurements of flux density are described for five planets, Mars, Jupiter, Saturn, Uranus, and Neptune, across the six \Planck\ High Frequency Instrument frequency bands (100--857\GHz) and these are then compared with models and existing data. In our analysis, we have also included estimates of the brightness of Jupiter and Saturn at the three frequencies of the \Planck\ Low Frequency Instrument (30, 44, and 70\GHz). The results provide constraints on the intrinsic brightness and the brightness time-variability of these planets. The majority of the planet flux density estimates are limited by systematic errors, but still yield better than 1\% measurements in many cases. Applying data from \Planck~HFI, the Wilkinson Microwave Anisotropy Probe (\wmapn), and the Atacama Cosmology Telescope (\actn) to a model that incorporates contributions from Saturn's rings to the planet's total flux density suggests a best fit value for the spectral index of Saturn's ring system of $\beta _\mrm{ring} = 2.30\pm0.03$ over the 30--1000\GHz\ frequency range. Estimates of the polarization amplitude of the planets have also been made in the four bands that have polarization-sensitive detectors (100--353\GHz); this analysis provides a 95\,\% confidence level upper limit on Mars's polarization of 1.8, 1.7, 1.2, and 1.7\,\% at 100, 143, 217, and 353\GHz, respectively. The average ratio between the \Planck-HFI measurements and the adopted model predictions for all five planets (excluding Jupiter observations for 353\GHz) is 0.997, 0.997, 1.018, and 1.032 for 100, 143, 217, and 353\GHz, respectively. Model predictions for planet thermodynamic temperatures are therefore consistent with the absolute calibration of \Planck-HFI detectors at about the three-percent-level. We compare our measurements with published results from recent cosmic microwave background experiments. In particular, we observe that the flux densities measured by \Planck~HFI and \wmap agree to within 2\,\%. These results allow experiments operating in the mm-wavelength range to cross-calibrate against \Planck\ and improve models of radiative transport used in planetary science.

}
\keywords{Cosmology: observations, cosmic background radiation --- Planets and satellites: general} 
\maketitle

\section{Introduction}
\label{sec:introduction}

This paper presents \Planck\ High Frequency Instrument (HFI) measurements of the flux densities of Mars, Jupiter, Saturn, Uranus, and Neptune at mm and sub-mm wavelengths.\footnote{\Planck\ (\url{http://www.esa.int/Planck}) is a project of the European Space Agency (ESA) with instruments provided by two scientific consortia funded by ESA member states and led by Principal Investigators from France and Italy, telescope reflectors provided through a collaboration between ESA and a scientific consortium led and funded by Denmark, and additional contributions from NASA (USA).} The HFI beam does not resolve the planets and thus the flux densities reported here are whole-disc averages. These observations were performed over a 27-month period during the 30-month operational lifetime of the \Planck\ HFI, spanning August 2009 to January 2012. As part of the nominal raster-scan strategy, approximately 20 planet observations were made, each lasting roughly a week.

This paper also reports on Jupiter and Saturn brightness measurements from the \Planck\ Low Frequency Instrument (LFI). For Jupiter, those brightness estimates are based on data accumulated over the entire operational lifetime of \Planck\ LFI, whereas the Saturn brightness estimates are based only on data from the first year of LFI observations.

Observations of the microwave flux density of planets help to inform radiative transfer modelling, which in turn constrains a combination of atmospheric thermal structure, chemical abundances, and surface/subsurface temperature distributions and emissivity properties. Planet observations can also be used to cross-calibrate other experiments with \Planck, providing a test of the absolute calibration of both instruments. Importantly, point source flux density reconstruction offers one of the few viable checks on the beam solid angle contained within far sidelobes, since a significant sidelobe contribution dilutes the perceived flux density. Finally, sufficiently precise planet models can be used to bracket the absolute calibration of \Planck\ detectors.

The primary goals of this paper are: (1) to accurately report on the planet flux densities; (2) compare results with existing models and most relevant data sets in order to constrain instrument properties, including absolute calibration; and (3) to quantify the limitations of these measurements.

\Planck\ orbits \lton, the Earth-Sun Lagrange point outside Earth's orbit that has an identical sidereal period. The colocation of the Earth and the Sun on the sky as viewed from \lto make it an optimal location for satellites conducting full-sky surveys. \Planck\ essentially rotated around its symmetry axis at 1 rpm while stepping azimuthally (or in the ecliptic plane) by 2.5\arcmin\ every hour. This ensured that the satellite's solar panels were pointed directly at the Sun at all times, therefore, maintaining a stable thermal environment and minimizing stray radiation \citep{Dupac2005}. In the time between these azimuthal steps, the satellite would trace out approximately 60 circles on the sky. Additionally, the spin axis precessed with a 7\pdeg5 amplitude over the duration of a ``survey" to cover the poles and smooth out the pixel hits \citep{tauber2010a}. Residual drifts and nutations were minimal and were accounted for in pointing reconstruction \citep{planck2011-1.1}. With this scan strategy, \Planck\ observed the whole sky (including the five planets outside Earth's orbit) in approximately 6\,months per survey. 

For a small patch of the sky, it is common to define a Cartesian coordinate system with axis aligned parallel and perpendicular to the primary scan direction of the satellite. In this system, the axes are usually referred to as the co- and cross-scan directions \citep{planck2013-p03c}. Using this terminology, we can make the following statement regarding \Planck's scan strategy: samples separated in the cross-scan direction by more than 2--3\arcmin\ will be separated temporally by at least one hour. Therefore, given the size of the \Planck\ beams, all planet observations spanned many hours. Note that the full field of view of the HFI focal plane in the cross-scan direction is just under 4\deg.

The absolute calibration of the HFI 100--353\GHz\ bands is derived from the cosmic microwave background dipole induced by the orbit of the \Planck\ spacecraft around the Sun, and is known to much better than 1\,\%; however as we will see, systematic errors prevent us from reaching this level of precision in some of the planetary flux densities.  The calibration of the HFI 545 and 857\GHz\ channels, on the other hand, uses Uranus and Neptune observations \citep{planck2014-a09,Bertincourt2016}.

The Second \Planck\ Catalogue of Compact Sources (PCCS2, \cite{planck2014-a35}) describes flux density reconstruction for Galactic and extragalactic objects seen in \Planck\ maps. In the standard HFI processing pipeline, planets and other moving solar system objects are masked from subsequent analysis in the time-ordered data, and therefore do not appear in the sky maps. Of the sources described in the PCCS2, only a handful illuminate the \Planck\ reflectors with flux density that exceeds that of Neptune and Uranus. Mars, Jupiter, and Saturn, on the other hand, are the brightest compact objects observed by \Planck, and the two gas giants outshine any PCCS2 object by at least an order of magnitude. 

\begin{figure}[t!]
\begin{center}
\includegraphics[width=\columnwidth]{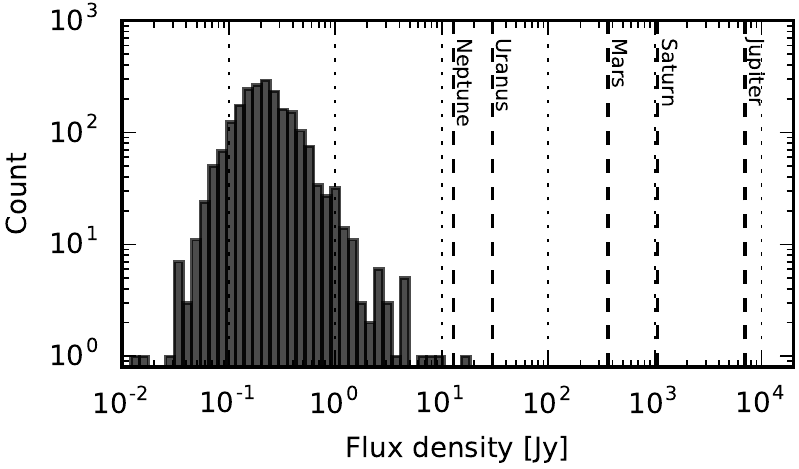}
\caption[Flux density histogram at 217\GHz]{\label{fig:flux_density217}Histogram of the flux density of all 217-GHz\ PCCS2 sources together with vertical lines indicating the flux of the five planets discussed in this paper. It is clear that at 217\GHz, Mars, Saturn, and especially Jupiter, are much brighter than any of the PCCS2 sources. Here the Mars data point corresponds to best estimates for the 22 December 2011 flux density. The majority of the 217-GHz\ PCCS2 sources have flux densities in the 0.1--1.0\,Jy range. Note that both axes are logarithmic on this figure.
}
\end{center}
\end{figure}

At 143 GHz, the five planets appear in timelines with a signal-to-noise ratio ranging from 3 (Neptune) to 1200 (Jupiter). The average sample density of each observation spans approximately 9--16 samples per arcmin$^2$, with variations caused by a combination of scan strategy and the apparent motion of the planets. Table \ref{tab:planet_obs} describes some properties of the planet observations for a single 143-GHz\ channel. Figure~\ref{fig:flux_density217} shows a histogram of the 217-GHz\ flux densities of sources from the PCCS2, as well as our estimates for the flux density of the five planets. From this figure it is clear that the planets offer a unique view of the instrument, allowing us to constrain its temporal, spatial, and frequency response. 

This paper is organized as follows. In Sect.~\ref{sec:PFM} we describe the details of the flux density analysis, including statistical and systematic uncertainty estimates. In Sect.~\ref{sec:planet_flux} we provide results on flux densities for Mars, Jupiter, Saturn, Uranus, and Neptune. We also investigate the contribution of Saturn's rings to the planet's total flux density and search for indications of diurnal variations in flux density for both Mars and Uranus. In Sect.~\ref{sec:comparison} we compare the planet flux density results with public results from \wmap and \act and look at the overall agreement between our measurements and model predictions. This section also discusses limits on polarization fraction of the five planets determined using \Planck-HFI measurements at 100--353\,\GHz. We provide our final conclusions in Sect.~\ref{sec:conc}.

\section{Planet flux density measurements}
\label{sec:PFM}

\begin{table}[tb!]
\begingroup
\newdimen\tblskip \tblskip=5pt
\caption{Properties of the \Planck~HFI planet observations specifically for detector 1b at 143\GHz. Here ``diameter" represents the planet diameter as viewed from \lto averaged over all observations, while ``sample density" refers to the average sample density, accounting for flagging, within a 40\arcmin\ wide field of view centred on the planet. The S/N given is the ratio between a fit to the peak signal registering in the timeline and the root mean square (rms) noise.}
\label{tab:planet_obs}
\vskip -5mm
\footnotesize
\setbox\tablebox=\vbox{
   \newdimen\digitwidth
   \setbox0=\hbox{\rm 0}
   \digitwidth=\wd0
   \catcode`*=\active
   \def*{\kern\digitwidth}
   \newdimen\signwidth
   \setbox0=\hbox{+}
   \signwidth=\wd0
   \catcode`!=\active
   \def!{\kern\signwidth}
   \newdimen\pointwidth
   \setbox0=\hbox{\rm .}
   \pointwidth=\wd0
   \catcode`?=\active
   \def?{\kern\pointwidth}
\halign{\hbox to 2.0cm{#\leaderfil}\tabskip 0pt&
        \hfil#\hfil\tabskip 0.5em&
        \hfil#\hfil\tabskip 1.0em&
        \hfil#\hfil\tabskip 1.0em&
        \hfil#\hfil\tabskip 0pt\cr
\noalign{\doubleline}
\noalign{\vskip -2pt}
\omit\hfil Planet\hfil& No.\ obs.& Diameter& Sample density& S/N\cr
\omit& & [arcsec]& [per arcmin$^{2}$]&\cr
\noalign{\vskip 3pt\hrule\vskip 5pt}
Mars&    3& *8?*& 16 / 12 / 12 ? ** ? **& **70\cr
Jupiter& 5& 40?*& *9 / 13 / 14 / *9 / 11& 1200\cr
Saturn&  4& 17?*& 11 / 14 / 10 / 14 ? **& *230\cr
Uranus&  5& *3.5& 12 / 12 / 13 / 12 / *9& ***7\cr
Neptune& 4& *2.3& 10 / 13 / 10 / 13 ? **& ***3\cr
\noalign{\vskip 3pt\hrule\vskip 5pt}
}}
\endPlancktable
\endgroup
\end{table}

In this section, we describe the \Planck~HFI reconstruction of planet flux densities using a time-domain fit of the scanning-beam shape to the detector response from individual planet crossings. The peak of the reconstructed planet signal is then combined with information about the detector spectral response, planet ephemeris, and beam size to estimate the planet thermodynamic temperature, an intrinsic property of the planet. The detailed description of the algorithm used for the planet flux density analysis that is presented in the following subsections is meant to leave as little room for ambiguity as possible.

According to the International System of Units (SI) and International Astronomical Union (IAU) conventions, radiance has SI derived units $\mrm{W\, m^{-2}\,sr^{-1}}$. Similarly, spectral radiance has SI derived units $\mrm{W\, m^{-2}\,sr^{-1}\,Hz^{-1}}$ \citep{NISTGuide2008,IASManual1989}. The IAU states that flux density, $\mathrm{Jy}$ (or~$10^{-26}\,\mrm{W\, m^{-2}\,Hz^{-1}}$), can also be reported in the appropriate context. In this paper, we use flux density to refer to quantities with units $\mrm{W\, m^{-2}\,Hz^{-1}}$. The flux density of a source is obtained by calculating the product of the solid angle extended by the source on the sky and the spectral radiance of the source.

The planets are often used for calibration and general instrument characterization of CMB experiments. For example, planet flux densities measured by \wmap are described in \cite{weiland2010} and \cite{Bennett2013} and those measured by \act in \cite{Hasselfield2013} and \cite{,ACT_PlanckCross2013}. Many other CMB observatories have also reported, or calibrated against, planet flux densities (e.g., \cite{Goldin1997,Mauskopf2000,Runyan2003}). 

\Planck\ LFI reports on planet brightness and calibration in \cite{planck2013-p02b,planck2014-a06}. In this paper, we include estimates of Jupiter's thermodynamic temperatures at LFI frequencies, as reported in \cite{planck2014-a06}. The work presented in that paper does not consider planets other than Jupiter, since the main purpose of that work was to use Jupiter's high S/N ratio to compare LFI and \wmap absolute calibrations. For Saturn's thermodynamic temperature at LFI frequencies we use earlier results presented in \cite{planck2013-p02b}, but correct them by approximately 10\% to account for Saturn's oblateness, an effect not considered in the LFI analysis. The Saturn results only include data extending through the end of LFI's first full year of observations (2010), approximately 2.5 full sky surveys. The LFI Saturn brightness estimates will improve with the inclusion of more data. We choose not to use estimates provided in \cite{planck2013-p02b} for Mars, Uranus or Neptune, because of either low S/N (Uranus and Neptune) or systematic effects possibly related to pointing and beam reconstruction (in the case of Mars). Although the preliminary flux density results from LFI are discussed in this new paper, it is important to note that our main emphasis is HFI analysis of planet brightness.

In this paper we use the terms thermodynamic (blackbody) temperature and Rayleigh-Jeans (RJ) temperature, both of which can be used to estimate the effective temperature of a source over some frequency range. There is, however, a significant difference between how these two quantities are used to derive spectral radiance (see Sect.~\ref{sec:psffit}). In radio astronomy, the term brightness temperature is normally used to indicate RJ temperature and this term has also been used to indicate RJ temperature in the context of planet spectral radiance at mm-wavelengths \cite{page2003a,weiland2010,Bennett2013,Hasselfield2013}. However, some authors use brightness temperature to indicate thermodynamic (blackbody) temperature at mm-wavelengths \cite{Rather1974,Gibson2005}. We choose not to use the term brightness temperature in this paper to reduce chances of confusion.

\subsection{General analysis description}
\label{sec:beam_description}

The \Planck-HFI scanning beams are derived from a combination of Jupiter and Saturn observations and then extended using a diffraction model where the signal-to-noise ratio is low \citep{planck2014-a08}. In this paper, we use the acronym ``PSF'' (which stands for point spread function) to refer to the \Planck-HFI scanning beams. The PSF fit to planet timelines returns a signal peak amplitude, $\DeltaT_\mrm{p}$, which is used to estimate thermodynamic temperatures and spectral radiance. A PSF-fit method is preferred because the poor sampling in the cross-scan direction for individual planet observations does not allow a straightforward application of aperture photometry. For an experiment like \Planck, it is important to distinguish between the so-called optical, scanning, and effective beams. The distinction is made clear in Sect.~1 of \cite{planck2013-p03c} and also described in the \Planck\ Explanatory Supplement \cite{planck2014-ES}.

\begin{figure}[t!]
\begin{center}
\includegraphics[width=\columnwidth]{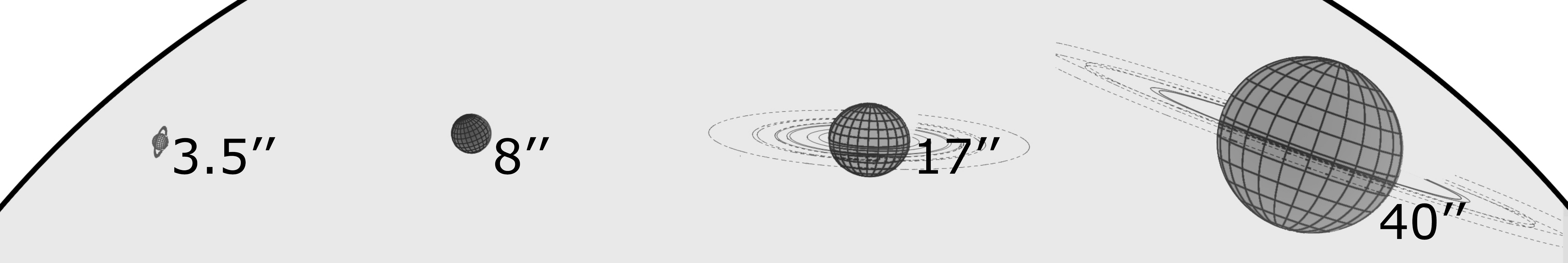}
\caption[Mars, Jupiter, Saturn, and Uranus]{Composite diagram showing four planets as viewed from the centre of Earth around the turn of the year 2010; the view from \lto would have been similar. From left to right, the figure shows Uranus, Mars, Saturn, and Jupiter. The planets are drawn on the same scale and the numbers to the right of each planet represent the approxiate apparent diameter (in arcmin) of the planets at this particular epoch. A fraction of a typical 143\GHz\ beam is shown by the black line and large grey region, the diameter of this circle corresponds to the FWHM of the beam. This suggests that the planets are point sources relative to the spatial response of the instrument. The planet diagrams have been extracted from \cite{Showalter2014}.
\label{fig:planets_2scale}
}
\end{center}
\end{figure}

In this section, we present a quantitative description of the analysis. Together, the pipeline description and tabulated results allow independent verification of the derived fluxes, up to an accurate estimate of signal peak amplitude, $\DeltaT_\mrm{p}$, a quantity that we extract from signal timelines. The \Planck-HFI scanning-beam solid angles, spectral response, and planet flux density estimates can be accessed on the \Planck\ Legacy Archive (PLA), the \Planck\ Explanatory Supplement, and on servers maintained by the Infrared Processing and Analysis Center (IPAC).\footnote{See \url{http://pla.esac.esa.int/pla/}, \url{https://wiki.cosmos.esa.int/planckpla2015/} and \url{http://irsa.ipac.caltech.edu/Missions/planck.html}. The planet data will be made available at the time of publication.}

\subsubsection{Description of PSF fit and function definitions}
\label{sec:psffit}

Within a region centred on the planet crossing, the PSF fit is a time-domain minimization of
\begin{equation}
\chi ^{2}= \sum _{i}\left( s_i - D (\theta _i,\phi _i) - \DeltaT_\mrm{p} \left[P (\theta _i,\phi _i) + g_{\mrm{NL}} P( \theta _i, \phi _i)^2 \right]\right)^{2} / \sigma_i^{2},
\label{eq:fit}
\end{equation}
where $D(\theta,\phi)$ and $P(\theta,\phi)$ are functional descriptions of the astrophysical background (everything except the planet) and the \Planck-HFI scanning beam, respectively, and $s_i$ represents the signal timeline, with $i$ indicating the sample index. The scanning beam is assumed to be constant throughout the mission, since no evidence to the contrary has been found.

The signal timelines used here have been processed in the same way as the scanning-beam planet data described in Appendix B of \cite{planck2014-a08}. The key differences from the main science timelines are a second deglitcher and baseline removal at 60 second intervals in the timelines. Since the satellite scans with one full revolution per minute, the baseline removal corresponds to the removal of a constant offset over the 360\deg\ circle. The term $g_\mathrm{NL}$ is a nonlinear gain term describing the response of the bolometer, necessary for observations of Jupiter above 217\GHz. We assume that $g_\mathrm{NL} = 0$ for all planet observations except Jupiter. The error term $\sigma_i^2$ is an estimate of the uncertainty in each time-ordered data sample, derived from the rms of the background-subtracted data more than a degree away from the planet. Because of nonlinearity and possible detector saturation during Jupiter observations, we mask parts of the timelines that fall within some minimum radius to the planet centre. This minimum radius is 10, 8, 5, 5, 5, and 5\arcmin\ at 100, 143, 217, 353, 545, and 857\GHz, respectively. The fit is performed using the Levenberg-Marquardt algorithm implemented in the {\tt scipy} package \citep{scipy}.

The signal timeline is calibrated in units of $\mrm{K_{CMB}}$ and $\mrm{MJy\, sr^{-1}}$ for 100--353 and 545--857\GHz, respectively \citep{planck2013-p03f,planck2014-a09}. Implicit in these two units, both of which can be related to signal intensity ($\mrm{W\, m^{-2}}$), are assumptions about the spectral energy distribution (SED) of the calibration source. 

The calibration to units of $\mrm{K_{CMB}}$, which is derived from comparing to a model of the CMB orbital dipole, assumes an SED for $\partial B(\nu,T)/\partial T |_{T = T_\mrm{CMB}}$, where
\begin{equation}
B(\nu,T) \equiv \frac{2h\nu ^3}{c^2}\frac{1}{e^{h\nu /kT}-1} \qquad\left[ \mrm{W\, m^{-2}\,sr^{-1}\,Hz^{-1}} \right]
\label{eq:Bnu}
\end{equation}
is the Planck blackbody function, $h$, $k$, and $c$ represent the Planck and Boltzmann constants and the speed of light in vacuum, respectively, and $T_\mrm{CMB} \equiv 2.7255$\,K is the temperature of the CMB monopole \citep{Fixsen1994, Fixsen2009}. In this paper, we refer to the parameter $T$ as thermodynamic temperature. Thermodynamic temperature should be distinguished from Rayleigh-Jeans temperature, $T_\mrm{RJ}$, which can also be used to calculate spectral radiance through the equation
\begin{equation}
\tilde{B}(\nu,T_\mrm{RJ}) \equiv \frac{2k\nu ^2}{c^2} T_\mrm{RJ}; \qquad\left[ \mrm{W\, m^{-2}\,sr^{-1}\,Hz^{-1}} \right].
\label{eq:Brj}
\end{equation}

The partial temperature derivative of the Planck blackbody function is 
\begin{eqnarray}
B^{\prime}(\nu,T) &\equiv& \pder{ B(\nu,T)}{T} \\ %\bigg| _{T = 2.7255} \\
&=& \frac{2\nu^2k}{c^2}\frac{x^2e^x}{(e^x-1)^2}, % \bigg| _{T = 2.7255},
\label{eq:dbdt}
\end{eqnarray}
where $x \equiv h\nu/kT$ and we define
\begin{equation}
b^{\prime}_\nu \equiv \partial B(\nu,T)/\partial T |_{T = T_\mrm{CMB}},
\label{eq:dbdt2}
\end{equation}
for consistency with \cite{planck2013-p03} and \cite{planck2013-p03d}. Spectral radiance (in units of $\mrm{W\, m^{-2}\,sr^{-1}\,Hz^{-1}}$) can be obtained by calculating the product $T \times B^{\prime}_\nu(\nu,T)$, where $T$ is the perceived source temperature relative to a $T_\mrm{CMB}$ blackbody. Section \ref{sec:colour_correction} gives the expression that can be used to convert between $\mrm{K_{CMB}}$ and $\mrm{W\, m^{-2}\,sr^{-1}}$.

The 545- and 857-GHz frequency bands were calibrated against models of Uranus and Neptune thermodynamic temperature \citep{planck2014-a09,Bertincourt2016}. The calibration adapted a reference in which the spectral radiance of a fiducial source, $S(\nu)$, follows $\nu S(\nu) = \mrm{constant}$ (discussed further in Sect.~\ref{sec:colour_correction}). This particular approach to calibration of the sub-mm bands was chosen to be analogous to that of the SPIRE instrument on the \herschel satellite \citep{Bendo2013,Swinyard2014}.

The 545- and 857-GHz beams are multi-moded in order to increase the throughput, and therefore increase the signal-to-noise ratio of those detectors at the cost of limiting the resolution \citep{Murphy2002,Murphy2010}. Neither scan strategy nor sample rate justify finer resolution at those frequencies. Unfortunately, the multi-moded nature of the sub-mm beams seriously complicates any analytical description of their spatial response \citep{Murphy1991}. In particular, a function basis with a Gaussian envelope does not easily capture the main-beam shape.

As part of the PSF fit to the planet timelines, we subtract best estimates for the astrophysical background, $D(\theta, \phi)$. Depending on the frequency, this background can have contributions from the CMB, dust, as well as synchrotron, and other Galactic emission \citep{planck2014-a12}. The background estimate is derived from a bilinear interpolation of the 2015 release maps at each frequency. 

In the following analysis, it is assumed that the correction, $g_\mrm{NL}$, captures, to first order, the large signal nonlinearity of the bolometers. Due to the wide dynamic range of the devices, this correction is only significant for observations of Jupiter at frequencies above 217\GHz; the $g_\mrm{NL}$ correction is therefore only applied for these observations. The validity of this correction is corroborated by a significant improvement in fit quality for these observations. 

We note that the ADC nonlinearity correction, which is discussed in detail in \cite{planck2014-a08} and \cite{planck2014-a09}, and required due to the limited range of the sampling of the ADC, is applied to all detector timelines and across all HFI frequency bands used in this analysis. This ADC correction reduces time variation in the gain down to the $2\times10^{-3}$ level \citep{planck2014-a10}. After implementing this correction, we assume that the calibration is constant in time.

\subsubsection{Calibration of sub-mm channels}
\label{sec:submmcal}

Calibration of the sub-mm channels uses the planet time-ordered-data processed for beam reconstruction in the 2015 public data release \citep{planck2014-a08}\footnote{\url{http://www.cosmos.esa.int/web/planck/pla}} which adds an improved planet timeline despiking algorithm and baseline drift removal compared to that used in \cite{Bertincourt2016} and \cite{planck2014-a09}.\footnote{These timelines have not been made available on the PLA.} Because of this, and due to differences in algorithms, the 545- and 857-GHz\ flux density for Uranus and Neptune described in this paper is not expected to agree perfectly with model predictions (see Sect.~\ref{sec:ura}). However, large discrepancies between the results presented here and the results found in \cite{planck2014-a09} would hint at poorly understood errors (and are not seen).

\subsubsection{Signal estimates}
\label{sec:signal_estimates}

The maximum planet signal amplitude, $\Delta T_\mrm{p}$, is influenced by a number of instrument and source-specific properties. Assuming an infinitely fast time response, a detector observing a point-like blackbody head-on should measure a background-removed signal, $s_i$, according to \citep{Kraus1950,page2003a}
\begin{eqnarray}
s_i &=& \iint \d\Omega\d\nu \tau^{\prime}(\nu) P(\theta_i,\phi_i) A_\mrm{eff} (\nu) B (\nu,T) \nonumber\\ 
&=& \frac{c^{2}}{\Omega _\mrm{b}} \iint \d\Omega\d\nu N \tau^{\prime}(\nu) P (\theta_i,\phi_i) B(\nu,T) / \nu^{2}\nonumber\\ 
&=& \frac{\Omega_{\mrm{p},i}}{\Omega _\mrm{b}}\int \d\nu \tau (\nu) B (\nu,T) \qquad \left[ \mrm{W\, m^{-2}\,sr^{-1}} \right].
\label{eq:P}
\end{eqnarray}
Here $\tau(\nu)$ is the \'etendue-normalized detector spectral response function (SRF), \citep{pajot2010,ade2010,planck2013-p03d}, $P(\theta,\phi)$ is the instrument scanning beam, normalized to unity at its peak, $\Omega _{\mrm{p},i}$ is the time-varying planetary solid angle as seen by the detector, and $\Omega _\mrm{b}$ is the scanning-beam solid angle, $\Omega _\mrm{b} = \int \d\Omega P(\theta,\phi)$. The first and second integrals are over solid angle covered by the planet disc and over frequency, respectively. The above derivation adopts the well known relationship between effective telescope area, number of radiation modes, frequency, and beam solid angle, namely $A_\mrm{eff}(\nu) = Nc^{2}/(\nu^{2}\Omega_\mrm{b})$ \citep{Hudson1974,Hodara1984}. We note that in cases where the number of radiation modes, $N$, is not identically equal to unity, $A_\mrm{eff}(\nu)$ can be a strong function of frequency. Finally, the SRF incorporates the detector throughput, such that
\begin{equation}
\tau (\nu) \equiv \tau ^{\prime} (\nu) Nc ^{2}/\nu^{2} = \tau ^{\prime} (\nu) A_\mrm{eff}(\nu)\Omega_\mrm{b}. 
\end{equation}
The last step of the derivation presented in Eq.\ (\ref{eq:P}) is obtained by assuming that $P(\theta,\phi) \simeq 1$ over the planet disc area so that the integral over solid angle simply yields $\Omega _\mrm{p}$.\footnote{In a worst case scenario, the normalized beam response will have fallen to 0.998 at Jupiter's limb when observed using a Gaussian beam with FWHM of 4\parcm5.} This is a good approximation for all five planets observed by \Planck\ (see Fig.~\ref{fig:planets_2scale}). The ratio $\Omega _\mrm{p} / \Omega _\mrm{b}$ is the beam correction factor often discussed in relation to flux estimates \citep{Ulich1976,Griffin2013,Swinyard2014}. Finally, we note that the signal amplitude obtained from viewing the planet off axis relative to the peak scanning-beam amplitude is simply the peak signal amplitude scaled by the relative change in scanning-beam response, $P(\theta,\phi)/P(0,0)$, where $P(0,0) = 1$ is the normalized peak response of the scanning beam and $P(\theta,\phi)$ is the off-axis response. Equation \ref{eq:P} can be modified to incorporate the nonlinearity correction, $g_\mrm{NL}$, by replacing $P(\theta_i,\phi_i)$ with $P(\theta_i,\phi_i) + g_\mrm{NL}P(\theta_i,\phi_i)^{2}$.

In deriving Eq.\ (\ref{eq:P}), we made the simplifying assumption that the beam response $P(\theta, \phi)$, and therefore beam solid angle, $\Omega _\mrm{b}$, would not vary with frequency. This is certainly incorrect \citep[see e.g.,][]{maffei2010}. However, as the \Planck~HFI scanning beam is calibrated using a combination of Jupiter and Saturn observations, whose SED closely resembles that of the Rayleigh-Jeans spectrum, it is reasonable to assume that the estimated scanning-beam solid angle, $\Omega _\mrm{b}$, properly represents a beam solid angle with frequency weighting appropriate for analysis of planet brightness \citep[see][for a discussion of beam colour corrections]{planck2014-a08}. 

So far, our notation has adopted a source thermodynamic temperature, $T$, that is frequency independent. However, models of planet thermodynamic temperature show significant variation over a typical 30\,\% frequency bandwidth of a \Planck-HFI detector \citep{planck2013-p03d}. To remedy this, we can allow the source thermodynamic temperature to be frequency-dependent. We define the frequency dependent thermodynamic temperature as the temperature $T(\nu)$ that is required to describe the source spectral radiance (defined by the Planck blackbody function):
\begin{equation}
B (\nu,T(\nu)) = \frac{2h\nu ^3}{c^2}\frac{1}{e^{h\nu /kT(\nu)}-1} \qquad\left[ \mrm{W\, m^{-2}\,sr^{-1}\,Hz^{-1}} \right].
\label{eq:brightness_temperature}
\end{equation}
In this paper, we choose to report monochromatic thermodynamic temperatures at a standard reference frequency, $\nu _\mrm{c}$, where $\nu _\mrm{c} \in$ \{100, 143, 217, 353, 545, 857\}\GHz. A colour correction is required to report the thermodynamic temperature at these reference frequencies (see Sect.~\ref{sec:colour_correction}).

An estimate for the CMB monopole has already been subtracted from the \Planck~HFI timelines. As the planets block the CMB monopole, this signal subtraction will lead to an underestimate for the absolute brightness of the planets \citep[see e.g.,][]{page2003a}. This will cause an approximately 1~K bias in thermodynamic temperature estimates at 100\GHz\ and successively lower bias at higher frequencies. We choose to report thermodynamic temperature before correcting for this effect, since this simplifies comparison with \cite{weiland2010}. The absolute thermodynamic temperature $T_\mrm{abs}$ can be calculated from the uncorrected thermodynamic temperatures, $T$, by solving the following transcendental equation
\begin{equation}
B(\nu,T_\mrm{abs}) = B(\nu,T) + B(\nu,T_\mrm{CMB}),
\end{equation}
where $B$ represents Planck's blackbody function and $T_\mrm{CMB} = 2.7255$\,K. Note that thermodynamic temperatures reported in this paper correspond to $T$, not $T_\mrm{abs}$. We also note that the sum of two blackbody functions is not a blackbody function.

The solid angle of the projected oblate spheroid face of each planet is derived as
\begin{equation}
\Omega_\mrm{p} = \pi r_\mrm{eq} r_\mrm{pp} / R^2,
\end{equation}
where $r_\mrm{eq}$ is the equatorial radius of the planet, $r_{\mrm{pp}}$ is the projected polar radius of the planet,  and $R$ is the distance between the \Planck\ spacecraft and the planet. The projected polar radius is given by:
\begin{equation}
r_{\mrm{pp}}  = r_{\mrm{pol}} \sqrt{\cos^2 (D_\mrm{w}) +  (r_{\mrm{eq}}/r_{\mrm{pol}})^{2} \sin^2 (D_\mrm{w})}.
\end{equation}
Here $D_\mrm{w}$ is the latitude of the planet coordinate system relative to \Planck\ (needed for this projection) and $r_{\mrm{pol}}$ is the true polar radius of the planet.

For a given detector, a single planet observation spans a few hours, during which the planet solid angle can vary by a non-negligible amount. Additionally, in the case of Mars, the observed thermodynamic temperature distribution varies with time, both on diurnal (Mars rotation) and seasonal timescales (see Sect.~\ref{sec:mars}). Variations in planet disc size are accounted for by dividing the signal amplitude with an estimate of time-variations in the planet solid angle
\begin{equation}
s_i = \tilde{s}_i \frac{\Omega _{\mrm{p},i}}{\Omega _\mrm{p,bw}},
\end{equation}
where $\tilde{s}_i$ is the uncorrected signal timeline, $\Omega _{\mrm{p},i}$ is an estimate for the planet solid angle at time $i$, and $\Omega _\mrm{p,bw}$ is an estimate for the planet solid angle at the beam-weighted average time of observation. Due to proximity, this effect is largest for Mars, with the planet solid angle changing by up to 0.07\,\% in an hour. We implement this correction in our analysis. For Mars, it can also be important to account for time-varying albedo by calculating
\begin{equation}
s_i = \tilde{s}_i \frac{\Omega _{\mrm{p},i}}{\Omega _\mrm{p,ref}} \frac{I _{\mrm{p},i}}{I _\mrm{p,ref}},
\end{equation}
where $\Omega _\mrm{p,ref}$ and $I _\mrm{p,ref}$ are, respectively, the predicted planet solid angle and flux density at a fixed reference time, and $\Omega _{\mrm{p},i}$ and $I _{\mrm{p},i}$ are the corresponding time-varying model predictions. We will provide results with and without this additional percent-level correction (see Sect.~\ref{sec:mars}). However, all Mars brightness measurements discussed in this paper include this correction.

% See test_cc.py
\begin{table}[t]
\begingroup
\newdimen\tblskip \tblskip=5pt
\caption{Band-average colour correction factors used in this analysis.  For Mars, $\kappa _1$ is insensitive to the survey number, to a very good approximation. We note that both $\kappa _1$ and $\kappa _2$ are unitless. Band-averages are calculated with uniform detector weights.}
\label{tab:unit_conversion}
\vskip -5mm
\footnotesize
\setbox\tablebox=\vbox{
   \newdimen\digitwidth
   \setbox0=\hbox{\rm 0}
   \digitwidth=\wd0
   \catcode`*=\active
   \def*{\kern\digitwidth}
   \newdimen\signwidth
   \setbox0=\hbox{+}
   \signwidth=\wd0
   \catcode`!=\active
   \def!{\kern\signwidth}
   \newdimen\pointwidth
   \setbox0=\hbox{\rm .}
   \pointwidth=\wd0
   \catcode`?=\active
   \def?{\kern\pointwidth}
\halign{\hbox to 2.0cm{#\leaderfil}\tabskip 1.0em&
        \hfil#\hfil&
        \hfil#\hfil&
        \hfil#\hfil&
        \hfil#\hfil&
        \hfil#\hfil&
        \hfil#\hfil\tabskip 0pt\cr
\noalign{\doubleline}
\noalign{\vskip -2pt}
\omit\hfil Freq.~[GHz]\hfil& $\kappa_1^{\mrm{Mar}}$& $\kappa_1^{\mrm{Jup}}$& $\kappa_1^{\mrm{Sat}}$& $\kappa_1^{\mrm{Ura}}$& $\kappa_1^{\mrm{Nep}}$& $\kappa _2$\cr
\noalign{\vskip 3pt\hrule\vskip 5pt}
100& 0.967& 0.968& 0.968& 0.974& 0.979&   N/A \cr
143& 0.999& 0.999& 0.999& 1.003& 1.003&   N/A \cr
217& 0.953& 0.955& 0.970& 0.966& 0.979&   N/A \cr
353& 0.952& 0.956& 0.959& 0.967& 0.953&   N/A \cr
545& 0.948& 0.942& 0.822& 0.964& 0.975& 1.013 \cr
857& 0.985& 1.001& 1.005& 0.993& 1.000& 0.995 \cr
\noalign{\vskip 3pt\hrule\vskip 5pt}
}}
\endPlancktable
\endgroup
\end{table}

\begin{figure*}[t]
\begin{center}
\includegraphics[width=18cm]{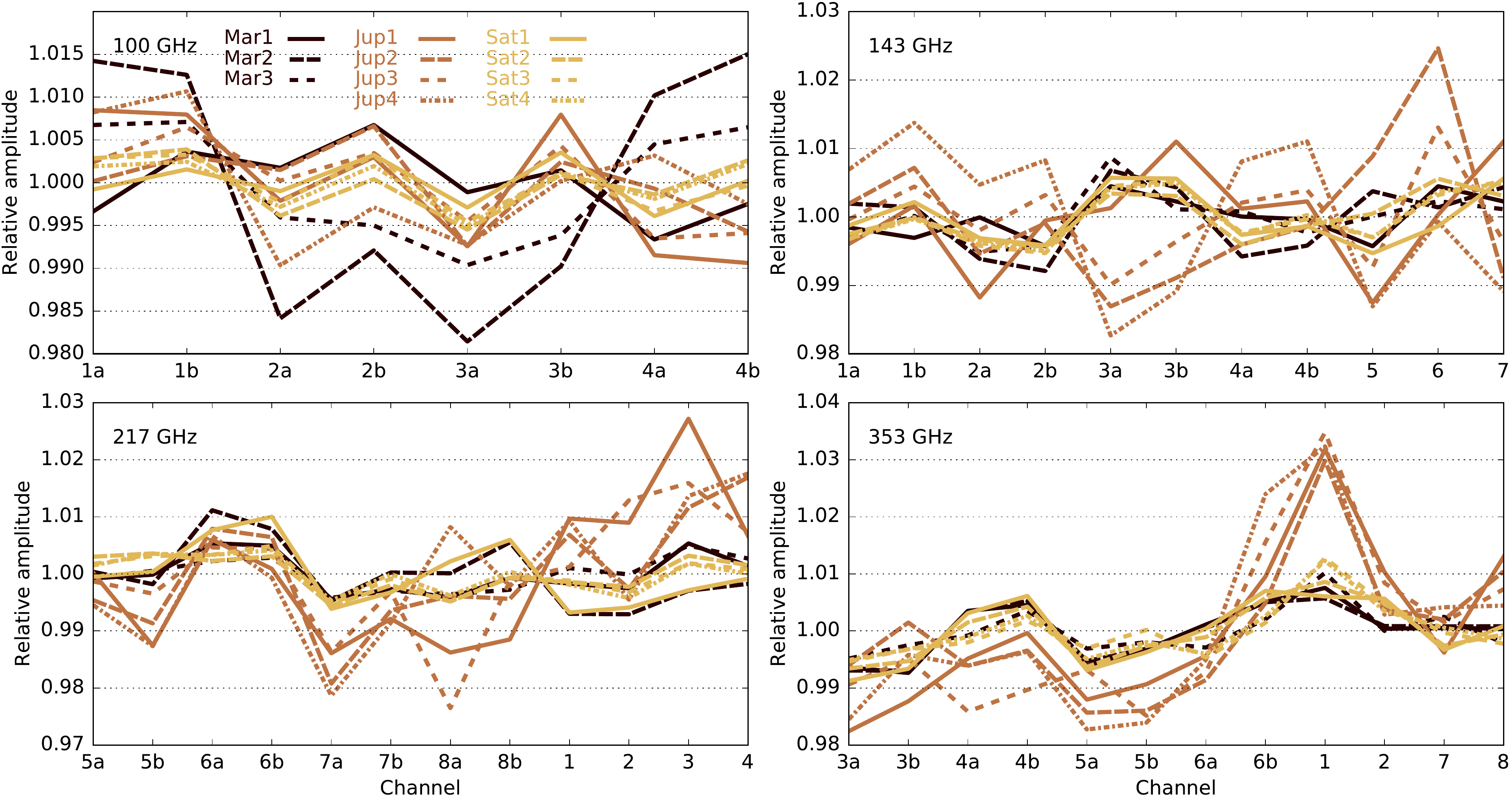}
\caption[In-band correlation]{Relative amplitude in derived flux density from observations of Jupiter, Saturn, and Mars at 100--353\GHz\ (Mars is limited by statistical error at 100\GHz). The flux density has been normalized to the band-mean. A clear correlation in relative flux amplitudes between planet observation periods is seen. Lines are used to guide the eye between flux values within a planet observation. For the top panel (100\GHz), the average standard deviation of the data is approximately 0.4\,\%.
}
\label{fig:band_correlation}
\end{center}
\end{figure*}

\subsubsection{Unit and colour conversion}
\label{sec:colour_correction}

Unit conversion can be obtained by combining estimates for spectral response, $\tau (\nu)$, with assumptions about the signal spectral energy distribution \citep[see][for a general discussion of unit conversions]{planck2013-p03d}. The unit conversion factor, $U_\mrm{C}$, that can be used to convert from the SI-unit for radiance $\mrm{W\, m^{-2}\,sr^{-1}}$ to units of $\mrm{K_{CMB}}$ is
\begin{equation}
U_\mrm{C} \equiv \int \d\nu \tau(\nu) b_\nu^{\prime} \qquad \left[ \mrm{W\, m^{-2}\,sr^{-1}}\,\mrm{K_{CMB}^{-1}} \right],
\end{equation}
where we have used the $b_\nu^{\prime}$ definition from Eq.\ (\ref{eq:dbdt2}). We can use this to convert our estimate for the $\mrm{K_{CMB}}$ peak signal amplitude, $\Delta T_\mrm{p}$, to the corresponding peak in radiance:
\begin{equation}
\Delta L_\mrm{p} = U_\mrm{C} \Delta T_\mrm{p} \qquad \left[ \mrm{W\, m^{-2}\,sr^{-1}} \right].
\label{eq:uc}
\end{equation}

We use colour corrections so that we can report flux density or spectral radiance, $S_\mrm{c}$, at a particular reference frequency, $\nu_\mrm{c}$. The colour correction incorporates the spectral response function of the detector in question as well as an assumption about the spectral energy distribution of the source. Assuming $M_\mrm{p}(\nu)$ is a model that accurately represents the spectral radiance of a planet at any frequency, the spectral radiance at a standard reference frequency is
\begin{equation}
S_\mrm{c} = \left( \int \d\nu \tau(\nu) \frac{M_\mrm{p}(\nu)}{M_\mrm{p}(\nu _\mrm{c})}\right)^{-1}\mkern-12mu\Delta L_\mrm{p} \equiv \kappa _1 \Delta L_\mrm{p}.
\label{eq:cc_planet_model}
\end{equation}
Despite being calibrated on Uranus and Neptune, this type of correction is also necessary for the 545 and 857\GHz\ bands, since the official calibration included a colour correction appropriate for a flat spectrum. However, because the 545 and 857\GHz\ bands implemented a calibration that differs from the rest of the \Planck\ bands, an additional correction is required
\begin{equation}
S_\mrm{c} = \left( \int \d\nu \tau(\nu) (\nu _\mrm{c}/\nu) \right) \kappa_1 \Delta L_\mrm{p} \equiv  \frac{\kappa_1}{\kappa_2} \Delta L_\mrm{p}.
\end{equation}
The band-average values for the colour correction described here are shown in Table \ref{tab:unit_conversion}.

Because of our non-negligible detector bandpass, estimates of planet spectral radiance, flux density, or thermodynamic temperature at a particular frequency are model dependent. Since we focus on comparing our flux density estimates with ESA model predictions, we report planet thermodynamic temperatures that can include some deviations from a perfect Rayleigh-Jeans like spectrum (Sects.~\ref{sec:mars}--\ref{sec:ura} describe models that are used for calculating these colour corrections). Other experiments, however, including \wmap and \act \citep{weiland2010,Hasselfield2013}, report an intrinsic temperature assuming that the planets SED can be approximated with a Rayleigh-Jeans-like spectrum. This essentially means that $M_\mrm{p} (\nu) \propto \nu^2$ is imposed in their estimates of planet temperature. Because no statement is made to the contrary, we also conclude that \wmap and \act assume a fixed RJ temperature across the bandpass. When comparing our results with measurements from \wmap and \actn, we make sure to account for this discrepancy (see Sect.~\ref{sec:wmap}). This approximation, although quite valid in the high-temperature or low-frequency limit, can for example induce a 1\,\% shift in the estimated flux density at 217\GHz\ for some of the colder planets.

\subsubsection{Summary of adopted methods}
\label{sec:bet}
For each planet observation we estimate the peak amplitude $\Delta T_\mrm{p}$ by minimizing the residual in Eq.\ (\ref{eq:fit}). We then use derivations from Sects.~\ref{sec:signal_estimates}--\ref{sec:colour_correction} to express $\Delta T_\mrm{p}$ in terms of instrument properties as well as planet thermodynamic temperature and size.

For the 100--353\GHz\ bands, we can now combine Eqs.\ (\ref{eq:P}), (\ref{eq:uc}), and (\ref{eq:cc_planet_model}) to write
\begin{equation}
U_\mrm{C} \Delta T_\mrm{p} = \frac{1}{\kappa _1} \frac{\Omega_\mrm{p}}{\Omega _\mrm{b}}\int \d\nu \tau (\nu) B (\nu _\mrm{c},T(\nu _\mrm{c})).
\label{eq:Tb1}
\end{equation}
No unit conversion is required for the 545 and 857\GHz\ bands, and the expression becomes
\begin{equation}
\Delta T_\mrm{p} = \frac{\kappa _2}{\kappa _1} \frac{\Omega_\mrm{p}}{\Omega _\mrm{b}}\int \d\nu \tau (\nu) B (\nu _\mrm{c},T(\nu _\mrm{c})).
\label{eq:Tb2}
\end{equation}
Equations\ (\ref{eq:Tb1}) and (\ref{eq:Tb2}) allow us to estimate the planet thermodynamic temperature at a standard reference frequency, $T(\nu _\mrm{c})$, by solving these transcendental equations. The corresponding flux density, $I_\mrm{p}$, can then be found by calculating
\begin{equation}
I_\mrm{p} = \Omega _\mrm{p} B (\nu _\mrm{c},T(\nu _\mrm{c})) \qquad \left[\mrm{W\, m^{-2}\,Hz^{-1}} \right].
\label{eq:flux_eq}
\end{equation} 

\subsection{Comparison with other fit methods}

The determination of the peak signal amplitude, $\Delta T_\mrm{p}$, is based on a least squares fit to the data (see Eq.\ \ref{eq:fit}). This method of determining the maximum likelihood value for the peak signal is by no means unique \citep[see for example the algorithm described in][]{planck2014-a09}. 

We have compared our results with a Gauss-Hermite (GH) reconstruction of the peak amplitude and found that the two methods agree quite well on average. Descriptions of the Gauss-Hermite reconstruction method can be found for example in \cite{hill2009}, \cite{huffenberger2010}, \cite{Monsalve2010}. Even with the elliptical Gaussian base parameters of the Gauss-Hermite functions fixed, the functions offer a versatile basis for deconstructing the signal independently of the PSF definition. Unfortunately, simulations have shown that a Gauss-Hermite reconstruction of the multi-moded sub-mm beam responses is significantly biased. We therefore limit the comparison between flux density estimates from PSF and GH peak reconstruction to 100--353\GHz.

After decomposing the Gauss-Hermite coefficients in the time-domain, a map of the planet is reconstructed at an arbitrarily high resolution to produce an estimate for $\Delta T_\mrm{p,GH}$. We compare the ratio of the nominal peak amplitude estimate and the one derived using the GH decomposition, $r_\mrm{peak} = \Delta T_\mrm{p} / \Delta T_\mrm{p,GH}$, over 100--353\GHz\ for all planet observations and find that the distribution has a mean and standard deviation of $\mu _\mrm{r} = 1.0000$ and $\sigma _\mrm{r} = 0.0056$, respectively. Limiting that statistic to Mars, Jupiter, and Saturn, where the peak determination is not significantly affected by statistical noise, the mean and standard deviation of the distribution is $\tilde{\mu} _\mrm{r} = 0.9992$ and $\tilde{\sigma} _\mrm{r} = 0.0004$, respectively. Since all planet flux density estimates are derived from $\Delta T_\mrm{p}$, we take this as evidence that the analysis does not suffer from significant representation bias.

\subsection{Statistical error estimates}
\label{sec:stat_err}

The \Planck-HFI planet flux estimates are affected by a number of telescope and detector properties, including spectral response, beam solid angle, and absolute calibration. Known error sources and their estimated amplitudes are described below. All estimates of statistical error are input into a simple Monte Carlo routine that propagates these errors through to a determination of the planet thermodynamic temperatures and flux densities. These statistical errors are then combined with estimates for systematic errors (see Sect.~\ref{sec:systematic_errors}).

\begin{figure*}
\begin{center}
\includegraphics[width=15cm]{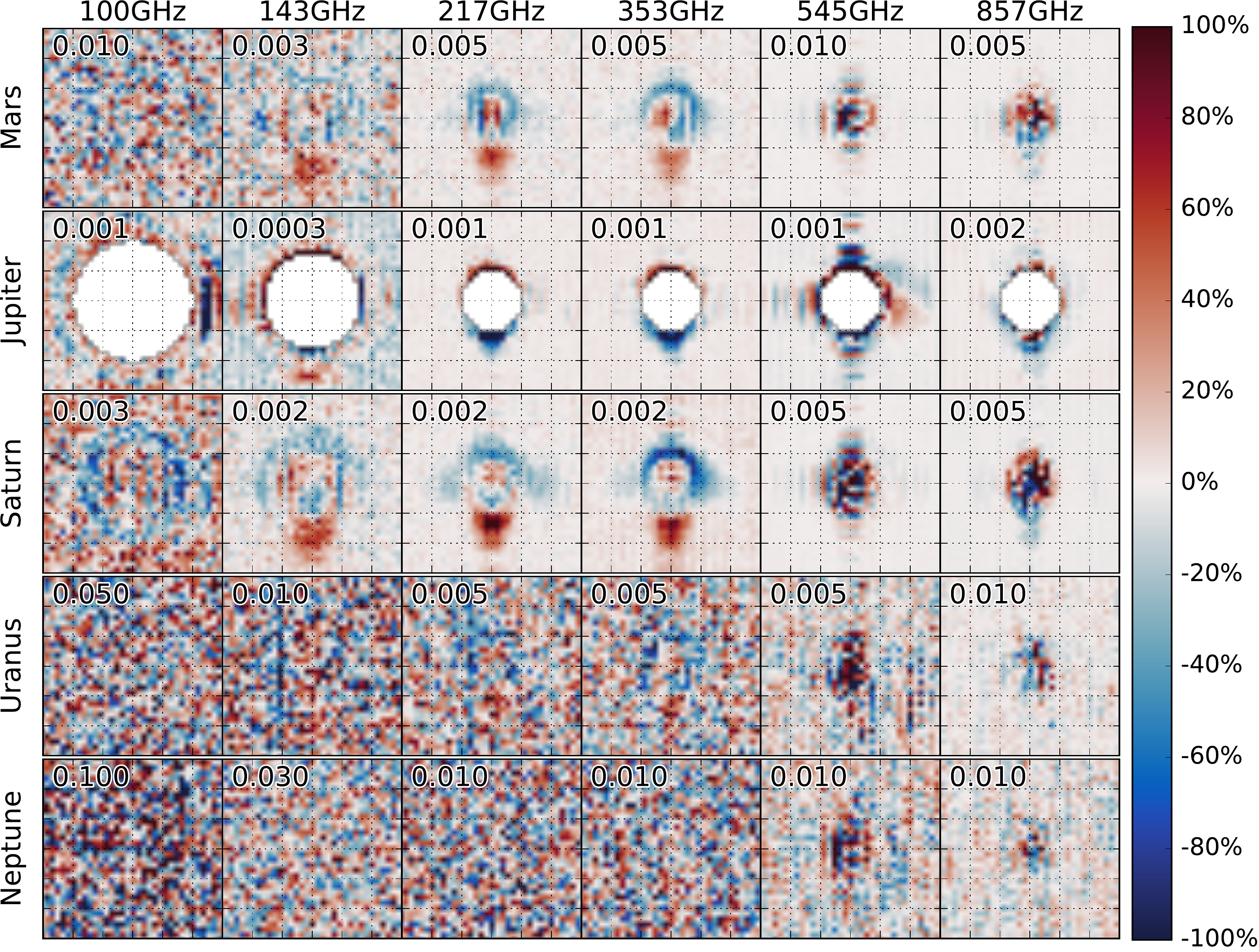}
\caption[Planet residuals]{Maps of the normalized and background-subtracted timeline residuals, $(s_i - D(\theta _i,\phi _i) - \DeltaT_\mrm{p} P (\theta _i,\phi _i))/\Delta T_\mrm{p}$, combining all available observations and detectors within a frequency band. The number shown in the top left corner of each panel represents the extent of the colour scale, e.g., a value of 0.01 means that the darkest colour corresponds to a 0.01 deviation from the peak response $\Delta T_\mrm{p}$, with red (blue) corresponding to positive (negative) deviation. We have masked out the centres of the Jupiter observations, since nonlinearity and saturation lead to a large residual. Each panel is 30\arcmin\ $\times$ 30\arcmin, and the co-scan direction is vertical on these plots (see discussion in Sect.~\ref{sec:introduction}).
}
\label{fig:fit_residuals}
\end{center}
\end{figure*}

\subsubsection{Planet solid angle}

The planet solid angles, $\Omega_\mrm{p}$, are estimated from the JPL ephemerides software \citep{Horizons} and subsequently corrected for planet oblateness (see Sect.~\ref{sec:signal_estimates}). The signal timestream is used to estimate the time at which each channel is centred on the planet. This corresponds to the time at which the peak signal is observed. Due to the raster-like scan-strategy of the \Planck\ satellite, a centred observation lasts for approximately one hour. Over this time, the solid angle extended by Mars as seen from \lto can change by up to 0.07\,\%. We account for this variation in our analysis (see Sect.~\ref{sec:signal_estimates}). For the error analysis, we conservatively assume a constant $\Delta\Omega _\mrm{p} = 0.05$\,\% fractional error in the planet solid angle estimate; this is conservative because JPL ephemerides are known with much greater accuracy. This source of error is likely negligible compared to other contributions.

\subsubsection{Absolute calibration and beam solid angle}

For detectors in the 100--353\GHz\ bands, the absolute calibration is based on a fit to the orbital dipole of the CMB. For the 545 and 857\GHz\ bands, however, the absolute calibration is referenced to ESA models of Uranus and Neptune thermodynamic temperature (see discussion in Sect.~\ref{sec:beam_description}). The absolute calibration of the instrument is described in detail in \cite{planck2014-a08}. The band-average statistical calibration error is found to be 0.09, 0.07, 0.16, 0.78, 1.1, and 1.4\,\% at 100, 143, 217, 353, 545, and 857\GHz, respectively. For the 545 and 857\GHz\ bands, an additional 5\,\% systematic calibration error is attributed to the planet flux models.

\begin{figure*}[t!]
\begin{center}
\includegraphics[width=18.6cm]{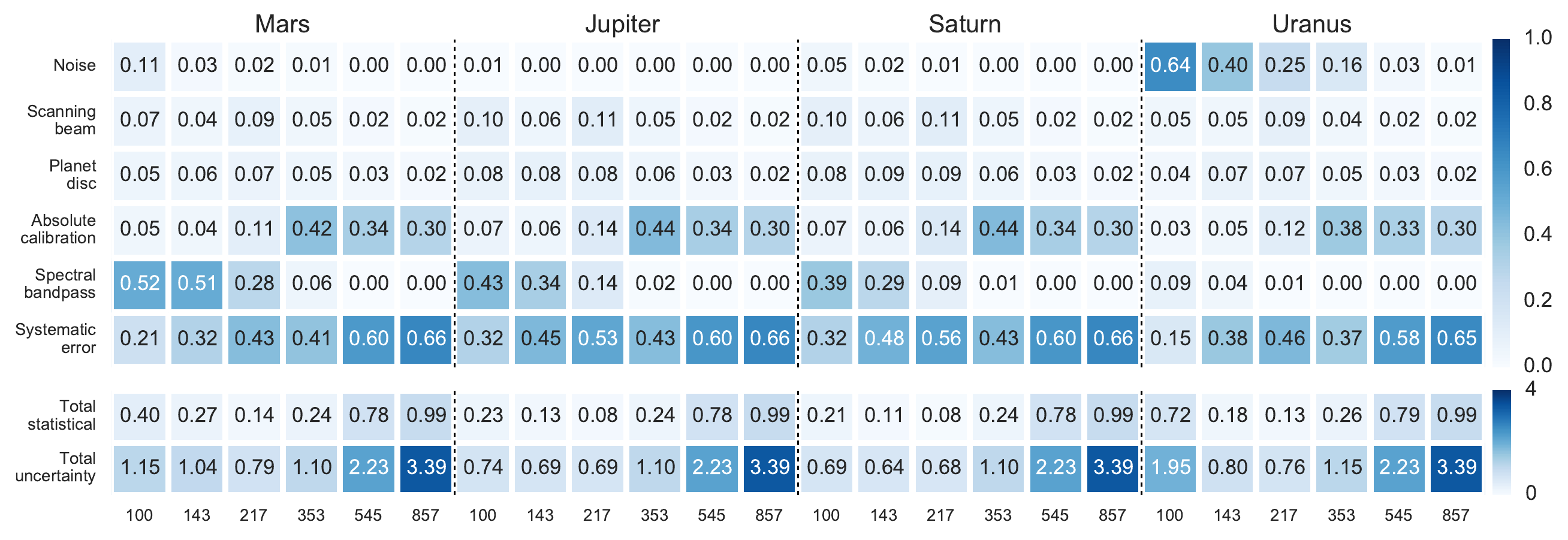}
\caption[Statistical and systematic error budget]{Normalized single-detector statistical and systematic uncertainties for Mars, Jupiter, Saturn, and Uranus at all six \Planck-HFI frequencies. The first six rows show the relative contribution of the different error terms described in Sects.\ \ref{sec:stat_err} and \ref{sec:systematic_errors}. From top to bottom, these six rows describe uncertainties related to: detector noise and astrophysical confusion; scanning-beam solid angle; planet-disc solid angle; absolute calibration; spectral response function; and an unknown systematic term described in Sect.\ \ref{sec:systematic_errors}. Note that the sum of the first six rows equals unity, but when combining errors, different terms are summed in quadrature. The last two rows show the total statistical uncertainty (in percent) and the total uncertainty in flux determination (in percent) appropriate for band-average flux density calculations, respectively. The total uncertainty is obtained by summing the total statistical and the total systematic uncertainties in quadrature. Note that statistical uncertainties (the first five rows) average down when we calculate band-averaged flux density estimates.
}
\label{fig:eb_all}
\end{center}
\end{figure*}

The absolute calibration of the sub-mm bands was recently discussed in \cite{planck2014-a10}. This paper states that the 545\GHz\ planet model calibrations have been compared with the absolute calibration obtained from observing the solar dipole and the first two peaks in the CMB angular power spectrum and that this analysis suggests 1.5\,\% agreement between the two calibration methods. Furthermore, the paper shows that the planet calibration agrees with the CMB calibration at the 1.5 and 2.5\,\% level for the 545 and 857\GHz\ bands, respectively. That analysis provides a crucial link between the CMB power and planet model predictions.

The cross-calibration of \Planck-HFI and the SPIRE instrument on \herschel is discussed in \cite{Bertincourt2016}. The relative calibration between SPIRE and \Planck-HFI is found to be $1.045\pm0.0085$ and $1.000\pm0.0080$ at 545 and 857\GHz, respectively.

Errors in determination of the scanning-beam solid angle, $\Omega_\mrm{b}$, are determined from Monte Carlo simulations of the hybrid B-spline beam. The band-average fractional scanning-beam errors are found to be 0.13, 0.07, 0.13, 0.09, 0.08, and 0.08\,\% at 100, 143, 217, 353, 545, and 857\GHz, respectively \citep{planck2014-a07}.

\subsubsection{Spectral bandpass}
The spectral bandpass of the entire HFI focal plane was estimated using a Fourier-transform spectrometer at Institut d'Astrophysique Spatiale (IAS) laboratory in Orsay, France \citep{planck2013-p03d}. By combining data from roughly 100 interferograms, those measurements obtained sub-percent accuracy at a resolution of approximately 0.5\GHz. With these per-frequency-bin spectral response measurements, Monte Carlo simulations can be used to obtain statistical errors in the derivation of unit or colour corrections. To perform these simulations, for each realization and frequency bin, the spectral response is modified by a Gaussian random variate before integrating over the spectral bandpass.

\subsection{Systematic uncertainties}
\label{sec:systematic_errors}

The uncertainties described above are largely expected to be random and uncorrelated. However, intra-frequency correlations display clear signs of systematic errors. Figure~\ref{fig:band_correlation} shows the relative amplitude of Mars, Jupiter, and Saturn derived from individual detectors within the 100--353\GHz\ bands. The 545 and 857\GHz\ intra-frequency correlations are omitted from Fig.~\ref{fig:band_correlation}; these can be reconstructed using the data accompanying this paper. If the flux determination was dominated by random uncorrelated errors, the relative planet flux densities would not be correlated between individual observations. Instead, a clear and repeatable pattern is observed for most planet observations shown.

Such a systematic effect could be caused by a number of sources, including an error in absolute calibration or scanning-beam solid angle. For the 100--353\GHz\ bands, a pure calibration error is ruled out as the sole cause of this effect by a number of intra-calibration checks \citep[see discussion in ][]{planck2014-a09}. Simulations of beam reconstruction suggest that the scanning beams are determined with high fidelity \citep{planck2014-a08}. From arguments presented in these papers, we are confident that a pure beam- or calibration-related systematic cannot be the sole cause of this effect. 

The significant temporal separation between planet observations also suggests that the systematic effect is stable in time, which argues against ADC nonlinearity \citep[see][for more discussion on ADC nonlinearity]{planck2014-a08} as well as Galactic sidelobe pickup. It is peculiar, however, to see that at 217 and 353\GHz, the in-band correlation for Jupiter is somewhat different than that of Mars or Saturn. This might hint at an inadequate determination of spectral response. Other possible contributions to this systematic effect include, but are not limited to: transfer function deconvolution residuals causing a mismatch in dipole and point source calibration; dynamical nonlinearity; beam effects; and calibration errors. It is interesting to see a similar effect in the \Planck-LFI observations of Jupiter \cite[see e.g., figure~16 in][]{planck2014-a06}. This might hint at a common cause of this systematic effect for both LFI and HFI.

We have tried to probe this effect by looking at differences in the relative scaling of detectors as a function of source SED. This can be done by comparing the in-band variations derived from the relatively thermal spectrum of the planets ($S \propto \nu ^\beta$ with $\beta\approx 2$) to that which is obtained by looking at objects in the PCCS2 that have a softer spectrum ($\beta < 2$). This procedure has provided some limits on the amplitude of this effect, but results are not conclusive enough to warrant a correction.

Figure~\ref{fig:fit_residuals} shows maps of the normalized and background subtracted timeline residuals, $(s_i - D(\theta _i,\phi _i) - \DeltaT_\mrm{p} P (\theta _i,\phi _i))/\Delta T_\mrm{p}$, for each planet and frequency band. These maps help quantify any discrepancy between the PSF fit and the raw data, since any residual amplitude that is not consistent with noise hints at a systematic difference. From these results it is clear that the Uranus and Neptune observations are noise-limited, except possibly at 545 and 857\GHz, whereas low-amplitude structure is apparent at all frequencies in residual maps for the other three planets. The relatively low amplitude of the residuals, however, suggests that the fits are accurate at the sub-percent level. The systematic effect that is shown in Fig.~\ref{fig:band_correlation} is not evident from the residual maps in Fig.~\ref{fig:fit_residuals}.

This systematic effect is comparable in amplitude to the combined statistical uncertainties described in the preceding sections. In view of our lack of understanding of the distribution of systematic errors, we are compelled to assign a systematic uncertainty corresponding to the standard deviation of the in-band variation. As an example, for Jupiter and Saturn, the mean standard deviation of the relative amplitude difference within the 100\GHz\ band is observed to be 0.42\,\%. We implement this value as an estimate for the $1\,\sigma$ systematic uncertainty. Using this method, the systematic uncertainty that we assign to each determination of $\Delta T_\mrm{p}$ is 0.42, 0.54, 0.63, 0.76, 1.93, and 3.08\,\% for detectors in the six HFI frequency bands 100, 143, 217, 353, 545, and 857\GHz, respectively. This uncertainty is shown in the last column of Table~\ref{table:PlanetTB}.  

Although we have found in this work an apparent systematic effect in the determination of point source brightness, there is no reason to suspect that this effect propagates to analysis pertaining to the CMB. This is because of \Planck's use of the orbital dipole for calibrating the channels used for analysis of the CMB signals \citep{planck2014-a09}. Based on the analysis presented in this paper, however, we believe that the calibration of compact sources in the 100--353\GHz\ frequency bands needs to incorporate a fractional systematic error of about 0.4--3.1\,\%.

\subsection{Combined Error Budget}
\label{sec:eb}

Figure \ref{fig:eb_all} shows the relative contributions of different statistical and systematic uncertainties to the total error budget of a single detector within a given band. Apart from the systematic uncertainties described in Sect. \ref{sec:stat_err}, all error terms are assumed to be statistical. These terms therefore average down when we calculate band-averaged quantities. Figure \ref{fig:eb_all} also shows the fractional (as percentages) total uncertainty contributions to flux density. It is clear that for all planets, flux density estimates at high frequencies (545 and 857\GHz) are limited by the systematic uncertainties. Detector noise and background confusion only limit flux density estimates of Uranus and Neptune and only at 100 and 143\GHz. From the figure it is also clear that uncertainties on scanning-beam solid angle and planet disc size are negligible compared to other error terms, whereas uncertainty in absolute calibration becomes relevant at frequencies above 217\GHz.

\section{Planet flux density results}
\label{sec:planet_flux}

We now describe the general planet flux and thermodynamic temperature results and compare them with existing models. Tables~\ref{tab:PlanetTB_LFI} and \ref{table:PlanetTB} highlight the main results of this paper. The LFI measurements are referenced to 28.4, 44.1, and 70.4\,\GHz\ \citep{planck2013-p02b,planck2014-a06}. For \Planck~HFI, the average bandwidths used to derive these measurements are 32.9, 45.8, 64.5, 101, 171, and 246\GHz\ at 100, 143, 217, 353, 545, and 857\GHz, respectively \citep{planck2013-p03d}.  

\begin{figure}
\begin{center}
\includegraphics[width=\columnwidth]{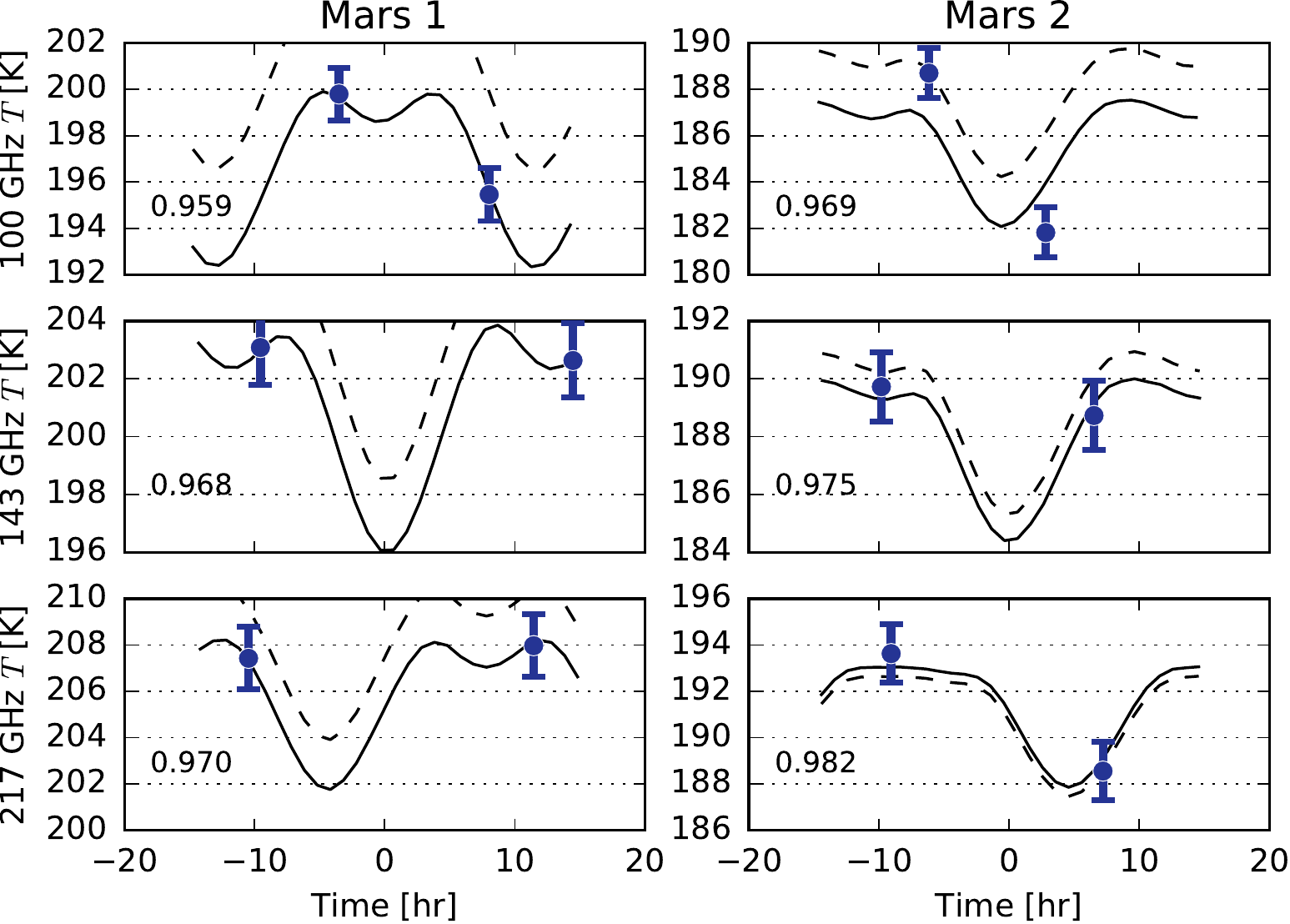}
\caption[Mars diurnal variations]{Estimates of the thermodynamic temperature of Mars compared to a model for diurnal variations. The panel columns correspond to the first two Mars observations while the rows represent estimates at 100, 143, and 217\GHz. The model output has been scaled by $\zeta _\mrm{P} = 0.980$ (dashed line) and a variable best-fit scale (solid line), which is annotated in each panel. The horizontal axis shows time relative to the mean observation time for that frequency band. Error bars show systematic and statistical uncertainties summed in quadrature.
\label{fig:mars_diurnal}
}
\end{center}
\end{figure}

\begin{figure}
\begin{center}
\includegraphics[width=\columnwidth]{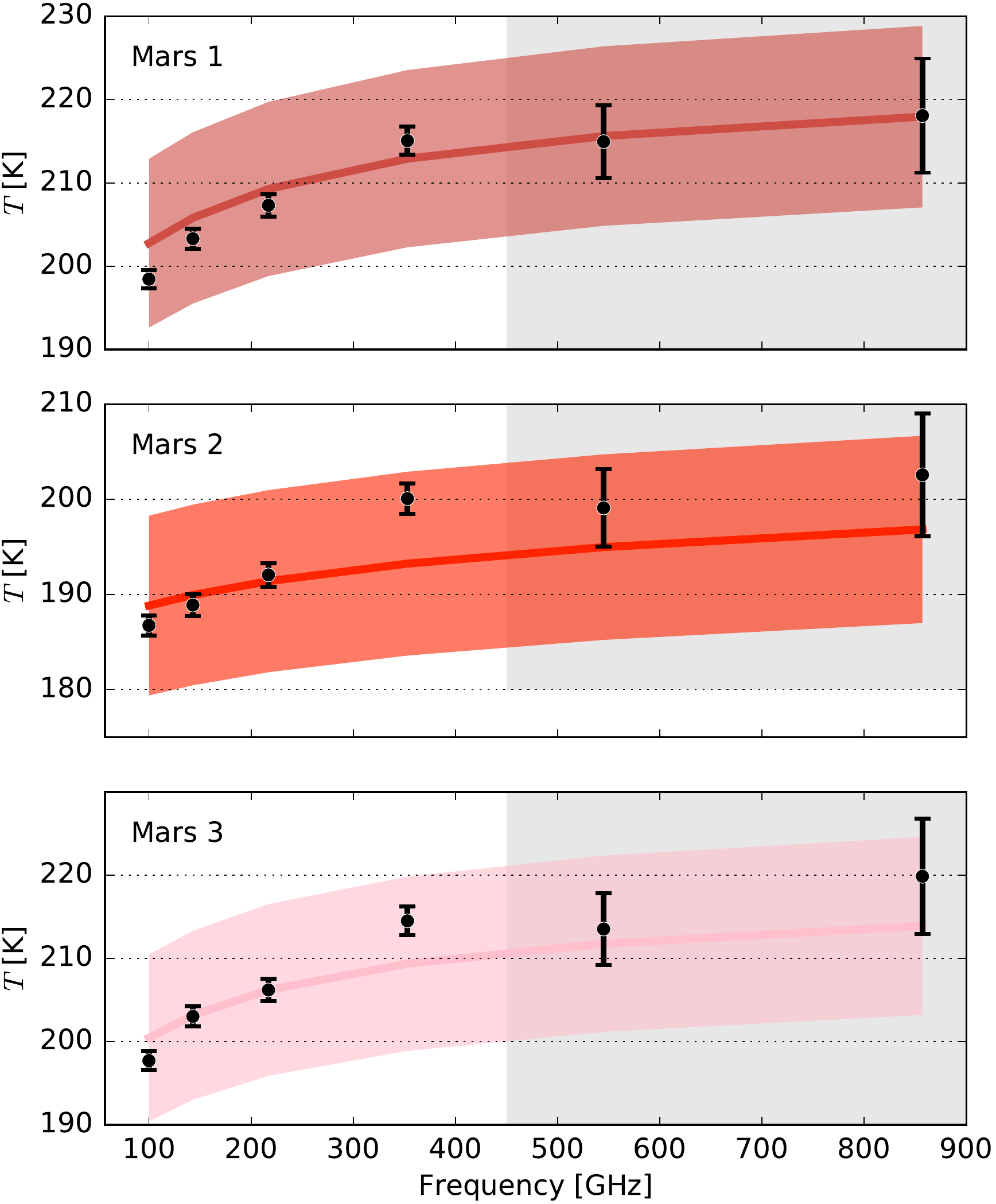}
\caption[Mars thermodynamic temperature]{Measured thermodynamic temperature for the three \Planck~HFI observations of Mars. Black points represent band averages and corresponding error estimates. The thermodynamic temperature predictions of the \cite{Lellouch2008} model (solid lines), scaled by $\zeta _\mrm{P} = 0.980$, and the corresponding 5\,\% absolute errors (coloured regions) are also shown. Both data and model results presented in this figure have been scaled to correspond to a single reference time (see discussion in Sect.~\ref{sec:mars}). The grey region is meant to remind the reader that the sub-mm bands are calibrated on models of Uranus and Neptune brightness (see Sect.~\ref{sec:submmcal}). The absolute thermodynamic temperature of Mars can be obtained by adding the spectral radiance from an occulted CMB monopole (see Sect.~\ref{sec:signal_estimates}).
\label{fig:pl_mars}
}
\end{center}
\end{figure}

The ESA planet brightness models that are mentioned in the following sections have been used for calibration of a number of astronomy experiments operating in the microwave range to the near-infrared. We reference publicly available discussions of these models as they are mentioned in the text (see Sects.~\ref{sec:jup}--\ref{sec:ura}). Digits following the \enquote{ESA} designation are used to indicate model revision numbers. Band-averages are calculated with uniform detector weights.

\begin{table}[tb!]
\begingroup
\newdimen\tblskip \tblskip=5pt
\caption{Band-average planet thermodynamic temperature, $T$, as measured by \Planck~LFI and reported in \cite{planck2013-p02b,planck2014-a06}. Measurements of Saturn's thermodynamic temperature are accompanied by two uncertainties, the first spans the systematic scatter in the measurements within the frequency band and the second is the scatter due to background confusion. In estimating this thermodynamic temperature, we have subtracted contributions from Saturn's rings; the quoted value therefore only refers to the thermodynamic temperature of the planet disc. For Jupiter, the systematic in-band scatter dominates any statistical noise, therefore the uncertainty from background confusion is omitted.}
\label{tab:PlanetTB_LFI}
\vskip -5mm
\footnotesize
\setbox\tablebox=\vbox{
   \newdimen\digitwidth
   \setbox0=\hbox{\rm 0}
   \digitwidth=\wd0
   \catcode`*=\active
   \def*{\kern\digitwidth}
   \newdimen\signwidth
   \setbox0=\hbox{+}
   \signwidth=\wd0
   \catcode`!=\active
   \def!{\kern\signwidth}
   \newdimen\pointwidth
   \setbox0=\hbox{\rm .}
   \pointwidth=\wd0
   \catcode`?=\active
   \def?{\kern\pointwidth}
\halign{\hbox to 2.0cm{#\leaderfil}\tabskip 1em&
        \hfil#\hfil&
        \hfil#\hfil\tabskip 0pt\cr
\noalign{\doubleline}
\noalign{\vskip -1pt}
\omit\hfil Planet\hfil& Freq.\ [GHz]& $T$ [K]\cr
\noalign{\vskip 3pt\hrule\vskip 5pt}
Saturn& 28.4& 138.9 $\pm$ 3.9 (syst.) $\pm$ 1.0 (stat.)\cr
Saturn& 44.1& 147.3 $\pm$ 3.3 (syst.) $\pm$ 1.1 (stat.)\cr
Saturn& 70.4& 150.6 $\pm$ 2.8 (syst.) $\pm$ 0.6 (stat.)\cr
\noalign{\vskip 3pt\hrule\vskip 5pt}
Jupiter&28.4& 146.6 $\pm$ 0.9 (syst.)\phantom{ $\pm$ 1.0 (stat.)}\cr
Jupiter&44.1& 160.9 $\pm$ 1.4 (syst.)\phantom{ $\pm$ 1.0 (stat.)}\cr
Jupiter&70.4& 173.3 $\pm$ 1.0 (syst.)\phantom{ $\pm$ 1.0 (stat.)}\cr
\noalign{\vskip 3pt\hrule\vskip 5pt}
}}
\endPlancktable
\endgroup
\end{table}

\begin{table*}[tb]
\begingroup
\newdimen\tblskip \tblskip=5pt
\caption{Band-average planet thermodynamic temperature and corresponding statistical uncertainty as measured by \Planck~HFI (survey 1--5). Estimates for spectral radiance at the nominal band frequencies can be found by inserting these temperatures together with an estimate for the planet solid angle into Eq.\ (\ref{eq:flux_eq}). The corresponding observation times can be extracted from the data products accompanying this publication or from Table 1 of \cite{planck2013-p03c}. Absolute thermodynamic temperatures of the planets can be obtained by adding the spectral radiance from an occulted CMB monopole (see Sect.~\ref{sec:signal_estimates}). The large seasonal variations in Mars thermodynamic temperature are evident from surveys 1 and 2. For surveys 1--5, error bars represent statistical uncertainty estimates. The error bars accompanying the last column (mean) represent both the statistical (first) and systematic (second) uncertainty. Mars measurements incorporate a scaling factor to account for time-variable brightness (see discussion in Sect.\ \ref{sec:signal_estimates}). Saturn measurements have not been corrected for contributions from rings. Band-averages are calculated with uniform detector weights.}
\label{table:PlanetTB}
\nointerlineskip
\vskip -3mm
\footnotesize
\setbox\tablebox=\vbox{
   \newdimen\digitwidth 
   \setbox0=\hbox{\rm 0} 
   \digitwidth=\wd0 
   \catcode`*=\active 
   \def*{\kern\digitwidth}
   \newdimen\signwidth 
   \setbox0=\hbox{+} 
   \signwidth=\wd0 
   \catcode`!=\active 
   \def!{\kern\signwidth}

\halign{\hbox to 0.8cm{#\leaderfil}\tabskip 2em&
        \hfil#\hfil&
        \hfil#\hfil&
        \hfil#\hfil&
        \hfil#\hfil&
        \hfil#\hfil&
        \hfil#\hfil&        
        \hfil#\hfil\tabskip 0pt\cr  
\noalign{\doubleline}
\omit&&\multispan6\hfil{\sc Thermodynamic temperature} [K]\hfil\cr
\noalign{\vskip -3pt}
\omit&&\multispan6\hrulefill\cr
\omit\hfil Planet\hfil&\omit\hfil Freq.\ [GHz]\hfil&\omit\hfil Survey 1 \hfil&\omit\hfil Survey 2 \hfil&\omit\hfil Survey 3\hfil&\omit\hfil Survey 4\hfil&\omit\hfil Survey 5\hfil&\omit\hfil Mean \hfil\cr
\noalign{\vskip 3pt\hrule\vskip 5pt}
Mars&100&198.4 $\pm$ 0.7&186.7 $\pm$ 0.7&\dots&\dots&197.7 $\pm$ 0.7&194.3 $\pm$ 0.5\,(stat.) $\pm$ 0.8\,(syst.)\cr
Mars&143&203.3 $\pm$ 0.6&188.9 $\pm$ 0.5&\dots&\dots&203.0 $\pm$ 0.5&198.4 $\pm$ 0.4 $\pm$ 1.1\cr
Mars&217&207.3 $\pm$ 0.3&192.1 $\pm$ 0.3&\dots&\dots&206.2 $\pm$ 0.3&201.9 $\pm$ 0.2 $\pm$ 1.3\cr
Mars&353&215.1 $\pm$ 0.5&200.1 $\pm$ 0.5&\dots&\dots&214.5 $\pm$ 0.5&209.9 $\pm$ 0.4 $\pm$ 1.6\cr
Mars&545&215.0 $\pm$ 1.7&199.1 $\pm$ 1.5&\dots&\dots&213.5 $\pm$ 1.5&209.2 $\pm$ 1.1 $\pm$ 4.0\cr
Mars&857&218.1 $\pm$ 1.7&202.6 $\pm$ 1.9&\dots&\dots&219.9 $\pm$ 1.8&213.5 $\pm$ 1.3 $\pm$ 6.6\cr
\noalign{\vskip 3pt\hrule\vskip 5pt}
Jupiter&100&172.8 $\pm$ 0.4&172.1 $\pm$ 0.4&173.1 $\pm$ 0.4&171.0 $\pm$ 0.4&\dots&172.3 $\pm$ 0.4 $\pm$ 0.7\cr
Jupiter&143&174.0 $\pm$ 0.2&172.5 $\pm$ 0.3&174.4 $\pm$ 0.2&172.3 $\pm$ 0.2&174.7 $\pm$ 0.2&173.6 $\pm$ 0.2 $\pm$ 0.9\cr
Jupiter&217&175.4 $\pm$ 0.1&174.7 $\pm$ 0.1&174.6 $\pm$ 0.1&175.2 $\pm$ 0.1&173.8 $\pm$ 0.1&174.7 $\pm$ 0.1 $\pm$ 1.1\cr
Jupiter&353&166.1 $\pm$ 0.4&166.0 $\pm$ 0.4&166.5 $\pm$ 0.4&165.9 $\pm$ 0.4&167.1 $\pm$ 0.4&166.3 $\pm$ 0.4 $\pm$ 1.3\cr
Jupiter&545&137.0 $\pm$ 0.9&138.2 $\pm$ 0.9&136.5 $\pm$ 0.9&135.1 $\pm$ 1.0&135.7 $\pm$ 1.0&136.5 $\pm$ 0.9 $\pm$ 2.6\cr
Jupiter&857&156.7 $\pm$ 1.2&163.8 $\pm$ 1.3&160.1 $\pm$ 1.3&158.3 $\pm$ 1.4&162.3 $\pm$ 1.4&160.3 $\pm$ 1.3 $\pm$ 4.9\cr
\noalign{\vskip 3pt\hrule\vskip 5pt}
Saturn&100&145.2 $\pm$ 0.3&148.3 $\pm$ 0.3&143.5 $\pm$ 0.3&145.9 $\pm$ 0.3&\dots&145.7 $\pm$ 0.3 $\pm$ 0.6\cr
Saturn&143&146.4 $\pm$ 0.2&148.6 $\pm$ 0.2&145.4 $\pm$ 0.2&147.7 $\pm$ 0.2&\dots&147.0 $\pm$ 0.2 $\pm$ 0.8\cr
Saturn&217&143.8 $\pm$ 0.1&145.4 $\pm$ 0.1&144.3 $\pm$ 0.1&146.0 $\pm$ 0.1&\dots&144.9 $\pm$ 0.1 $\pm$ 0.9\cr
Saturn&353&139.9 $\pm$ 0.3&140.4 $\pm$ 0.3&142.4 $\pm$ 0.3&143.1 $\pm$ 0.3&\dots&141.5 $\pm$ 0.3 $\pm$ 1.1\cr
Saturn&545&100.1 $\pm$ 0.6&*99.9 $\pm$ 0.7&105.0 $\pm$ 0.7&104.3 $\pm$ 0.7&\dots&102.4 $\pm$ 0.6 $\pm$ 2.0\cr
Saturn&857&112.1 $\pm$ 0.9&111.0 $\pm$ 0.8&120.0 $\pm$ 1.1&118.7 $\pm$ 1.0&\dots&115.5 $\pm$ 1.0 $\pm$ 3.6\cr
\noalign{\vskip 3pt\hrule\vskip 5pt}
Uranus&100&121.1 $\pm$ 0.8&118.1 $\pm$ 0.8&120.9 $\pm$ 0.8&121.6 $\pm$ 0.8&120.6 $\pm$ 1.0&120.5 $\pm$ 0.4 $\pm$ 0.5\cr
Uranus&143&107.6 $\pm$ 0.2&109.1 $\pm$ 0.2&108.5 $\pm$ 0.2&108.6 $\pm$ 0.2&108.4 $\pm$ 0.2&108.4 $\pm$ 0.1 $\pm$ 0.6\cr
Uranus&217&*98.3 $\pm$ 0.1&*98.5 $\pm$ 0.1&*98.6 $\pm$ 0.1&*98.7 $\pm$ 0.1&*98.5 $\pm$ 0.1&*98.5 $\pm$ 0.1 $\pm$ 0.6\cr
Uranus&353&*86.5 $\pm$ 0.2&*86.3 $\pm$ 0.2&*86.1 $\pm$ 0.2&*85.9 $\pm$ 0.2&*86.2 $\pm$ 0.2&*86.2 $\pm$ 0.1 $\pm$ 0.7\cr
Uranus&545&*74.0 $\pm$ 0.5&*73.5 $\pm$ 0.5&*73.2 $\pm$ 0.4&*73.5 $\pm$ 0.5&*75.1 $\pm$ 0.6&*73.9 $\pm$ 0.2 $\pm$ 1.4\cr
Uranus&857&*66.0 $\pm$ 0.5&*66.2 $\pm$ 0.5&*66.3 $\pm$ 0.5&*66.2 $\pm$ 0.5&\dots&*66.2 $\pm$ 0.2 $\pm$ 2.0\cr
\noalign{\vskip 3pt\hrule\vskip 5pt}
Neptune&100&118.2 $\pm$ 2.2&117.6 $\pm$ 1.9&117.3 $\pm$ 1.9&116.6 $\pm$ 1.9&\dots&117.4 $\pm$ 1.0 $\pm$ 0.5\cr
Neptune&143&105.8 $\pm$ 0.5&106.3 $\pm$ 0.4&107.0 $\pm$ 0.5&106.5 $\pm$ 0.4&\dots&106.4 $\pm$ 0.2 $\pm$ 0.6\cr
Neptune&217&*97.1 $\pm$ 0.3&*97.7 $\pm$ 0.2&*97.8 $\pm$ 0.3&*97.0 $\pm$ 0.2&\dots&*97.4 $\pm$ 0.1 $\pm$ 0.6\cr
Neptune&353&*82.2 $\pm$ 0.3&*82.8 $\pm$ 0.3&*82.7 $\pm$ 0.3&*82.6 $\pm$ 0.2&\dots&*82.6 $\pm$ 0.1 $\pm$ 0.6\cr
Neptune&545&*72.4 $\pm$ 0.5&*71.9 $\pm$ 0.4&*72.4 $\pm$ 0.5&*72.2 $\pm$ 0.4&\dots&*72.3 $\pm$ 0.2 $\pm$ 1.4\cr
Neptune&857&*65.2 $\pm$ 0.5&*65.5 $\pm$ 0.5&*65.3 $\pm$ 0.4&*65.1 $\pm$ 0.5&\dots&*65.3 $\pm$ 0.2 $\pm$ 2.0\cr
\noalign{\vskip 3pt\hrule\vskip 5pt}
}}
\endPlancktablewide                  
\endgroup
\end{table*}

\subsection{Mars}
\label{sec:mars}

The Martian orbital period corresponds to 687 Earth days and the planet spins around its axis approximately once every 24.6\,hours \citep{Horizons}. The 25\pdeg2 axial tilt is comparable to Earth's 23\pdeg4 value, but the relatively large orbital eccentricity makes the southern hemisphere experience greater seasonal variations. The perceived thermodynamic temperature is highly dependent on viewing location, since the Martian surface is far from homogeneous. Finally, dynamical factors such as dust storms can also affect the planet's surface temperature \citep{Zurek1982}. 

%see http://www-mars.lmd.jussieu.fr
%(http://www.imcce.fr)
A number of models predicting Mars thermodynamic temperature exist \citep[e.g.,][]{Golden1979,Rudy1987}. We have primarily considered the models of \cite{Lellouch2008} and \cite{weiland2010}. The Lellouch and Amri model uses surface and subsurface temperatures taken from the European Martian General Circulation Model \citep{Forget1999,Millour2015} and Martian ephemerides from IMCCE.\footnote{See \href{http://www-mars.lmd.jussieu.fr}{http://www-mars.lmd.jussieu.fr} and \href{http://www.imcce.fr}{http://www.imcce.fr}} A standard dust scenario (\enquote{Climatology}) is used. For each user-provided date, the model first computes the aspect of Mars. The disc is then split on a $100\times100$ grid, each of them having its own latitude, longitude, and local time. On each point of the grid, the usual radiative transfer equation \citep[e.g., Eq.\ (5) of][]{Rudy1987} is used. Radiative transfer in the surface and subsurface includes an absorption coefficient corresponding to a radio absorption length equal to 12 times the wavelength. In addition, the thermal emission of the surface includes an emissivity term, calculated from a Fresnel reflection model with a dielectric constant of 2.25 and for the relevant emission angle. The latter is calculated taking into account a surface roughness of 12\deg. Local fluxes calculated in this manner are finally convolved with a Gaussian beam to obtain beam-averaged fluxes and the corresponding Planck thermodynamic temperature. 

The Weiland model is an alternative version of a model that was originally constructed by Edward Wright \citep{Wright2007,weiland2010}. This updated version was used by the WMAP team and incorporates \lto viewing angles, as well as extending the spectral coverage down to WMAP frequencies.

The radio and microwave brightness of Mars has been estimated by a number of experiments. The following papers discuss the brightness of the planet, either as an absolute measurement or one that is relative to another planet or a model \cite{Wright1976,Rather1974,Rudy1987,Muhleman1991,Goldin1997,Sidher2000,Runyan2003,Swinyard2010,Perley2013,Muller2016}. The polarization properties of Mars are described in \cite{Perley2013}.

% Diurnal variations
Because of \Planck's scan strategy, detectors on the HFI focal plane observe a fixed point on the sky over the span of a week. Using the \cite{Lellouch2008} model for comparison, we appear to detect rotational variations in Mars brightness with high significance; Fig.~\ref{fig:mars_diurnal} shows estimates for thermodynamic temperature as a function of time spanning approximately one Martian day. These diurnal variations have been reported before and are consistent with predictions from models \citep[e.g.,][]{Sidher2000}. We correct for this in analysis, scaling measured values to a common observation time. These times correspond to unix time 1256545807, 1271109507, and 1324607821 for Mars observations 1--3, respectively, and roughly represent the time when detectors in the 353\GHz\ band were observing the planet head on.\footnote{Unix time is defined as the number of seconds that have elapsed since 00:00:00 Coordinated Universal Time (UTC), Thursday, 1 January 1970.} The corresponding Julian date (MJD) is 2455130.854, 2455299.416, and 2455918.609. Our estimates for the planet solid angle at those times are 45.616, 54.471, and 54.678~arcsec$^2$, respectively. This rescaling changes the in-band standard deviation in measured thermodynamic temperature of the second Mars observation at 217\GHz\ from 2.5 to 1.0\,K.

% Comparison of measurement results with model predictions
Figure~\ref{fig:pl_mars} compares the measured thermodynamic temperature of Mars with predictions of the \cite{Lellouch2008} model for the reference times stated previously. The \cite{Lellouch2008} model has been scaled by a constant factor, $\zeta _\mrm{P} = 0.980$, in order to improve consistency with \Planck-HFI measurements. This scale factor was found by minimizing the 100--353\GHz\ residuals between model predictions for the three reference times with the corresponding diurnal-variation-corrected \Planck\ thermodynamic temperature results. The greyed out region in Fig.~\ref{fig:pl_mars} is there to remind the reader that the sub-mm bands are calibrated on models of the brightness of Uranus and Neptune (see Sect.~\ref{sec:submmcal}).

Figures~\ref{fig:mars_diurnal} and \ref{fig:pl_mars} indicate that once the Lellouch~and~Amri model is rescaled downwards by about 2.0\,\%, it provides a good match of the absolute and relative (i.e., variations on diurnal and seasonal scales, and spectral dependence) Mars thermodynamic temperature as measured by HFI. This is consistent with the claimed 5\% absolute accuracy of the model.

A series of HFI end-of-life tests were undertaken in December 2011 \citep[see][]{planck2014-a08}. During these scans, the 100-, 143-, 217-, and 353-GHz frequency bands observed Mars at two different spin rates, the nominal 1.0~rpm and a faster 1.4~rpm. Although these tests were useful in further constraining the bolometer time-response functions, we have not considered these data in this paper.

\subsection{Jupiter}
\label{sec:jup}

There exists a rich literature on the mm and sub-mm flux densities of the Jovian planets \citep{Goldin1997,WeissteinThesis,Weisstein1996b,Burgdorf2003,Gibson2005,Desert2008}.

\begin{figure}
\begin{center}
\includegraphics[width=\columnwidth]{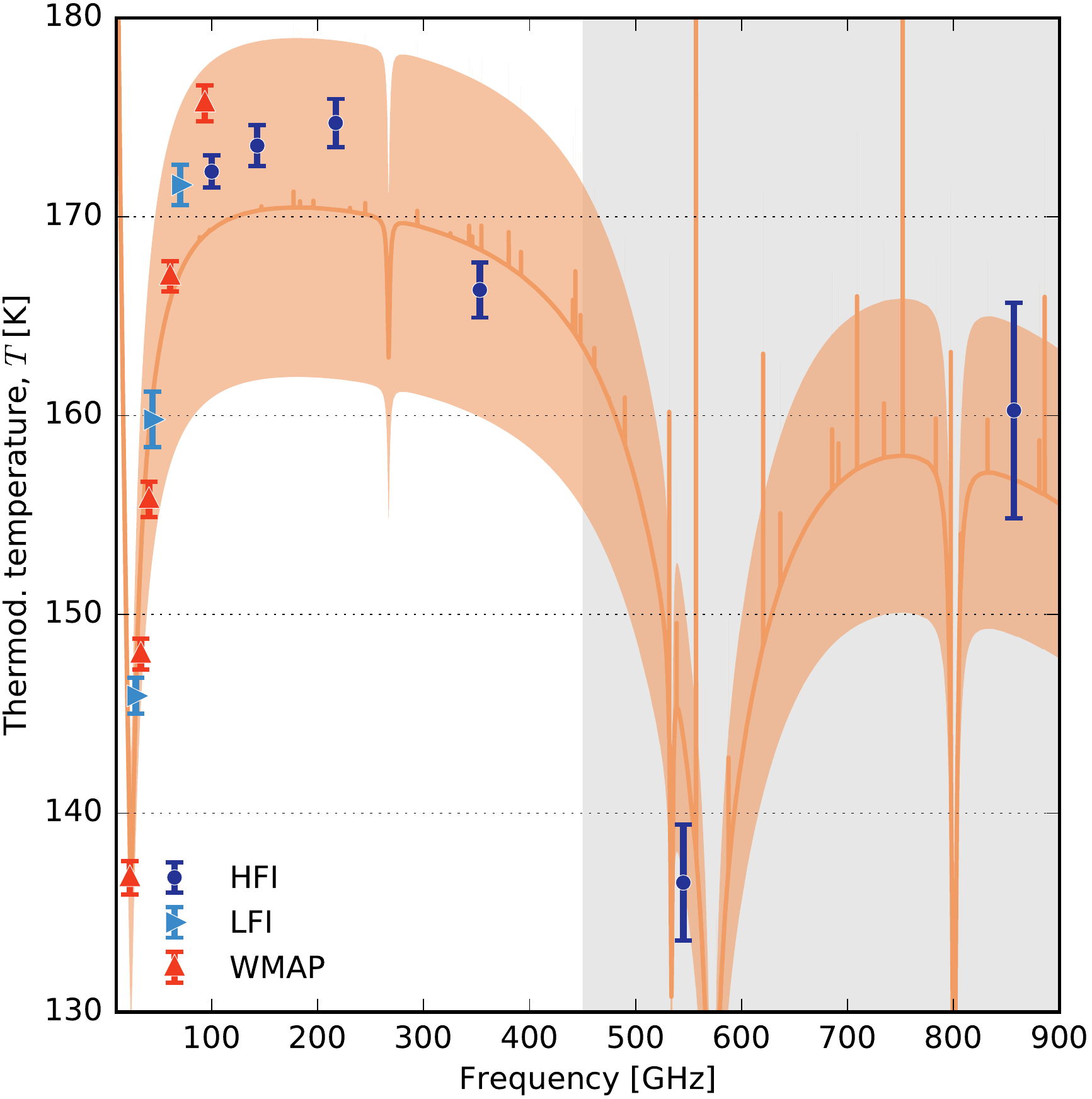}
\caption[Jupiter thermodynamic temperature]{Measured thermodynamic temperature for the five \Planck-HFI observations of Jupiter as well as \Planck-LFI results \citep{planck2014-a06}. Points and error bars represent average (maximum likelihood) thermodynamic temperature and associated errors (including systematic error). The orange line and region represent the ESA1 model predictions and the estimated absolute uncertainties. Note the ammonia absorption line at 572.5\GHz. Emission lines are due to stratospheric emission from $\mrm{H}_2\mrm{O}$, CO, CS, and HCN. \wmap measurements are included for comparison \citep{Bennett2013}. The grey region is there to remind the reader that the sub-mm bands are calibrated on models of the brightness of Uranus and Neptune (see Sect.~\ref{sec:submmcal}).}
\label{fig:pl_jupiter}
\end{center}
\end{figure}

Despite having significantly lower millimetre wavelength thermodynamic temperature than Mars, Jupiter's apparent size makes it the brightest planet on the sky as seen by \Planck. Although high-frequency detectors are driven to saturation, in particular at 545 and 857\GHz, we attempt to estimate the planet brightness across all frequencies by masking the centre of the planet crossing. For frequencies above 217\GHz\, we also apply a nonlinearity correction (see discussion in Sects.~\ref{sec:psffit} and \ref{sec:signal_estimates}). Fig.~\ref{fig:pl_jupiter} shows the thermodynamic temperature at all \Planck-HFI frequencies derived from the five Jupiter observations available and compares them with the ESA1 model predictions.

The act of masking the planet centre dramatically reduces the statistical power in the Jupiter data set, but we also see that it significantly affects the flux density estimate. The masking radius was chosen so that the signal amplitude outside the masked region would be comparable to the signal amplitude seen in Saturn observations. Although we have not attempted to estimate the amplitude of any nonlinearity and saturation biases, we expect some increase in error, especially at 545 and 857\GHz, because of this effect. Note that we do not account for any contamination in flux density determination due to Jupiter's rings and moons.

In order to improve consistency between the ESA1 model predictions and the \wmap and \Planck\ measurements at frequencies below 353\,\GHz, a roughly 3\% upwards scaling of the model is required. If this model rescaling is necessary, we note an interesting discrepancy between \Planck-HFI measurements and model predictions at 353 and 545\,\GHz\ (see Fig.~\ref{fig:pl_jupiter}). We further note that the 353\GHz\ thermodynamic temperature measurement of approximately ($166.3\pm1.7$)\,K is quite stable in time (see Table \ref{table:PlanetTB}).

\begin{table*}[th!] % table* is a two-column table.  Drop the * for one column.
\begingroup
\newdimen\tblskip \tblskip=5pt
\caption{Parameters used in the fit to a ring system model of Saturn. Columns are: season number; date range; position in Galactic coordinates ($l$,$b$); ring inclination  angle $B$; planet range $r$; planetary solid angle unobscured by the disc, $\Omega_{\mrm{ud}}$; the total unobscured ring area, $\Omega_{\mrm{ring}}$; the solid angle obscured by rings, $\Omega_{\mrm{cusp}}$; and the total solid angle of the planet disc, $\Omega _\mrm{P}$.}
\label{table:SaturnSeasons}
\nointerlineskip
\vskip -3mm
\footnotesize
\setbox\tablebox=\vbox{
   \newdimen\digitwidth 
   \setbox0=\hbox{\rm 0} 
   \digitwidth=\wd0 
   \catcode`*=\active 
   \def*{\kern\digitwidth}
   \newdimen\signwidth 
    \setbox0=\hbox{-} 
   \signwidth=\wd0 
   \catcode`!=\active 
   \def!{\kern\signwidth}
\halign{\hbox to 2cm{#\leaderfil}\tabskip 1em&# \tabskip 2em&\hfil#\hfil \tabskip 2em&\hfil#\hfil \tabskip 2em&\hfil#\hfil \tabskip 2em&\hfil#\hfil \tabskip 2em&\hfil#\hfil \tabskip 2em&\hfil#\hfil \tabskip 2em&\hfil#\hfil \tabskip 2em&\hfil#\hfil\tabskip 0em\cr                            % Template goes here.
\noalign{\doubleline}
                                    % Table headings go here.
\omit\hfil Season\hfil& \omit\hfil Date range \hfil &
\omit\hfil $l$ \hfil & \omit\hfil $b$ \hfil & \omit\hfil $B$ \hfil & \omit\hfil $r$ \hfil & \omit\hfil $\Omega_{\mrm{ud}}$ \hfil & \omit\hfil $\Omega_{\mrm{cusp}}$ \hfil & \omit\hfil $\Omega_{\mrm{ring}}$ \hfil & \omit\hfil $\Omega_{\mrm{P}}$\hfil\cr
\omit&&\omit\hfil [\deg] \hfil & \omit\hfil [\deg] \hfil & \omit\hfil
[\deg] \hfil & \omit\hfil [AU] \hfil & \omit\hfil [arcsec$^2$] \hfil & \omit\hfil [arcsec$^2$] \hfil & \omit\hfil [arcsec$^2$] \hfil & \omit\hfil[arcsec$^2$]\hfil\cr
\noalign{\vskip 3pt\hrule\vskip 5pt}
1 &04 Jan -- 08 Jan 2010&286.0&*62.2&*6.03&9.08&218.28&2.36&*80.70&237.40\cr
2 &11 Jun -- 17 Jun 2010& 271.6&*62.5&*2.45&9.57&206.71&0.20&*29.52&213.82\cr
3 &18 Jan -- 22 Jan 2011& 310.3&*58.2&12.56&9.19&169.41&8.27&169.44&232.72\cr
4 &29 Jun -- 05 Jul 2011& 298.4&*60.9&*9.40&9.73&182.19&4.66&110.11&207.48\cr
\noalign{\vskip 3pt\hrule\vskip 5pt}
}
}
%\endPlancktable                    % ends one-column \halign
\endPlancktablewide                 % ends two-column \halign
\endgroup
\end{table*}                        % table* is a two-column table.  Drop the * for one column.

\subsection{Saturn}
\label{sec:sat}

Saturn's flux determination is complicated by the presence of extended rings. High resolution images of Saturn and its rings at 1- and 3-mm wavelength are provided in \cite{Dunn2005}.

\begin{table}[tb!]
\begingroup
\newdimen\tblskip \tblskip=5pt
\caption{Best-fit parameters of two models for the Rayleigh-Jeans temperature of Saturn and its rings. The two-component model treats disc and ring contributions separately, whereas the single-component model ``system" represents the mean temperature of the ring-disc system. Treating the band-average results listed in Table \ref{table:PlanetTB} as individual measurements of four Saturn observational seasons, the two- and single-component models have two and three degrees of freedom (dof), respectively.}
\label{tab:saturn_two_component}
\vskip -5mm
\footnotesize
\setbox\tablebox=\vbox{
   \newdimen\digitwidth
   \setbox0=\hbox{\rm 0}
   \digitwidth=\wd0
   \catcode`*=\active
   \def*{\kern\digitwidth}
   \newdimen\signwidth
   \setbox0=\hbox{+}
   \signwidth=\wd0
   \catcode`!=\active
   \def!{\kern\signwidth}
   \newdimen\pointwidth
   \setbox0=\hbox{\rm .}
   \pointwidth=\wd0
   \catcode`?=\active
   \def?{\kern\pointwidth}
\halign{\hbox to 2.0cm{#\leaderfil}\tabskip 0pt&
        \hfil#\hfil\tabskip 1.0em&
        \hfil#\hfil&
        \hfil#\hfil\tabskip 2.0em&
        \hfil#\hfil\tabskip 1.0em&
        \hfil#\hfil\tabskip 0pt\cr
\noalign{\doubleline}
\noalign{\vskip -2pt}
\omit\hfil Freq.\hfil& Ring& Disc& $\chi^{2}$& System& $\chi^{2}$\cr
\omit\hfil [GHz]\hfil& [K]& [K]& & [K]& \cr
\noalign{\vskip 3pt\hrule\vskip 5pt}
100& $15.7\pm1.5$& $148.5\pm0.8$& 6.2& $143.2\pm0.3$& 25.8\cr
143& $18.4\pm1.8$& $147.6\pm0.9$& 5.3& $143.6\pm0.4$& *9.1\cr
217& $21.3\pm2.0$& $142.1\pm1.0$& 4.5& $139.7\pm0.5$& *3.5\cr
353& $25.3\pm2.4$& $133.0\pm1.3$& 3.0& $133.1\pm0.6$& *5.8\cr
545& $23.5\pm4.5$& $*86.7\pm2.3$& 0.7& $*89.8\pm1.0$& *5.2\cr
857& $31.3\pm7.9$& $*89.9\pm4.0$& 0.4& $*95.8\pm1.8$& *4.7\cr
\noalign{\vskip 3pt\hrule\vskip 5pt}
}}
\endPlancktable
\endgroup
\end{table}
As is evident from Fig.~\ref{fig:planets_2scale}, \Planck~HFI (4\parcm5 minimum resolution) does not resolve Saturn (radius approximately 8\arcs), meaning that the flux seen from Saturn is an integrated, whole-disc signal.  However, the four \Planck-HFI observations of Saturn occurred at different ring inclination angles, $B$, making it possible to separate emission from the planetary disc and emission from the rings (see Table~\ref{table:SaturnSeasons}). 

\begin{figure}[t!]
\begin{center}
\includegraphics[width=\columnwidth]{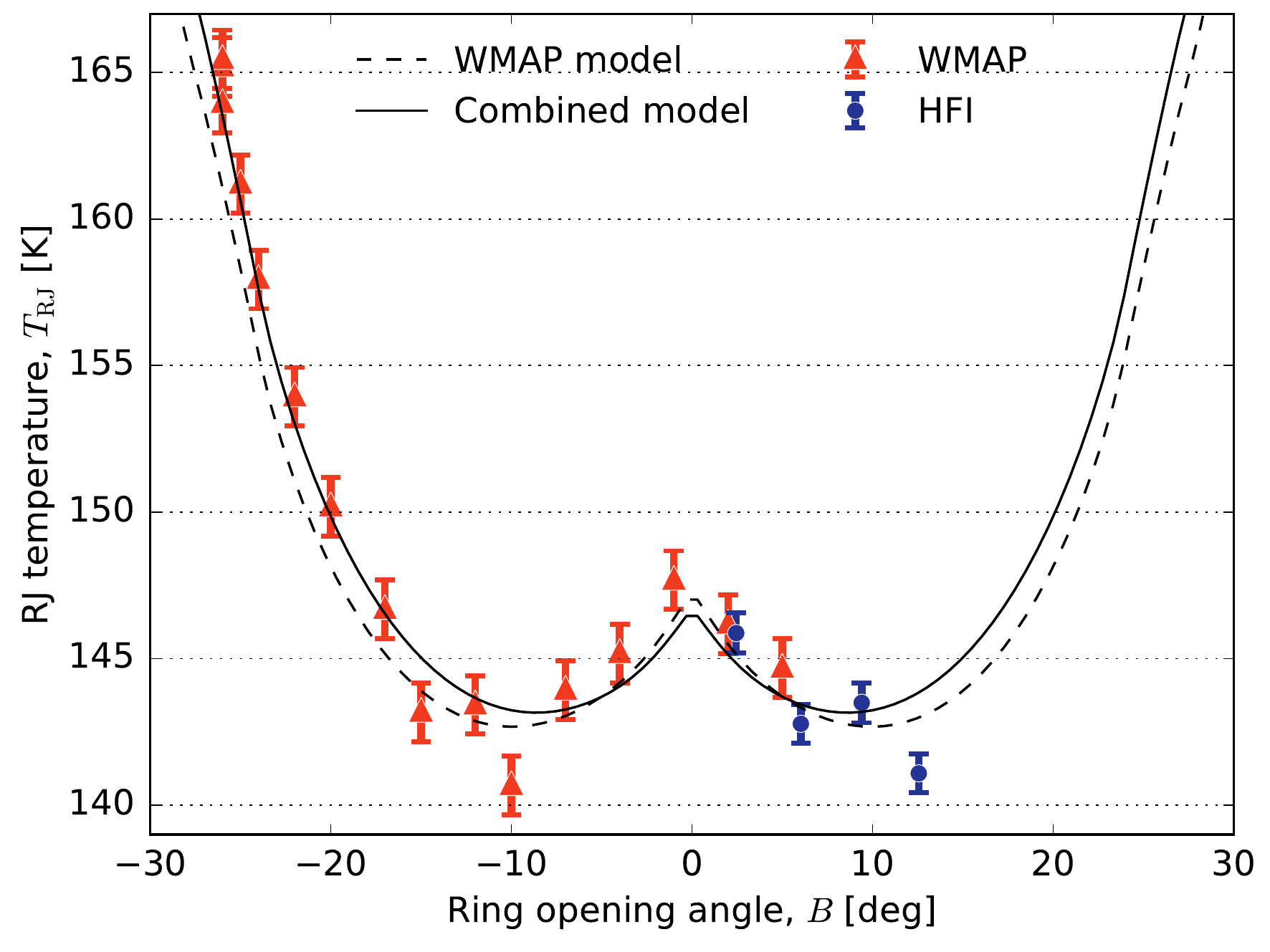}
\caption[Saturn ring model]{Rayleigh-Jeans temperature of the combined Saturn ring-disc system at 100\GHz\ with the \wmap measurements scaled from 94\GHz\ to match \Planck\ (0.1\,\% upwards scaling). The best-fit ring-disc models for the two data sets are also shown. We use RJ temperature instead of thermodynamic temperature for the Saturn ring-disc system for ease of comparison with \wmap and \act results. The HFI error bars represent a combination of statistical and systematic error. The systematic error is expected to be strongly correlated between observations. We note that the third \Planck-HFI observation of Saturn, corresponding to an inclination angle of +12.6\deg, appears somewhat anomalous.
}
\label{fig:pl_saturn_rings}
\end{center}
\end{figure}

During the four \Planck\ observations of the planet, Saturn's ring inclination angle spanned +3 to +13\deg\ as viewed from \lton. Saturn's equinox was in August 2009, so the first two observations were nearly edge on. All \Planck\ observations of Saturn are during northern spring, whereas \wmap observed Saturn during its northern winter, corresponding to primarily negative inclination angles. Together, \wmap and \Planck\ measured the 100\GHz\ brightness of Saturn over a 39\deg range of inclination angles, from --26 to +13\deg. 

Following the notation of \cite{weiland2010} we construct a model for the frequency and viewing-angle dependent Rayleigh-Jeans temperature that incorporates contributions from both the planet disc and the rings:
\begin{eqnarray}
T_\mrm{RJ}(\nu,B) &=& \frac{T_{\mrm{disc}(\nu)}}{\Omega _\mrm{P}}  \left[  \Omega_{\mrm{ud}}  + \textstyle \sum_{i=1}^{7} e^{-\tau_i
    |\csc B|} \Omega_{\mrm{od},i}\right] \nonumber \\
    &+& \frac{T_{\mrm{ring}(\nu)}}{\Omega _\mrm{P}}  \textstyle \sum_{i=1}^{7} \Omega_{\mrm{r},i},
\end{eqnarray}    
where $\Omega_\mrm{P}$ is the solid angle of the planet disc and $T_{\mrm{disc}}$ and $T_{\mrm{ring}}$ are the planetary disc and ring Rayleigh-Jeans (RJ) temperatures, respectively. Note that we use RJ temperature instead of thermodynamic temperature in this section for ease of comparison with \wmap and \act results and for simplified modelling. The index $i$ refers to one of the seven ring components used at microwave wavelengths (we use the value in table 10 of \cite{weiland2010}, following \cite{dunn2002}). Here, $\tau_i$ is the optical depth of the ring component, $\Omega_{\mrm{ud}}$ is the planetary disc solid angle unobscured by rings, $\Omega_{\mrm{od},i}$ is the solid angle of the disc that is obscured by the $i$th ring component and $\Omega_{\mrm{r},i}$ is the solid angle of the $i$th ring component. The portion of the disc solid angle obscured by the rings is 
\begin{equation}
\Omega_{\mrm{cusp}} = \textstyle \sum_{i=1}^{7} e^{-\tau_i  |\csc B|} \Omega_{\mrm{od},i},
\end{equation}
and the total ring area is
\begin{equation}
\Omega_\mrm{ring} =  \textstyle \sum_{i=1}^{7} \Omega_{\mrm{r},i}.
\end{equation}
In this model, all rings are assumed to have the same RJ temperature and the opacity of the rings is fixed a priori \citep{dunn2002}. We do not explore models where these properties are allowed to vary.

\begin{figure}[t!]
\begin{center}
\includegraphics[width=\columnwidth]{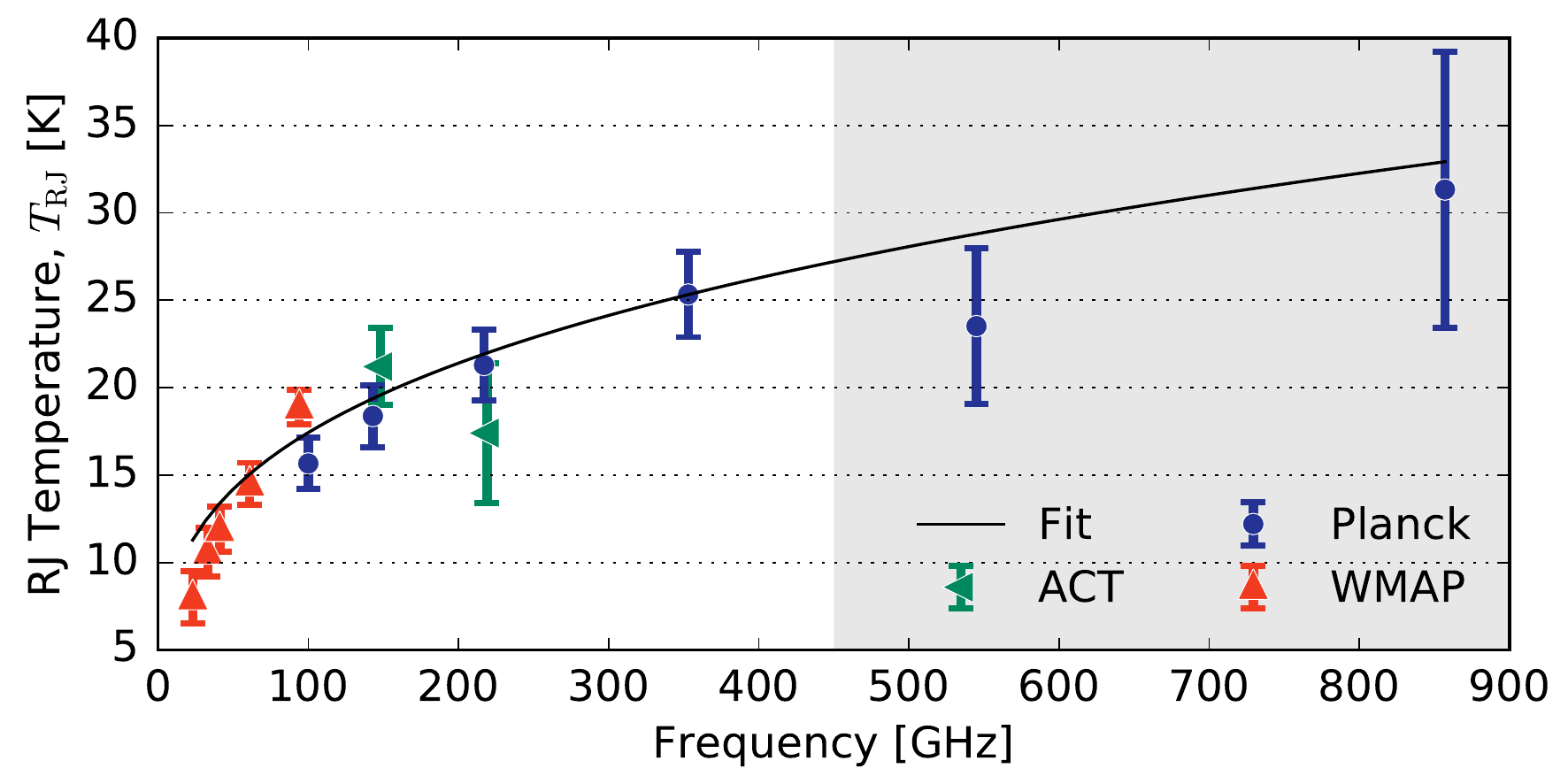}
\caption[Saturn two-component model]{Rayleigh-Jeans temperature of Saturn's ring as a function of frequency, as estimated by \wmapn, \actn, and \Planck~HFI, plotted together with a power-law model fit to all measurements. The best-fit model (solid line) suggests a spectral index of $\beta _\mrm{ring} = 2.30\pm0.03$ compared to a spectral index of $\beta = 2$ for a perfect RJ source. The grey region is there to remind the reader that the sub-mm bands are calibrated on models of the brightness of Uranus and Neptune (see Sect.~\ref{sec:submmcal}).}
\label{fig:pl_saturn_two_component}
\end{center}
\end{figure}

Figure~\ref{fig:pl_saturn_rings} compares the 100-GHz\ Saturn RJ temperatures measured by \wmap and \Planck\ and the best-fit models to the two independent data sets, as well as the combined data set \citep[\wmap data are extracted from table 8 in][]{Bennett2013}. The \wmap results have been scaled up from 94 to 100\GHz\ using the ESA2 Saturn thermodynamic temperature model, corresponding to a 0.1\,\% increase in RJ temperature. 

The best-fit two-component model is listed in Table \ref{tab:saturn_two_component}. The two-component model significantly improves the fit to the data over a simpler single-disc model that neglects effects from the ring geometry. We note, however, that the systematic uncertainty that we expect to be strongly correlated between observations dominates the statistical uncertainty in flux determination of Saturn at 100\GHz. Table \ref{tab:saturn_two_component} reports a goodness of fit parameter, $\chi ^{2}$, as well as the corresponding degrees of freedom assuming that the errors are uncorrelated. If we instead account for the correlated error in our model comparison the two-component model is still significantly favoured over the simpler single component model, however, in that scenario the third (January 2011) observation of Saturn becomes significantly at odds with the other three (see the discrepant HFI data point in Fig.~\ref{fig:pl_saturn_rings}).

On 5 December 2010, a planetary-scale disturbance erupted in Saturn's northern hemisphere, leading to large temperature perturbations (up to 50~K at 0.5~mbar) in Saturn's stratosphere \cite[e.g.,][]{Fletcher2012}. Perturbations in the troposphere were much more subdued, although \cite{Achterberg2014} reported an increase of about 3~K in the far-IR (20--200~$\mrm{\mu m}$, probing the upper troposphere near 400 mbar) thermodynamic temperature at the storm's latitude. In addition, 2.2-cm thermodynamic temperatures measured by the Cassini spacecraft three months into the storm were observed to increase more significantly (from 148\,K to 166\,K) at the storm latitude, likely due to a strong reduction in the relative humidity of ammonia there \citep{Laraia2013}. Although this information is not quite sufficient to quantitatively estimate the change of the disc-averaged continuum thermodynamic temperature at the storm's epoch, it seems clear that if anything, the effect of the storm would be an increase in these continuum temperatures.

\begin{figure}
\begin{center}
\includegraphics[width=\columnwidth]{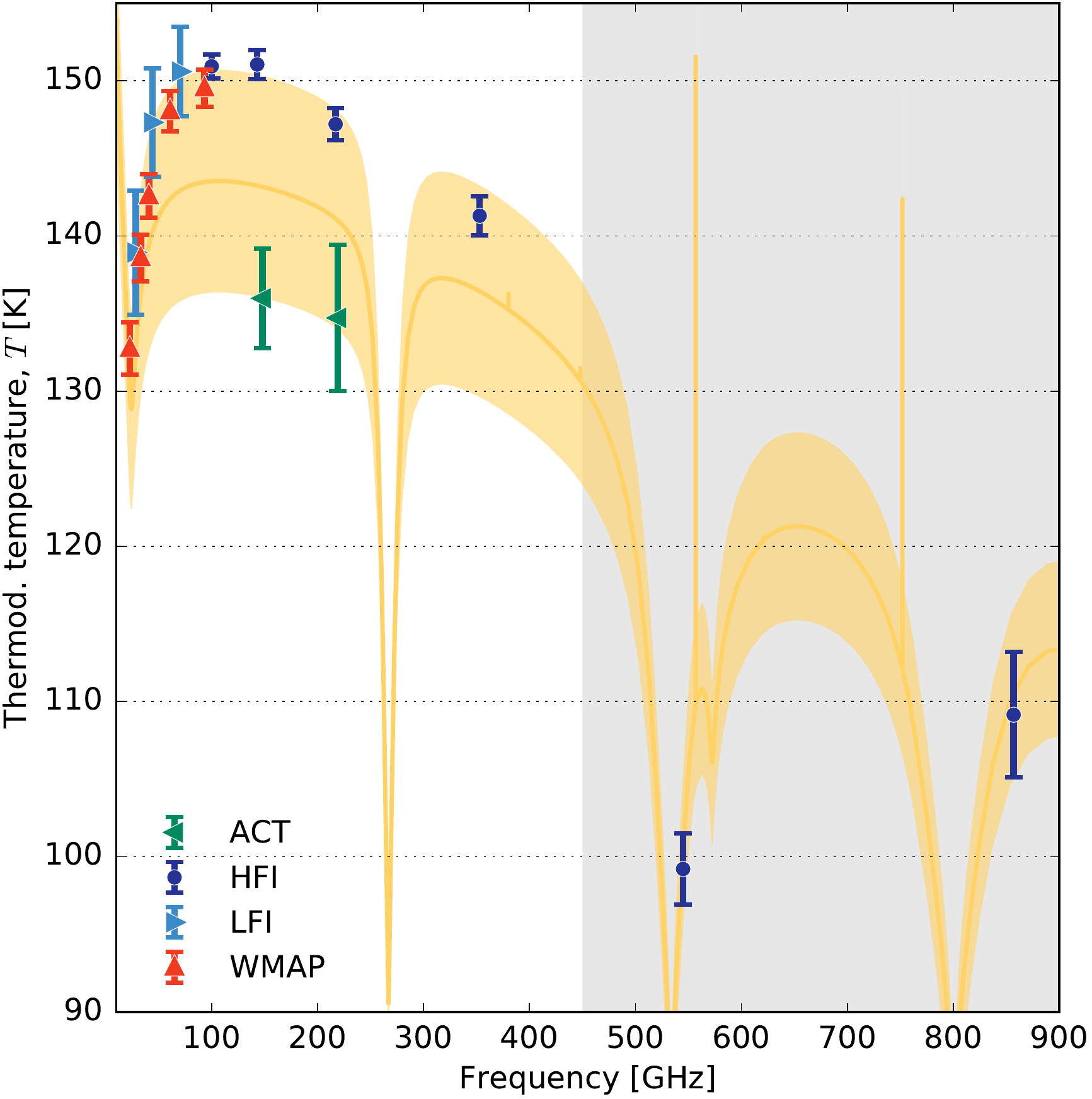}
\caption[Saturn thermodynamic temperature]{
Thermodynamic temperature of Saturn's disc as calculated from the four \Planck-HFI observations compared with the ESA2 model output. The LFI data at 28, 44, and 70\,\GHz\ are taken from \cite{planck2013-p02b}, but corrected for disc oblateness and ring contributions. The three prominent absorption lines, including the one at 530\GHz, are phosphine (PH$_3$) related \citep{WeissteinThesis}. Emission lines are due to stratospheric emission from $\mrm{H}_2\mrm{O}$. \wmap and \act measurements are included here for comparison \citep{Bennett2013,Hasselfield2013,ACT_PlanckCross2013}.
}
\label{fig:pl_saturn}
\end{center}
\end{figure}

The third HFI observation of Saturn took place in early January 2011, shortly after this disturbance is thought to have begun. The fourth observation took place in late June and early July that same year. Given our expectations for the thermal effects of the storm, and since our measurement uncertainty is dominated by systematic error which we expect to be strongly correlated between observations, we are surprised to see that the fourth Saturn observation appears more consistent with the two-component model than the third observation. 

Figure~\ref{fig:pl_saturn_two_component} shows determinations of the Rayleigh-Jeans ring temperature as a function of frequency as estimated by \wmapn, \actn, and \Planck~HFI. Here, all three experiments have employed the same two-component model described in \cite{weiland2010}. We adopt the following model to describe the Rayleigh-Jeans temperature as a function of frequency: 
\begin{equation}
T_\mrm{RJ} = T_0 \left( \frac{\nu}{\nu_0} \right)^{\beta_\mrm{ring}-2},
\end{equation}
with $\nu_0 \equiv 23$\GHz, and the data being used to fit for $T_0$ and $\beta_\mrm{ring}$. The RJ-temperature can be used to calculate the spectral radiance of Saturn's rings according to Eq.\ (\ref{eq:Brj}). The best-fit model, assuming errors are uncorrelated, suggests a spectral index of $\beta _\mrm{ring} = 2.30\pm0.03$, with $T_0 = (11.3\pm0.6$)\,K. The goodness of fit for this models is $\chi ^{2} / \mrm{dof} = 13.5/11$. 

Figure~\ref{fig:pl_saturn} shows the observation-averaged Saturn thermodynamic temperature as a function of frequency using the disc contribution from the two-component ring-disc model, and compares those results with the ESA2 model. In order to accommodate \Planck-LFI measurements of Saturn's total system brightness, we have used the best-fit ring-disc model to subtract contributions from the planet's rings from the \Planck-LFI total system brightness measurements. The LFI data points presented in Figure~\ref{fig:pl_saturn} therefore represent best estimates for disc brightness only. The LFI measurements can be improved by incorporating data that extend past the first 2.5~surveys; however, we note that these preliminary LFI measurements appear to be in excellent agreement with the \Planck-HFI measurements.

The \wmap and \Planck\ measurements of Saturn's disc brightness in the 100--353\GHz\ frequency range seem to suggest that a roughly 5\% upward scaling, or other adjustment at 100-353 GHz, of the ESA2 model is necessary. If this overall absolute scaling is needed, the \Planck-HFI results at 545 and 857\GHz\ become somewhat discrepant with model predictions. The dip in thermodynamic temperature near 545\GHz, observed in both Jupiter and Saturn, is understood, and is due to absorption features from PH$_3$ and to a lesser extent NH$_3$. It is worth noting that the 545-GHz\ observations of Saturn are subject to significant colour corrections (see Table \ref{tab:unit_conversion}).

\subsection{Uranus and Neptune}
\label{sec:ura}

The dimmest of the Jovian planets, Uranus and Neptune are often used as calibrators for CMB experiments that probe relatively small angular scales. The thermodynamic temperature of Uranus and Neptune at millimetre wavelengths is discussed in \cite{Muhleman1991}, \cite{Griffin1993}, \cite{Serabyn1996}, \cite{Sayers2012}, and \cite{Dempsey2013}. 

The calibration of the sub-mm channels was initially performed using results from FIRAS \citep{planck2011-1.7}. Peculiar discrepancies with planet models and other inconsistencies in the FIRAS calibration, however, caused the \Planck~HFI collaboration to decide in favour of a calibration using planet models \citep{planck2013-p03f}. A detailed description of the calibration approach for the 545- and 857-GHz\ bands can be found in \cite{planck2014-a09} and \cite{Bertincourt2016}. This new calibration procedure has the advantage of being similar to that used by instruments on-board \Planck's sister experiment, \herscheln. Since the sub-mm channels are calibrated using models of Uranus and Neptune, the planet flux density results at these frequencies can only be used to check for self-consistency in flux reconstruction and relative differences between models of different planets.

Approximately 30-K variation in Uranus thermodynamic temperature at 3.5-cm wavelengths, spanning a 36 year period, is reported in \cite{Klein2006}. The evidence for brightness variations at 90 and 150\GHz\ is discussed in \cite{Kramer2008} and \cite{Hasselfield2013}. Like \actn, we see no evidence for variation in the thermodynamic temperature of Uranus and Neptune over the five \Planck-HFI observations that span a 750-day period (see Table~\ref{table:PlanetTB}). Instead, we find that our results for the thermodynamic temperature of these two ice giants are remarkably stable across observations.

A model of Uranus flux, attributed to Griffin and Orton \citep{Griffin1993}, has recently been incorporated into the set of available ESA models \citep{Moreno2014}. This model is discussed in the \act analysis of Uranus's flux \citep{Hasselfield2013}. The Griffin and Orton model of Uranus is referred to as ``ESA5" by scientists working on the calibration of SPIRE, an instrument on the \herschel satellite \citep{Moreno2014}. 

\begin{figure}[t!]
\begin{center}
\includegraphics[width=\columnwidth]{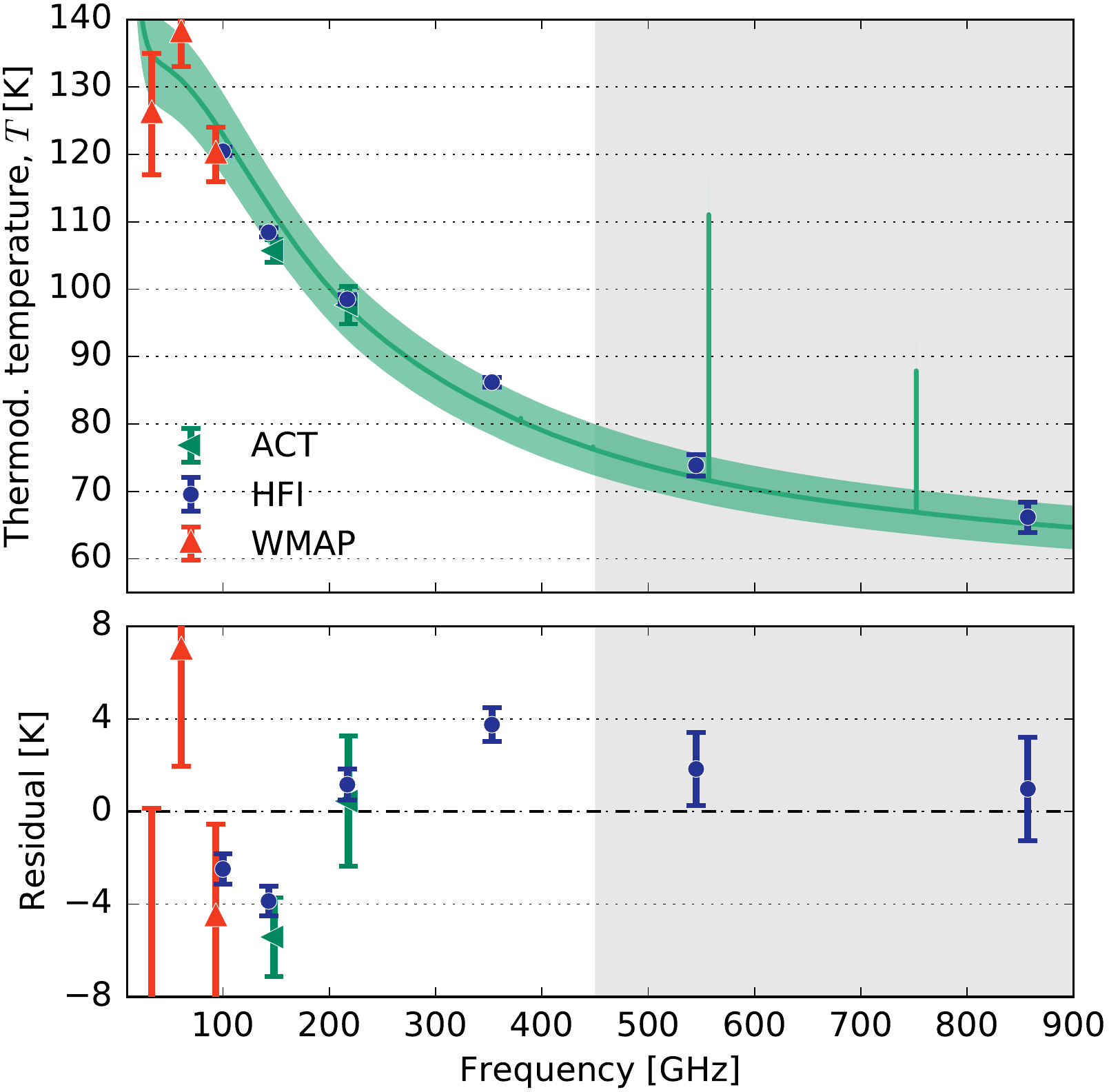}
\caption[Uranus thermodynamic temperature]{{\it Top:} Measured Uranus thermodynamic temperature derived from five \Planck-HFI observations, compared to predictions of the ESA2 model.  Emission lines are due to stratospheric emission from $\mrm{H}_2\mrm{O}$. We also show data from \wmap \citep{Bennett2013} and \act \citep{Hasselfield2013,ACT_PlanckCross2013}. {\it Bottom:} Difference (residuals) between the ESA2 model and the measurements.}
\label{fig:pl_uranus}
\end{center}
\end{figure}

\begin{figure}[t!]
\begin{center}
\includegraphics[width=\columnwidth]{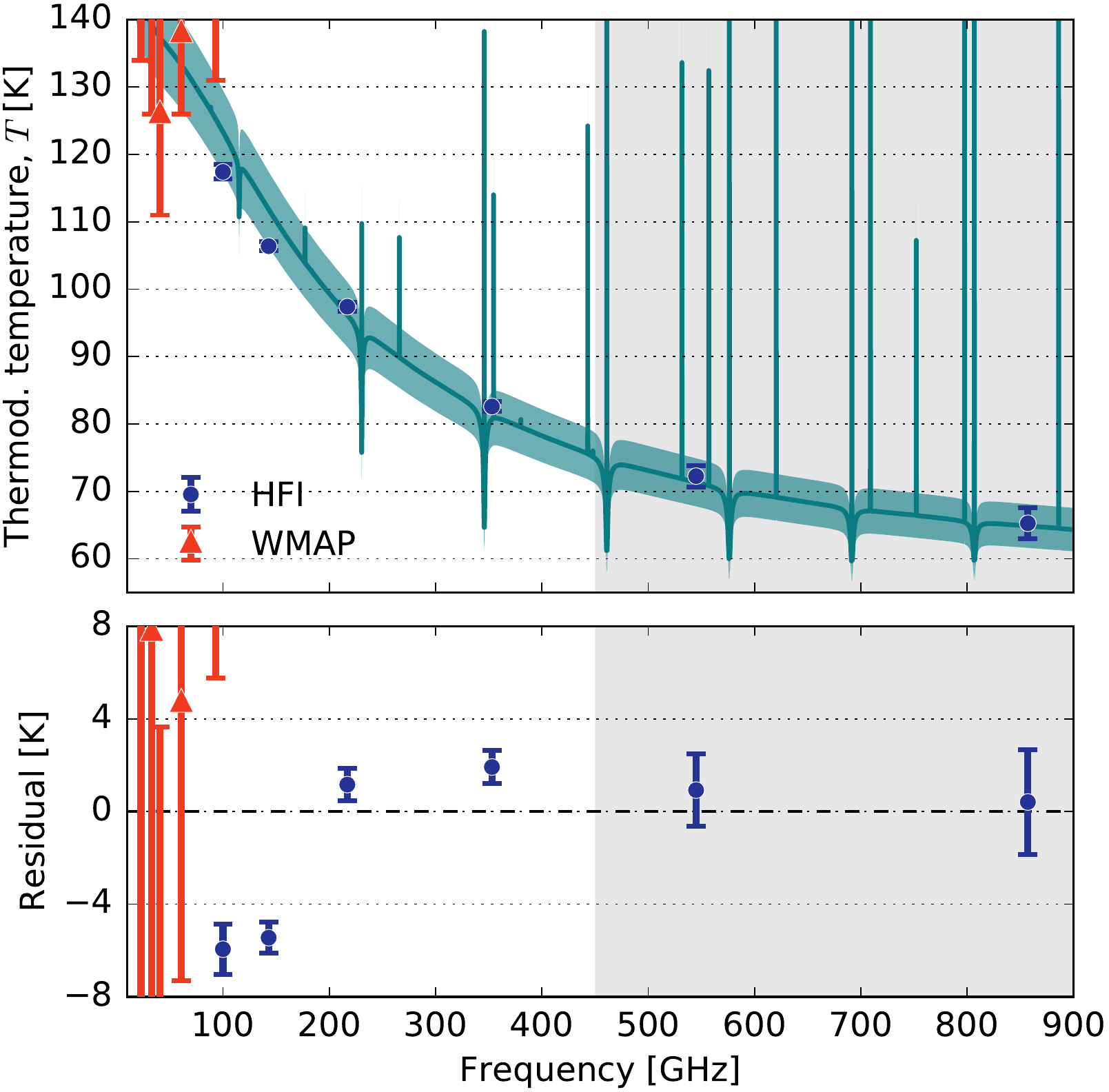}
\caption[Neptune thermodynamic temperature]{{\it Top:} Measured thermodynamic temperature for the four \Planck-HFI observations of Neptune compared to the ESA5 model. Emission lines are due to stratospheric emission from $\mrm{H}_2\mrm{O}$, CO, and HCN. We also show results from \wmap \citep{Bennett2013}. {\it Bottom:} The difference (residuals) between the ESA5 model and the measurements.}
\label{fig:pl_neptune}
\end{center}
\end{figure}

The two ice giants are quite dim at 100 and 143\GHz\ (see Table \ref{tab:planet_obs} and Fig.~\ref{fig:fit_residuals}). Unlike Mars, Jupiter, and Saturn, where instrument calibration dominates uncertainty, statistical error is significant in the flux determination of Uranus and Neptune. Figures~\ref{fig:pl_uranus} and \ref{fig:pl_neptune} show a comparison of the predicted thermodynamic temperatures of these planets with the most up-to-date ESA models at the time of writing, namely ESA2 for Uranus and ESA5 for Neptune. We note an interesting discrepancy between the \Planck-HFI measurements and model predictions for Uranus and Neptune at 100 and 143\,\GHz. This might be due to increased atmospheric absorption in the two planets at those frequencies.

\section{Comparative tests and polarization limits}
\label{sec:comparison}

\subsection{Comparison with \textit{Planck}-HFI calibration}
\label{sec:calibration_comparison}


\begin{figure}
\begin{center}
\includegraphics[width=\columnwidth]{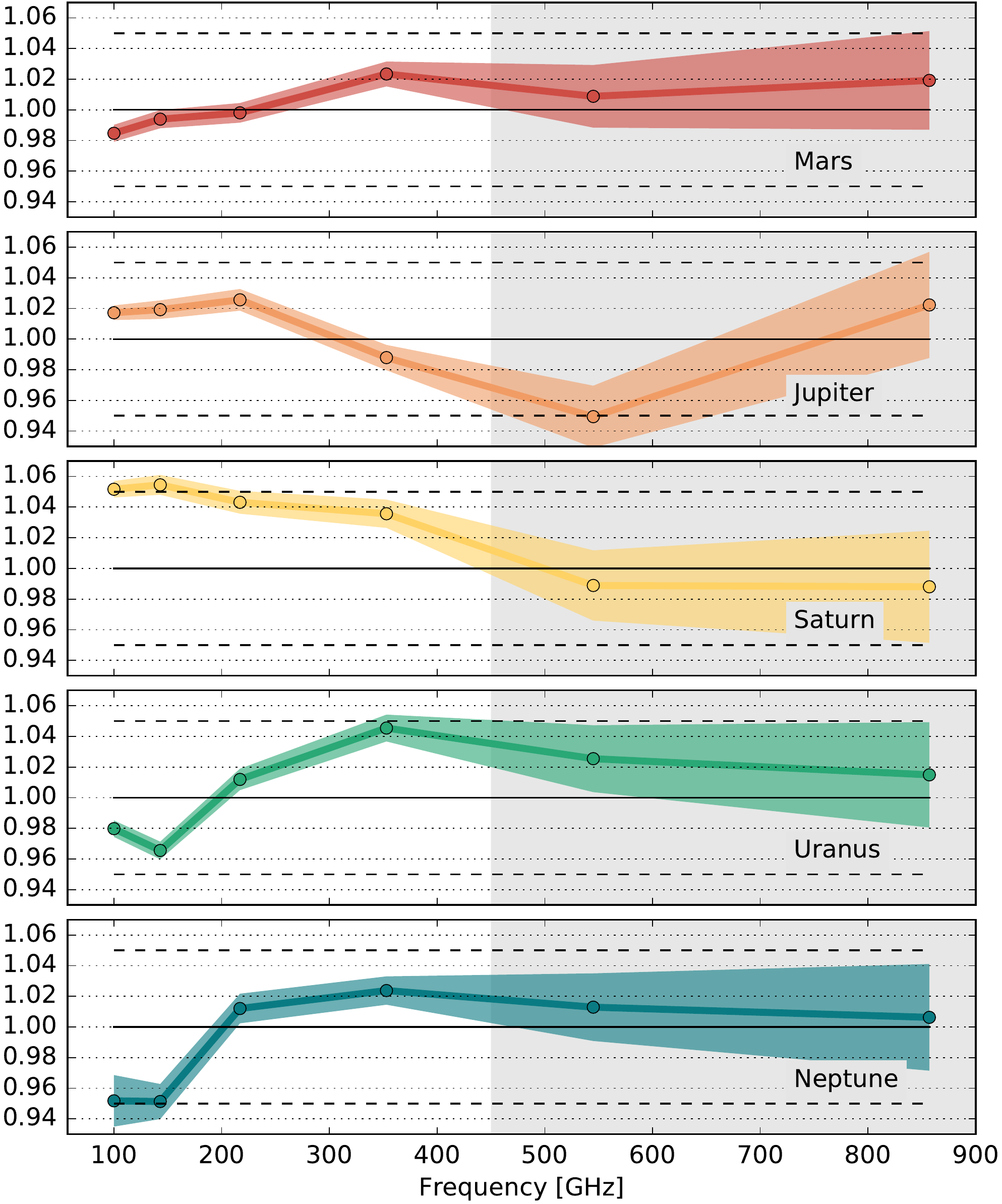}
\caption[Flux comparison]{Average ratio of measured flux (numerator) and model predicted flux (denominator) as a function of frequency for all five planets and all \Planck-HFI frequencies. For Mars, the predicted thermodynamic temperature has been scaled by ${\zeta _\mrm{P}=0.980}$. The thin solid line is the estimated flux density, combining data from all planet observations, while the larger coloured region indicates the combined systematic and statistical uncertainties. Dashed horizontal lines represent the absolute 5\,\% model error envelopes, we note that relative model errors between different frequencies are expected to be less than 5\,\%.}
\label{fig:flux_comparison}
\end{center}
\end{figure}

Figure~\ref{fig:flux_comparison} compares the ratio between measured spectral flux density and what is predicted by the adopted models for the five planets. Note that, unlike the thermodynamic temperature, this ratio is linearly dependent on parameters that commonly affect the calibration, such as the beam solid angle, gain, and colour correction. The dashed horizontal lines represent the 5\,\% absolute model errors and the coloured region represents the statistical and systematic measurement uncertainty summed in quadrature.

The average ratio between the \Planck-HFI measurements and the model predictions for all five planets (excluding Jupiter observations for frequencies above 217\GHz) is 0.997, 0.997, 1.018, 1.032, 1.009, and 1.007, for 100, 143, 217, 353, 545, and 857\GHz, respectively. Since the 545 and 857\GHz\ frequency bands derived their absolute calibration from models of the thermodynamic temperature of Uranus and Neptune, it is more appropriate to exclude those planets from such a comparison. However, since Jupiter observations at the highest frequencies are possibly affected by poorly captured detector nonlinearities and because the \cite{Lellouch2008} model for Mars likely requires a 2--4\,\% overall scaling, we are left with only the Saturn model for such a comparison. We observe a 1.1\,\% and 1.2\,\% agreement with the Saturn model at 545 and 857\GHz, respectively.

\subsection{Comparison with \wmap and \actn}
\label{sec:wmap}
Assuming constant brightness, planet observations provide an approach for cross-calibrating millimetre-wavelength observatories that is independent of the CMB. We compare the absolute calibration of \Planck~HFI with \wmap at 100\GHz\ using Mars, Saturn, and Jupiter. The uncertainties on the Uranus and Neptune thermodynamic temperatures reported by \wmap are too large to provide a useful constraint \citep{Bennett2013}. Extensive comparison of \Planck-LFI and \wmap absolute calibration using Jupiter is presented in \cite{planck2014-a06}. This analysis suggests sub-percent level agreement in the absolute calibration of the two experiments.

%sqrt((0.8/172.3)^2+(0.9/175.7)^2)
%sqrt(0.4^2+0.7^2) = 0.8
% 172.1/175.7 = 0.980
Jupiter is an ideal candidate for transferring \wmapn's dipole calibration to other instruments \citep{hill2009}. Assuming the intrinsic brightness of the planet is stable in time, we use our estimate for the 100-GHz observation-averaged thermodynamic temperature to compare with the \wmap brightness estimates at 94\GHz. The seasonally averaged Jupiter thermodynamic temperature at 94\GHz, as reported by \WMAP, is $T_\mrm{W} = {(175.7\pm0.9)}$\,K \citep{Bennett2013}. The HFI measurement at 100\GHz\ is $\widetilde{T}_\mrm{P} = (172.3\pm0.8)$\,K, combining statistical and systematic $1\sigma$ error estimates. Using the ESA1 model to scale HFI predictions down to 94\GHz, we obtain $T_\mrm{P} = (172.1\pm0.8)$\,K. The ratio is $T_\mrm{P}/T_\mrm{W} = 0.980\pm0.007$, where we have summed the errors from different experiments in quadrature. Calculating the corresponding ratio in predicted spectral radiance using the Planck blackbody function yields the same result. This suggests some tension between the two experiments at the 2.9\,$\sigma$ level.

Unlike Jupiter, Mars displays both seasonal and diurnal variations in brightness. To correct for this, we use the time-dependent model from \cite{Lellouch2008} to provide a temporal link between the \wmap and \Planck-HFI observations. For the 100--353\GHz\ channels, we find that a global rescaling of the Lellouch and Amri model predictions by $\zeta _\mrm{P} = 0.980$ minimizes the residual between the model predictions and the three available HFI measurements of Mars thermodynamic temperature. This same procedure suggests that a $\zeta _\mrm{W} = 0.968$ rescaling of the Lellouch and Amri model minimizes the residual between model predictions and \wmap results \citep{weiland2011}. The ratio of scaling factors, $\zeta _\mrm{P} / \zeta _\mrm{W} = 1.012$, suggests that \wmap and \Planck\ HFI are consistent in their absolute calibration at the percent-level. Of course, this cross-calibration procedure puts stringent requirements on the temporal stability of the validity of the Lellouch and Amri model. It is worth noting that if we fit the Lellouch and Amri model only to the 100\GHz\ results from \Planck\ HFI, we find  $\tilde{\zeta} _\mrm{P} = 0.965$, which suggest even better agreement with \wmapn. The five percent planet flux density modelling uncertainty dominates the error in this consistency check.

Saturn flux determinations are complicated by the presence of rings that can affect the effective thermodynamic temperature of the planet. From all four \Planck-HFI observations of Saturn at 100\GHz, the estimate for the Rayleigh-Jeans temperature of the Saturn disc component is $\Upsilon_\mrm{P} = (148.5\pm1.3)$~K. The \wmap estimate for the disc RJ temperature at 94\GHz\ is $\tilde{\Upsilon}_\mrm{W} = (147.3\pm1.2)$~K, which we can scale up to 100\GHz\ using the ESA2 model to find $\Upsilon_\mrm{W} = (147.5\pm1.2)$~K. The ratio is therefore $\Upsilon_\mrm{P} / \Upsilon_\mrm{W} = 1.007\pm0.010$\, suggesting the two experiments are in good agreement.

These comparisons suggest that at 100\GHz, \wmap and \Planck\ HFI agree at the 0.980, 0.996, and 1.007 level for Jupiter, Mars, and Saturn, respectively. It is important to note that these calibration comparisons are performed using sources that have quite different spectra from that of the CMB. Comparison between the calibrations of the two satellites on the CMB dipole yield sub-percent agreement \citep{planck2014-a01,planck2014-a09}.

% 108.4/(1.002/0.99745)
% 98.5/(1.049/1.036)
A measurement by the \act collaboration \citep{Hasselfield2013,ACT_PlanckCross2013}, quotes an average thermodynamic temperature for Uranus of $T_\mrm{A,148} = (105.7\pm2.2)$\,K and $T_\mrm{A,218} = (97.6\pm2.8)$\,K at 148 and 218\GHz, respectively. These estimates are derived assuming that the planet SEDs can be well approximated by a Rayleigh-Jeans form. If we apply the same approximation to the \Planck-HFI results at those frequencies we find $T_\mrm{P,148} = (107.9\pm0.8)$\,K and $T_\mrm{P,218} = (97.3\pm1.2)$\,K at 147 and 217\GHz, respectively. These results are clearly consistent.

% 145.7/132.5
% ((1.1/145.7)^2+(3.2/132.5)^2)^0.5
% 140.0/129.6
% ((1.2/140.0)^2+(4.7/129.6)^2)^0.5
On the other hand, the \act collaboration also provides an estimate of Saturn's thermodynamic temperature using the model originally described in \cite{weiland2010}. These measurements suggest $\mathcal{T}_\mrm{A,148} = (132.5\pm3.2)$\,K and $\mathcal{T}_\mrm{A,218} = (129.6\pm4.7)$\,K at 148 and 218\GHz, respectively. The \Planck-HFI estimate for the disc RJ temperature is $\mathcal{T}_\mrm{P,147} = (145.7\pm1.1)$\,K and $\mathcal{T}_\mrm{P,218} = (140.0\pm1.2)$\,K at 147 and 217\GHz, respectively. Summing errors in quadrature suggest 3.9\,$\sigma$ and 2.2\,$\sigma$ difference between the results at 147 and 217\GHz, respectively.

\subsection{Polarized flux density of planets}
HFI's bands at 100, 143, 217, and 353\GHz\ include polarization-sensitive detectors. To search for polarized flux from the planets, we decompose the total flux from the planet into the $I$, $Q$, and $U$ Stokes parameters for each individual observation within a band. We then use estimates for polarization sensitivities, efficiencies, and the measured flux from each polarization-sensitive detector to obtain the best fit Stokes parameters. The parametrization is
\begin{equation}
F_i = I + \rho_i \left( Q \cos 2\psi_i + U \sin 2 \psi_i  \right) , 
\end{equation}
where the quantities $F_i$ are the total flux density measured for bolometer $i$, and $\psi_i$ and $\rho_i$ are the angles of polarization sensitivity and polarization efficiencies of the bolometers, respectively \citep{rosset2010}. For the majority of the planet observations, the degree of polarization (defined as $p \equiv \sqrt{Q^2+U^2}/I $) is consistent with zero to within the errors in the total flux measurements. The systematic relative calibration factor between detectors (see discussion in Sect.~\ref{sec:systematic_errors}) represents the largest limitation to this analysis.
 
\begin{table}[tb!]
\begingroup
\newdimen\tblskip \tblskip=5pt
\caption{68 and 95\,\% confidence limits on polarization fraction, $p$ (in percent), as derived using the approach of \cite{Simmons1985}.}
\label{tab:pol_frac}
\vskip -5mm
\footnotesize
\setbox\tablebox=\vbox{
   \newdimen\digitwidth
   \setbox0=\hbox{\rm 0}
   \digitwidth=\wd0
   \catcode`*=\active
   \def*{\kern\digitwidth}
   \newdimen\signwidth
   \setbox0=\hbox{+}
   \signwidth=\wd0
   \catcode`!=\active
   \def!{\kern\signwidth}
   \newdimen\pointwidth
   \setbox0=\hbox{\rm .}
   \pointwidth=\wd0
   \catcode`?=\active
   \def?{\kern\pointwidth}
\halign{\hbox to 2.0cm{#\leaderfil}\tabskip 1.0em&
        \hfil#\hfil&
        \hfil#\hfil&
        \hfil#\hfil&
        \hfil#\hfil\tabskip 0pt\cr
\noalign{\doubleline}
\noalign{\vskip -1pt}
\omit\hfil Planet\hfil& 100\GHz& 143\GHz& 217\GHz& 353\GHz\cr
\noalign{\vskip 3pt\hrule\vskip 5pt}
Mars&    1.2 / 1.8& 1.1 / 1.7& 0.8 / 1.2& 1.1 / 1.7\cr
Jupiter& 1.0 / 1.3& 1.0 / 1.4& 1.1 / 1.4& 1.4 / 2.0\cr
Saturn&  0.8 / 1.2& 0.6 / 1.0& 0.8 / 1.1& 1.2 / 1.8\cr
Uranus&  2.6 / 3.6& 1.5 / 2.0& 1.2 / 1.6& 1.3 / 2.0\cr
\noalign{\vskip 3pt\hrule\vskip 5pt}
}}
\endPlancktable
\endgroup
\end{table}

Uncertainties in the determination of $Q$ and $U$ inevitably bias estimates of the polarization fraction $p \propto \sqrt{Q^2+U^2}$. To combat this, we use the formalism of \cite{Simmons1985} and \cite{Montier2014} to construct 95\,\% confidence upper limits of the degree of polarization of the planets. We do not specifically assess possible bias in our determination of polarization fraction due to the relative calibration errors. Instead, both statistical and systematic errors are propagated into the estimation of the $Q$ and $U$ covariance matrix. Table~\ref{tab:pol_frac} lists the 68 and 95\,\% upper confidence levels on the polarization fraction for the four polarization-sensitive frequency bands. 

\section{Conclusions}
\label{sec:conc}
We have provided measurements of the flux densities and thermo\-dynamic temperatures of Mars, Jupiter, Saturn, Uranus, and Neptune. These measurements span a decade in frequency, from 30--857\GHz, allowing for improved constraints on models of planetary thermodynamic temperature. We report on the data reduction scheme, as well as statistical and systematic error characterization. We detect time-variation in the thermodynamic temperature of Mars at high significance. Overall, we observe acceptable agreement with the ESA models for planet thermodynamic temperatures, however, below 545\GHz\ we do in some cases observe frequency scaling that deviates from model predictions at a significant level. Finally, comparisons with \wmap measurements show agreement in point source calibration at the two-percent-level. The largest discrepancy between \wmap and \Planck\ HFI is in the brightness determination of Jupiter.

The planet flux density measurements are limited by systematic uncertainties and these are expected to affect the brightness determination of all \Planck-HFI compact sources at the 0.4--3.1\,\% level. These systematic uncertainties do not affect the calibration of CMB diffuse emission at the same level of significance, since that uses the cosmological dipole as primary calibrator, resulting in much lower uncertainties.

These results shed light on the fidelity of the \Planck-HFI instrument characterization. We hope that current and future ground based CMB experiments that use planets to characterize their instruments will benefit from these results.

\acknowledgements{
The Planck Collaboration acknowledges the support of: ESA; CNES, and CNRS/INSUIN2P3-INP (France); ASI, CNR, and INAF (Italy); NASA and DoE (USA); STFC and UKSA (UK);
CSIC, MINECO, JA, and RES (Spain); Tekes, AoF, and CSC (Finland); DLR and MPG (Germany);
CSA (Canada); DTU Space (Denmark); SER/SSO (Switzerland); RCN (Norway); SFI (Ireland);
FCT/MCTES (Portugal); ERC and PRACE (EU). A description of the Planck Collaboration
and a list of its members, indicating which technical or scientific activities they
have been involved in, can be found at \href{http://www.cosmos.esa.int/web/planck/planck-collaboration}{\texttt{http://www.cosmos.esa.int/web/planck/planck-collaboration}}. The authors thank Lyman Page for useful discussion. J.E. Gudmundsson acknowledges support by Katherine Freese through a grant from the Swedish Research Council (Contract No. 638-2013-8993).
}

\bibliographystyle{aat}
\bibliography{Kuiper_bib,Planck_bib}

\raggedright % Helps to keep the margins from overflowing
\end{document}